\documentclass[ reprint, twocolumn, superscriptaddress, amsmath,amssymb, aps,prb]{revtex4-2}

\usepackage{graphicx}
\usepackage{dcolumn}
\usepackage{bm}
\usepackage{lmodern}
\usepackage[usenames,dvipsnames]{color}
\usepackage[normalem]{ulem}
\usepackage[colorlinks = true,
            linkcolor = Black,
            urlcolor  = Blue,
            citecolor = Black,
            anchorcolor = Blue]{hyperref}

\usepackage{multirow}
\usepackage{nicefrac}
\usepackage{siunitx}
\usepackage{stmaryrd}
\usepackage{subcaption}
\usepackage{textcomp}
\usepackage{braket}

\begin{document}

\title{Generative design of functional organic molecules for terahertz radiation detection}

\author{Zsuzsanna Koczor-Benda}
\email{zsuzsanna.koczor-benda@warwick.ac.uk}
\affiliation{Department of Chemistry, University of Warwick, Coventry, CV4 7SH, United Kingdom}

\author{Shayantan Chaudhuri}%
\affiliation{Department of Chemistry, University of Warwick, Coventry, CV4 7SH, United Kingdom}
\affiliation{School of Chemistry, University of Nottingham, Nottingham, NG7 2RD, UK}
\author{Joe Gilkes}%
\affiliation{Department of Chemistry, University of Warwick, Coventry, CV4 7SH, United Kingdom}
\affiliation{Centre for Doctoral Training in Modelling of Heterogeneous Systems, University of Warwick, Coventry, CV4 7AL, UK}
\author{Francesco Bartucca}
\affiliation{Department of Chemistry, University of Warwick, Coventry, CV4 7SH, United Kingdom}
\author{Liming Li}
\affiliation{Department of Chemistry, University of Warwick, Coventry, CV4 7SH, United Kingdom}
\author{Reinhard J. Maurer}%
 \email{r.maurer@warwick.ac.uk}
\affiliation{Department of Chemistry, University of Warwick, Coventry, CV4 7SH, United Kingdom}
\affiliation{Department of Physics, University of Warwick, Coventry, CV4 7AL, United Kingdom}

\date{\today}

\begin{abstract}
Plasmonic nanocavities are molecule-nanoparticle junctions that offer a promising approach to upconvert terahertz radiation into visible or near-infrared light, enabling nanoscale detection at room temperature. However, the identification of molecules with strong terahertz-to-visible frequency upconversion efficiency is limited by the availability of suitable compounds in commercial databases. Here, we employ the generative autoregressive deep neural network, G-SchNet, to perform property-driven design of novel monothiolated molecules tailored for terahertz radiation detection. To design functional organic molecules, we iteratively bias G-SchNet to drive molecular generation towards highly active and synthesizable molecules based on machine learning-based property predictors, including molecular fingerprints and state-of-the-art neural networks. We study the reliability of these property predictors for generated molecules and analyze the chemical space and properties of generated molecules to identify trends in activity. Finally, we filter generated molecules and plan retrosynthetic routes from commercially available reactants to identify promising novel compounds and their most active vibrational modes in terahertz-to-visible upconversion.
\end{abstract}

\maketitle

\section{Introduction}
Terahertz (THz) radiation has applications in numerous fields, including medical diagnostics, security screening, communications, and astronomy~\cite{TonouchiNP07, DhillonJPD17}. Historically, the development of both powerful and affordable light sources, and efficient THz detectors, has been technologically challenging. 

Nanoscale, room-temperature detection of terahertz and mid-infrared radiation is enabled by molecular optomechanical devices utilizing the enhancement of electronic fields in plasmonic nanocavities to convert terahertz radiation into visible or near-infrared light~\cite{RoelliNN16, RoelliPRX20}. These nanocavities can be assembled on silicon-based photonic integrated circuits ~\cite{redolat2024synthesis}, opening possibilities for low-cost fabrication and multiplexed detection. To enhance the light-matter interaction, molecules are typically placed between two metallic nanoantennas~\cite{RoelliPRX20, xomalis_detecting_2021, chen_continuous-wave_2021}. One of the two antennas focuses terahertz radiation at the design frequency over the molecular sample volume to enhance the absorption of terahertz radiation via the surface-enhanced infrared absorption~\cite{NeubrechCR17} mechanism. The second optical antenna confines visible or near-infrared light to volumes below 100~\si{\nm\cubed}, which induces surface-enhanced Raman scattering~\cite{StilesARAC08} of molecules within the plasmonic nanocavity. Absorption of THz radiation by molecules within the nanocavity results in the vibrational excitation of a specific normal mode, which leads to an increase in the measured Raman anti-Stokes intensity of the same normal mode, similar to resonant sum-frequency generation spectroscopy~\cite{HumbertMaterials19}. For centrosymmetric molecules, simultaneous activity in absorption and Raman scattering is not possible. Even in asymmetric molecules, it is rare to have vibrational modes that can efficiently upconvert the THz radiation signal, as this requires a large change in both electronic dipole moment and in polarizability along the vibrational mode. Vibrational modes of organic molecules in the THz frequency range are often delocalized across several functional groups or across molecules, which makes it challenging to use chemical intuition to suggest promising candidates or define molecular design rules. This makes it necessary to use quantum chemical calculations in connection with computational screening or design approaches to identify good candidate molecules and their active vibrational modes~\cite{Koczor-BendaPRX21,Koczor-BendaJPCA22}. Such computational predictions motivate more detailed experimental investigations for the fabrication and application of new molecular optomechanical devices~\cite{redolat2024synthesis}.

Machine learning (ML) methods can facilitate the design and discovery of new functional materials by enabling the fast computational screening of large structural databases~\cite{Gomez-BombarelliNM16, SahuJMCA19, SaekiJJAP20}. ML-based screening has previously been used to identify promising candidates for THz radiation detection from commercially available compound databases~\cite{Koczor-BendaPRX21}. However, a drawback of this approach was that there was a limited search pool of molecules that have an affinity to the gold surfaces of the nanoantennas used in detector prototypes. Self-assembled monolayers of thiol-containing molecules have been shown to have high stability and reproducibility on gold surfaces~\cite{Checkik99}, which are often used in plasmonic devices. It is therefore prudent to focus on thiol-containing molecules that are commercially available or easily synthesizable. These requirements pose a challenge for high-throughput screening methods as the number of thiol compounds within large commercial databases is relatively low, with only around \num{150000} out of more than 20 million compounds in the eMolecules database and \num{32000} out of 8 million compounds from the MolPort database identified in \citet{Koczor-BendaPRX21} being monothiols, respectively.

An alternative solution for accelerating the discovery of promising molecules is generative deep learning, which in the past has been used for the property-driven design of functional organic molecules~\cite{WestermayrNCS23, GebauerNC22_cG-SchNet, JoshiJPCB21_3D-Scaffold, Sanchez-LengelingScience18, Gomez-BombarelliACSCS18, MeyersDDT21}. Most proposed generative deep learning models use text-based or two-dimensional (2D) molecular representations~\cite{Arús-PousJC20, KongFP22}. G-SchNet is a generative autoregressive deep neural network that has the advantage of being able to generate molecules in three-dimensional (3D) space~\cite{GebauerNeurIPS19_G-SchNet}. Previous studies have shown that G-SchNet can be iteratively biased to generate molecules satisfying certain target properties. \citet{WestermayrNCS23} coupled G-SchNet with a neural network that predicts molecular quasiparticle energies~\cite{westermayr_machine_2021} to bias molecular generation towards small fundamental gaps, low ionization potentials, or high electron affinities, while conserving low synthetic complexity of the molecules. \citet{GebauerNC22_cG-SchNet} developed conditional G-SchNet, which, in addition to structures, trains on electronic property and structural motif labels in order to condition molecular generation.

In this paper, we perform property-driven generative design of functional organic molecules for THz radiation detection using G-SchNet by driving the generative model to create novel molecules with high frequency-upconversion efficiency, affinity to gold surfaces, and synthetic accessibility. 
To predict the upconversion properties of molecules, we use the target property P introduced in \citet{Koczor-BendaPRX21}, which is based on the total spectral intensity in a wide frequency window (30--1000~\si{\per\cm}) relevant for THz and mid-infrared applications. To increase the pool of candidates for this application, we train G-SchNet models on a dataset of around \num{30000} thiol-containing molecules and generate hundreds of thousands of monothiolated molecules by iterative biasing. We analyze chemical trends in the generated databases and identify functional groups that correlate with high upconversion intensity. Previously used ML predictors of the frequency upconversion efficiency based on molecular fingerprints~\cite{Koczor-BendaPRX21} become unreliable as the property-driven generative biasing workflow explores novel molecules beyond the training dataset. We replace them with more transferable equivariant graph neural network (GNN) models that make use of the 3D molecular conformations that G-SchNet generates. To train these models, we use calculations based on density functional theory (DFT) for P values contained in Molecular Vibration Explorer~\cite{Koczor-BendaJPCA22}, which are available for around 2800 gold-thiolate molecules, and extend this database with new DFT calculations on generated molecules. Finally, highly spectroscopically active compounds are identified by generative design and further validated with quantum chemistry calculations and retrosynthetic route planning to identify promising, novel compounds for THz radiation detection.

\section{Methods}

\subsection{Generative machine learning}

\paragraph*{Training dataset.} A training dataset of \num{29246} monothiolated molecules was compiled from the eMolecules~\cite{eMolecules} commercial molecular database, that was previously used by \citet{Koczor-BendaPRX21}. This database contains over 20 million readily available or custom synthesized compounds from over 15 suppliers, aimed mainly at drug discovery applications \cite{eMolecules}. This training dataset was selected to ensure that the generative model creates molecules that are chemically similar to known synthesizable compounds, thus facilitating the search for viable candidates. The eMolecules database was first filtered for monothiols based on the corresponding SMARTS pattern. Charged molecules and duplicates were removed, resulting in an initial pool of \num{147623} molecules containing the following elements: hydrogen (H), boron (B), carbon (C), nitrogen (N), oxygen (O), fluorine (F), silicon (Si), phosphorus (P), sulfur (S), chlorine (Cl), selenium (Se), bromine (Br), tin (Sn), and iodine (I). 
In contrast to \citet{Koczor-BendaPRX21}, molecular size and number of rotatable bonds were not restricted, resulting in a larger pool of molecules. Initial 3D structures for the unique monothiolated molecules were created from Simplified Molecular Input Line Entry System (SMILES) strings \cite{SMILES} and relaxed with the MMFF94 Merck molecular force field~\cite{halgren1996merck} using the RDKit package~\cite{RDKit}. To maximize chemical diversity, a Smooth Overlap of Atomic Positions (SOAP)~\cite{SOAP} descriptor with a local region cut-off of \SI{4.0}{\angstrom}, 4 radial basis functions, and a maximal degree of spherical harmonics of 3 was calculated for each molecule (resulting in \num{6384} features), using the DScribe package~\cite{dscribe}. After singular value decomposition with 500 components, 
\num{30000} clusters were identified with $k$-means clustering using the scikit-learn~\cite{scikit-learn} library. For each cluster, the molecule closest to the cluster center was selected. Molecules that had already been calculated in the THz database were removed (604 duplicates), resulting in the final training set of \num{29246} molecules. Structure optimization was performed with the xTB software package using the GFN2-xTB parametrization~\cite{GFN2-xTB}, based on which the final database of 3D geometries for the generative model was constructed. For a discussion on using the computationally less expensive xTB method instead of commonly used DFT for structure optimisation, and a comparison of generated unrelaxed and relaxed structures see section S10 in the ESI.

\paragraph*{Training workflow}
The schnetpack-gschnet~\cite{schnetpack-gschnet, SchüttJCP23_SchNetPack2.0} package was used to train G-SchNet models on the aforementioned training database. Each G-SchNet model was trained using a SchNet~\cite{SchüttNeurIPS17_SchNet} neural network with 128 features, 9 interaction blocks, a cut-off of \SI{10}{\angstrom} and 25 centers for the radial basis expansion of distances. A learning rate of 0.0001 was used and 5 random atom placements per molecule per batch were drawn. For all trained G-SchNet models, data were randomly split (as implemented within schnetpack-gschnet) 80\%/10\%/10\%  for training, validation and testing, respectively. Approximately \num{100000} molecules were generated with each trained model, with a maximum molecular size of 60 atoms. Non-unique, disconnected, or invalid (incorrect valency) generated molecules were discarded. Molecules were filtered to only contain one thiol group, which can act as the linker to the gold nanoantenna in a THz radiation detector device. The number of molecules generated and remaining after filtering are summarized in the ESI (Table~SI). 

\paragraph*{Iterative biasing of G-SchNet.} The generation of molecules with desired properties was achieved by an iterative workflow similar to the one proposed by \citet{WestermayrNCS23} Herein, in each iteration, the G-SchNet model is trained, molecules are generated, molecules are filtered with a property prediction model, and a new training dataset is built that contains the original and a subset of the novel generated molecules with selected properties above or below a certain threshold value. As a result, molecular generation is iteratively biased towards molecules with desired properties. In each iteration, G-SchNet was trained (from scratch) with the modified dataset. The sizes of the training databases for each of the six biasing iterations are detailed in Table~SII in the ESI. 

In each iteration, molecules were selected according to two properties: THz upconversion efficiency, predicted with a previously trained Kernel Ridge Regression (KRR) model~\cite{Koczor-BendaPRX21}, and the SCScore metric of synthetic complexity~\cite{SCScore}. The upconversion efficiency figure of merit, P, is defined as the logarithm of the orientation-averaged upconversion intensity ($I^c_m$) summed over all $M$ vibrational frequencies in the 1--30~\si{\tera\hertz} frequency window (30--1000~\si{\per\cm}):~\cite{Koczor-BendaPRX21}
\begin{equation}
    \mathrm{P} = \log\left( \sum_{m\in M} \braket{I^c_m} \right)
    \label{eq:P}
\end{equation}
Higher P values correspond to greater total frequency upconversion intensity of vibrations in the selected frequency range. $I^c_m$ is based on the absorption and Raman scattering intensities of vibrational mode $m$ (a full definition of $I^c_m$ can be found in section~S2 of the ESI). $I^c_m$ was calculated using DFT for a simplified model of the molecule-metal interface in \citet{Koczor-BendaPRX21} resulting in the Gold database of Molecular Vibration Explorer \cite{Koczor-BendaJPCA22}, and used as training data for the KRR model \cite{Koczor-BendaPRX21}. We also discuss the details of these DFT calculations in the next section. 

The SCScore neural network by \citet{SCScore} was trained on 12 million reactions from the Reaxys~\cite{Reaxys} database. The SCScore correlates with the number of reaction steps required to synthesize the molecule from reasonable starting materials and ranges between 1 and 5, where higher numbers indicate reduced synthesizability~\cite{SCScore}. Canonical SMILES~\cite{SMILES} representations of molecules generated using Open Babel~\cite{Open_Babel} were used as input for the KRR predictor and the SCScore calculator.  To simultaneously bias molecular generation towards large P (high THz upconversion efficiency) and low SCScore (S, low synthetic complexity) values, molecules with properties satisfying both $\mathrm{P} \geq \overline{\mathrm{P}} + 0.5\sigma_\mathrm{P}$ and $\mathrm{S} \leq \overline{\mathrm{S}} - 0.5\sigma_\mathrm{S}$ were appended to the training dataset for the subsequent training iteration, where $\overline{\mathrm{X}}$ and $\sigma_\mathrm{X}$ are the mean average and standard deviation, respectively, of property X. 

\paragraph*{Reference calculations and property predictors.}
As reference data for the ML models, a database of about \num{2800} gold-thiolate molecules, available from Molecular Vibration Explorer~\cite{Koczor-BendaJPCA22}, was used, henceforth referred to as the `THz database'. This database was originally compiled in \citet{Koczor-BendaPRX21} and contains P values calculated with Kohn-Sham DFT~\cite{Hohenberg-Kohn, Kohn-Sham}, using the B3LYP~\cite{becke1988density,lee1988development} hybrid generalized gradient approximation, the DFT-D3~\cite{grimme2010consistent} dispersion correction, the Karlsruhe basis set with split valence polarization (def2-SVP)~\cite{weigend2005balanced}, and a tight energy convergence threshold. The molecules were modeled as gold-thiolates to consider the most immediate chemical effects of the metal-molecule interface. This choice of modeling was also validated against surface-enhanced Raman spectroscopy measurements in \citet{griffiths2021resolving}, \citet{boehmke2024uncovering}, and \citet{wright2021mechanistic}. \citet{Koczor-BendaPRX21} validated the computational approach in detail against Raman and infrared measurements for powder, solution and nanoparticle-on-mirror constructs of a set of test molecules and found that individual spectral features as well as surface-enhanced Raman spectroscopy intensities integrated over a wide spectral window correlate well with measurements. However, for an accurate modeling of low-frequency vibrational features (below 200~\si{\per\cm}), considering the metal facets as well as molecule-molecule interactions becomes necessary~\cite{boehmke2024uncovering}, which increases computational costs. To enable a fast computational assessment of a large number of molecules, and benefit from the existing, openly available database we follow the approach of \citet{Koczor-BendaPRX21}.
To assess the accuracy of ML property predictors along the biasing iterations, additional reference calculations at the same level of theory were performed whereby the thiol group in each molecule was modified to a gold-thiolate group. The Gaussian16~\cite{g16} software package was used to run DFT calculations and analysis tools from Molecular Vibration Explorer~\cite{Koczor-BendaJPCA22} were used to calculate P values.
The pretrained KRR model from \citet{Koczor-BendaPRX21} was used to predict P values; additionally, PaiNN~\cite{Schütt21ICML_PaiNN} and MACE~\cite{BatatiaNeurIPS22_MACE} equivariant GNN models were trained on the P values of the DFT-optimized structures of the THz database. Full details of training and hyperparameter optimization, as well as learning curves, are provided in the ESI (Table SIII and SIV, Figures S1-S3). The PaiNN and MACE predictions of P values are based on the unrelaxed 3D structures of generated molecules, following \citet{WestermayrNCS23}. Section 10 of the ESI discusses the effect of using unrelaxed structures on the predicted P values for a subset of generated molecules.

\subsection{Dimensionality reduction and clustering}
To visualize the chemical space spanned by molecules within various datasets and to create inputs for subsequent cluster analysis, dimensionality reduction via principal component analysis (PCA) was applied. The inputs for PCA were one of two applied molecular descriptors, henceforth referred to as bonding and structural descriptors. Structural descriptors were averaged SOAP~\cite{SOAP} descriptors, obtained using the DScribe~\cite{HIMANEN2020} package, which results in a \num{50820}-dimensional description of molecules that encodes the average atomic environment around each atom. To obtain bonding descriptors from molecules, the Open Babel~\cite{Open_Babel} and RDKit~\cite{RDKit} software packages were used to extract as many interesting features as possible relating to molecular bonding. These ranged from simple quantities, such as the number of different elements within the molecule, to complex quantities such as the molecular aromaticity, resulting in a 403-dimensional bonding descriptor. Descriptor vectors were calculated for each molecule of the training database and used as inputs for PCA. To visualize the chemical space spanned by the training database in comparison with the spaces spanned by the generated molecules, the descriptor for generated molecules was represented using the same principal components as obtained from the training database. For clustering, a mixture of the balanced iterative reducing and clustering using hierarchies (BIRCH)~\cite{BIRCH} data mining algorithm and agglomerative clustering~\cite{DBSCAN} was used to allow for uneven cluster sizes. Clustering was performed across the first three principal components of the bonding and structural descriptors, in addition to the PaiNN-predicted P values, weighted to achieve an approximately equal contribution of the first principal components of each descriptor and the predicted P value across all clusters.

\subsection{Retrosynthetic planning}
The AiZynthFinder~\cite{GenhedenJC20_AiZynthFinder} software was used for the retrosynthetic planning of select molecules. The retrosynthesis algorithm is based on a Monte Carlo tree search that recursively breaks down a molecule to existing precursor molecules~\cite{GenhedenJC20_AiZynthFinder} based on a stock from compounds available within the ZINC~\cite{SterlingJCIM15_ZINC15} database. The tree search itself is guided by a policy that suggests possible precursors by utilizing a neural network trained on a library of known reaction templates. The employed policy~\cite{ThakkarCS20} was trained on US patent office data~\cite{Lowe17}, as available within AiZynthFinder. The SMILES strings of molecules with successful retrosynthetic routes were cross-referenced against the PubChem~\cite{PubChem, KimNAR25} database using the PubChemPy~\cite{PubChemPy} package.

\section{Results and Discussion}

\subsection{Analysis of generated molecules} 

\begin{figure*}
    \centering
    \includegraphics[width=0.95\linewidth]{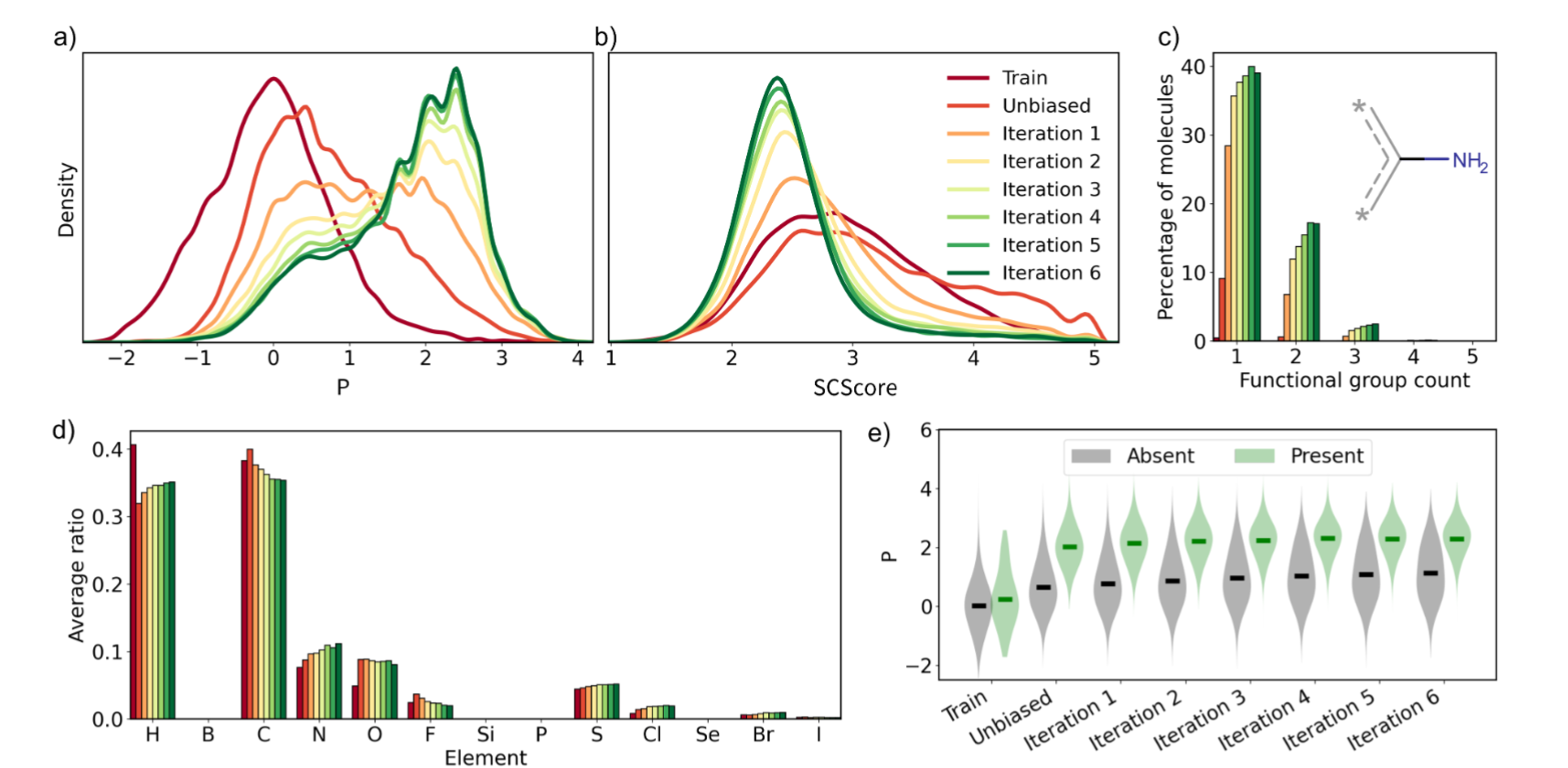}
    \caption{Distribution of (a) predicted P values and (b) SCScore for molecules used for training G-SchNet (Thiol database) and molecules generated in the biasing iterations. (c) Increase in relative occurrence and number of aromatic amine groups in molecules through the biasing iterations, (d) the average elemental composition of training and generated molecules; and (e) the distribution and mean average of P values predicted by the KRR model for molecules in which an aromatic amine is absent or present.}
    \label{fig:biasing}
\end{figure*}

The G-SchNet generative model is initially trained on the original dataset and used to generate novel and `unbiased' molecules. A subset of the generated molecules is selected according to their predicted THz upconversion efficiency (high P value) and synthetic complexity (low SCScore) and added to the dataset. This process is repeated in six successive iterations during which properties of the generated molecules are driven towards the desired ranges (Figure~\ref{fig:biasing}a and b). Iterative biased generation of molecules successfully leads to molecules with higher P and lower SCScore in later iterations when compared to the training dataset (`Train') and the unbiased initial generation (`Unbiased'). Further shifts in property values after iteration 5 were not significant and biasing was stopped after Iteration 6.

The composition of generated compounds differs significantly from the training set, as shown by the elemental composition of molecules in Fig.~\ref{fig:biasing}d. The differences are largest between the training set and the unbiased generated molecules, which highlights the fact that G-SchNet, without biasing or conditioning, does not fully reproduce the chemical features of its training set. This shortcoming has been previously observed by \citet{WestermayrNCS23} and  \citet{gebauer_autoregressive_2024}. This effect is more significant for models trained on diverse datasets featuring many elements and molecular sizes than for models trained on small and simple molecules (such as QM9~\cite{RuddigkeitJCIM12, RamakrishnanSD14}). The unbiased generated molecules feature a significantly reduced proportion of hydrogen atoms compared to the training dataset, which suggests increased numbers of unsaturated bonds and heteroatomic groups. The proportion of hydrogen atoms slightly increases through the subsequent biased iterations. Nitrogen atoms also become more prevalent in generated sets, while the proportion of carbon and fluorine atoms decreases. There is a shift of the size distribution of molecules to smaller values, as shown in Fig.~S4a. While unbiased generation creates significant numbers of molecules with 30--60 atoms, generated molecules in later iterations have, on average, about 20 atoms.  A significant number of molecules generated by the unbiased model have an SCScore above 4 (Fig.~\ref{fig:biasing}b), which was also observed by \citet{WestermayrNCS23}. We note that all training molecules are commercially available so the SCScore metric does not fully reflect their accessibility but rather was used as an indicative metric by which we filter out generated molecules that are overly complex. For the most promising generated candidate molecules, we perform comprehensive retrosynthetic planning analysis to assess their synthesizability more accurately (\textit{vide infra}).

As the training database only contained monothiols, the proportion of thiols in generated molecules is high, around 65\% in the unbiased case, which increases in subsequent iterations to around 85\%, as shown in Fig.~S5 in the ESI. It is interesting to see that the frequency of certain functional groups is significantly increased throughout the biasing iterations. An example of this is the aromatic amine group, which is present in only 0.5\% of training molecules, but found in 9.8\% of molecules generated by the unbiased G-SchNet model (Fig.~\ref{fig:biasing}c). By Iteration 6,  58.7\% of generated molecules contain one or more aromatic amine groups. Simultaneously, the number of instances of this functional group per molecule also increases with iterations, as shown in Fig.~\ref{fig:biasing}c, with some of the generated molecules having as much as five aromatic amine groups. This functional group was identified by \citet{Koczor-BendaPRX21} to correlate with high P values according to the ML predictor and as shown in Fig.~\ref{fig:biasing}e, the presence of this functional group also correlates with significantly higher predicted P values. We note that the sudden increase in the presence of this and other functional groups between the training and the unbiased generated molecules could explain the significant shift in the predicted P value distribution between the two sets in Fig. \ref{fig:biasing}a.


\subsection{Evaluation and improvement of property predictors}

As shown above, generated molecules significantly differ in chemical composition from the training molecules. This raises the question of whether the KRR predictor of the THz upconversion efficiency metric, P, provides transferable prediction accuracy for the novel, generated molecules -- a crucial prerequisite for targeted property-driven molecular design. To assess this, DFT structure optimizations and vibrational spectrum calculations were performed on randomly selected molecules from the Thiol database that was used to train the G-SchNet model and from the dataset generated in Iteration 6. Table~\ref{tab:ml_mae} shows the performance of the KRR predictor on these molecules. The mean absolute error (MAE) on the Thiol database is similar to the MAE on the test set of the THz database, while the MAE increases significantly for molecules generated in Iteration 6. In particular, the KRR model severely underestimates the P values of high-P molecules, as shown in the ESI (Figure~S6), which suggests that the true P values of molecules generated in the biasing workflow reach much higher values than what is predicted in Fig.~\ref{fig:biasing}a. 

As the KRR predictor uses SMILES strings as input and is based on 2D Morgan fingerprints, it does not benefit from the information contained in the 3D structures generated by G-SchNet. As the THz upconversion efficiency sensitively depends on the molecular conformation and vibrational frequencies, this limits the expressiveness and prediction accuracy of the model. We therefore trained two equivariant GNN models with 3D atom-wise embeddings on the same THz dataset, namely the MACE and PaiNN models. Table~\ref{tab:ml_mae}  compares the MAE of the different ML models for the reference DFT-calculated P values, determined for the DFT-optimized structures of test molecules from the THz dataset. Both MACE and PaiNN provide improved predictions compared to the EN and KRR models of \citet{Koczor-BendaPRX21}, with PaiNN providing the best prediction. PaiNN also learns faster than MACE from less data, as shown by the learning curves in the ESI (Fig.~S3); for this reason, the PaiNN predictor was used for all subsequent analyses. When testing the PaiNN model on the molecules generated in Iteration 6, the MAE is larger with 0.73 (Table \ref{tab:ml_mae}). PaiNN also underestimates the P values of high-P value molecules, as shown in the ESI (Fig.~S6), though this is slightly less pronounced than with KRR. Therefore, all tested models show reduced prediction accuracy when applied to the iteratively biased datasets, suggesting that the models are forced to predict outside of the chemical space spanned by the training data. This severely limits their ability to act as a transferable property predictor that drives molecule generation. The deterioration of the model accuracy for the THz upconversion efficiency is more significant than what was observed by \citet{WestermayrNCS23} for electronic property prediction. We hypothesize that this is due to the integrated nature of the THz upconversion metric P and its sensitive dependence on collective low-frequency molecular vibrations and the molecular polarizability. 

\begin{table}
    \centering
    \begin{tabular}{c|ccc}
    & \multicolumn{3}{c}{Dataset} \\
        Model           & THz & Thiol &Iteration 6\\
        \hline
        EN \cite{Koczor-BendaPRX21}  & 0.60 & -- & -- \\
        KRR \cite{Koczor-BendaPRX21}          & 0.59 & 0.62 & 0.89 \\
        MACE            & 0.46 & -- & -- \\
        PaiNN           & 0.41 & 0.53 & 0.73 \\
    \end{tabular}
    \caption{Performance of different ML models for P prediction, reported as mean absolute error for test molecules from the THz database, Thiol database, and molecules generated in Iteration 6. EN and KRR models are taken from \citet{Koczor-BendaPRX21} with predictions based on SMILES strings of molecules. In the case of MACE and PaiNN, predictions are based on DFT-optimized molecular structures.}
    \label{tab:ml_mae}
\end{table}

To alleviate the problem of underestimated high P values and the lack of transferability of the PaiNN predictor across the biased generation runs, the PaiNN predictor was retrained on a random subset of DFT-calculated P values from molecules generated in Iteration 6 and molecules from the Thiol database.  A committee of 5 PaiNN models was trained on different train/validation splits, and the mean average and standard deviation of their predictions were analyzed (ESI, Fig.~S7). The standard deviation of predictions was found to not correlate strongly with the absolute error of the prediction, indicating that the uncertainty of predictions cannot be used in an active learning-type workflow for augmenting the training set in a data-efficient way. After retraining, the mean average of the prediction becomes significantly more accurate for high P values, as shown in the ESI (Fig.~S8). The retrained PaiNN model achieves an MAE of 0.43 in P prediction on the Iteration 6 dataset which is consistent with the MAE previously achieved on the validation set when training on only the THz dataset (Table~1). 

\begin{figure}
    \centering
    \includegraphics[width=\linewidth]{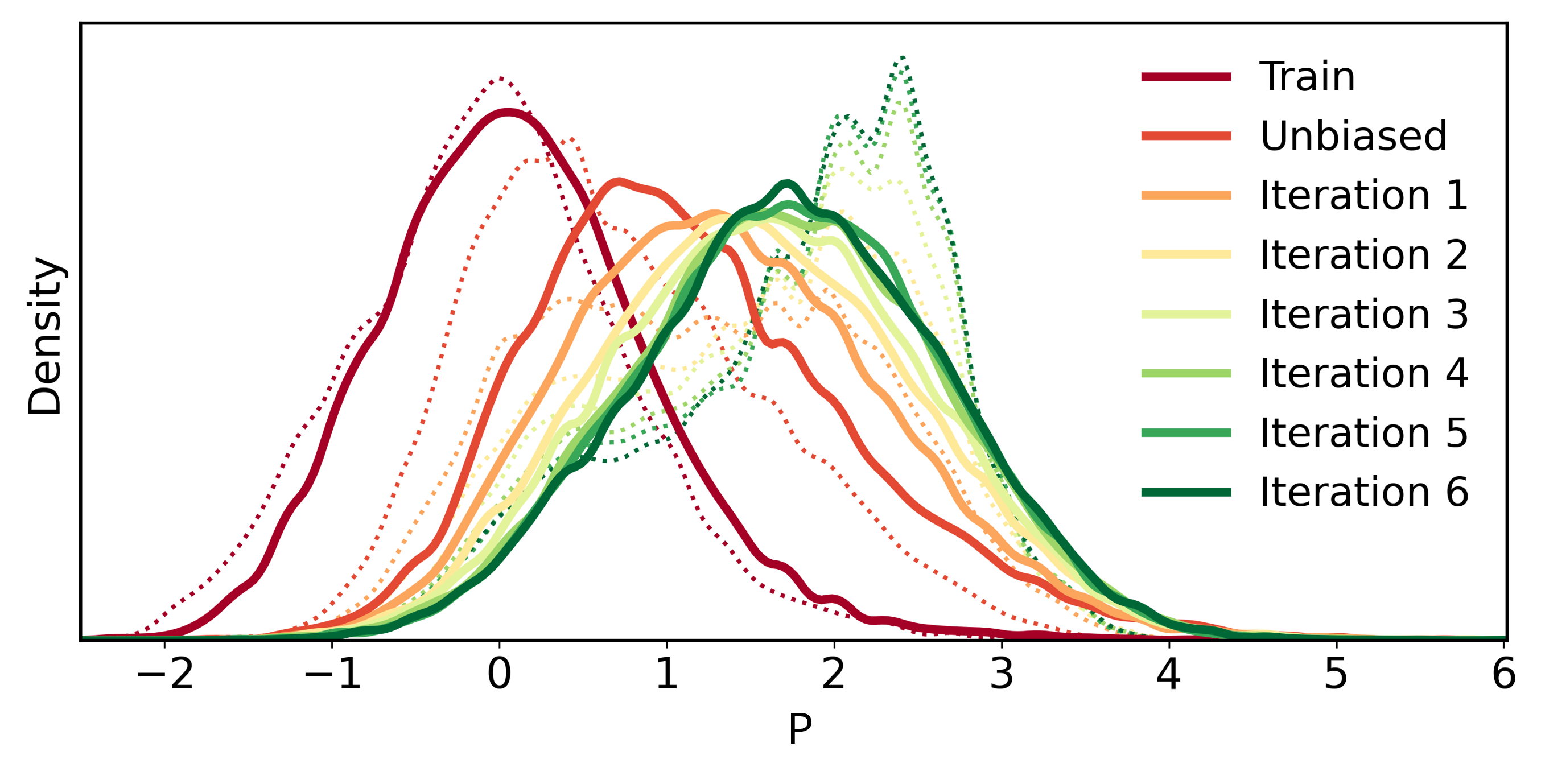}     
    \caption{Distribution of PaiNN predictions (full lines) and original KRR predictions (dotted lines) for P values on all training and generated molecules. In the case of PaiNN, the distributions show the mean predicted P value by a committee of 5 PaiNN models that were trained on the original THz database augmented by randomly selected molecules from the G-SchNet training database (Thiols) and molecules generated in Iteration 6.}
    \label{fig:painn_P_distribution}
\end{figure}

Equipped with a robust and transferable P predictor, new P values were predicted using the committee of 5 PaiNN models for all molecules in the training and generated molecule datasets (Fig.~\ref{fig:painn_P_distribution}). Compared to the KRR predictions, the distribution of PaiNN-predicted P values for the generated molecules shifts to significantly higher values, with the highest predicted P value reaching 7.30. The presence of specific functional groups can be analyzed alongside the PaiNN predictions for P values. This analysis (ESI, Fig.~S9), indicates that some of the promising features identified by \citet{Koczor-BendaPRX21}, such as the aromatic amine group (Fig.~\ref{fig:biasing}d), correlate with higher P values in the generated molecules as well as in the training set of commercial thiols. 

\subsection{Analysis of the chemical space of generated molecules}

\begin{figure*}
    \centering
    \includegraphics[width=\linewidth]{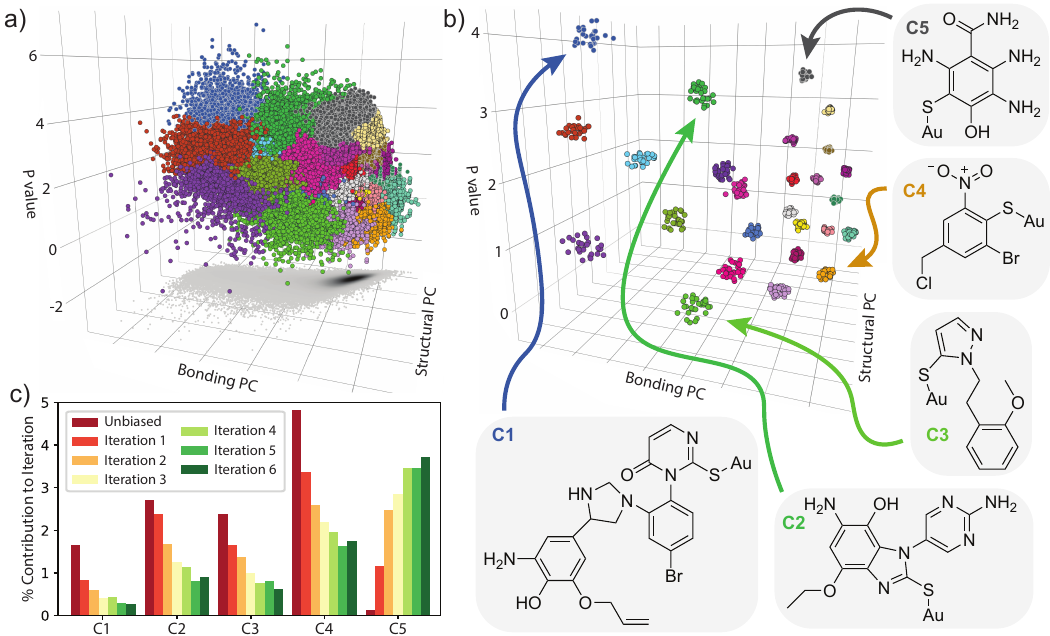}
    \caption{Latent chemical space clustering results for all generated molecules. Shown are: (a) generated molecules in the latent space formed by the first principal components (PCs) of the bonding and structural descriptors, separated vertically by their predicted P values and clustered with respect to these axes. The bottom plane depicts the density of points within the principal component space, with darker areas indicating regions of high density;  (b) subsamples of clusters around their centroids to reveal the 20 most representative molecules for each cluster, with illustrative examples from five such subclusters (C1--C5) shown; (c) separation of molecules in their respective clusters from (a) into contributions from each biasing iteration to reveal trends in the types of molecules that are prioritized and penalized during iterative biasing.}
    \label{fig:clustering}
\end{figure*}

\begin{table}[h]
    \centering
    \begin{tabular}{cccc}
        Subcluster & Average P value & SCScore & Number of atoms \\
        \hline
        C1 & 4.1 & 3.9 -- 4.9 & 50 -- 59 \\
        C2 & 3.2 & 3.3 -- 3.4 & 35 -- 40 \\
        C3 & 0.1 & 2.7 -- 3.8 & 28 -- 33 \\
        C4 & -0.2 & 1.6 -- 2.9 & 17 -- 20 \\
        C5 & 3.4 & 2.2 -- 3.0 & 21 -- 25 \\
    \end{tabular}
    \caption{Statistics for the generated molecules in the chosen subclusters shown in Fig.~\ref{fig:clustering}, including PaiNN-predicted P values.}
    \label{tab:clustering-stats}
\end{table}

Structural and bonding descriptors were calculated for all generated molecules. Principal components of these descriptors span a latent representation of the chemical space covered by the molecules. A heat map of the distribution of molecules in this latent space is projected into the basal plane of Fig.~\ref{fig:clustering}a, where it is clear that molecular generation is prioritized in a specific region of latent space. Previous efforts at biasing G-SchNet have shown significant localization in such latent chemical spaces as biasing iterations proceed~\cite{WestermayrNCS23}. This can be visualized by separating out the contributions of each iteration, as shown in the ESI (Fig.~S10). However, unlike in \citet{WestermayrNCS23}, in this work, we did not find a clear correlation between the progression of biasing iterations and the occupied chemical space decreasing in size; while there was an initial decrease in the covered area for the molecules of the unbiased generation, the molecules in successive iterations did not localize any further to one particular area of chemical space. This is because we retain original molecules in each biasing iteration, but will likely also relate to the P value biasing target being less related to specific changes in functional groups and chemical composition.
The P value is likely more closely related to several features that can appear across a diverse range of molecules.

To better resolve the types of molecules that were being generated in different areas of the latent space, the heat map in Fig.~\ref{fig:clustering}a was expanded through the inclusion of the PaiNN-predicted P values and was clustered as previously described. These clusters are also shown in Fig.~\ref{fig:clustering}a, with data points corresponding to their counterparts in the heat map. Many of the clusters span a wide range of P values and a large area of latent space, indicating that there is little correlation between the latent space and the THz radiation sensitivity of each molecule, again signifying that the P value is a complex biasing target. This leads to inefficiency in the biasing procedure, as structurally similar molecules can result in dramatically different P value predictions. The high-density region of the heat map results in many closely packed clusters, while the lower-density regions are inhabited by fewer large clusters. We note that while the sheer number of data points makes it difficult to see all the clusters, it is clear that some generated molecules with high P values, clustered near the top of Fig.~\ref{fig:clustering}a, have the potential to perform very well for THz radiation detection.

To perform further analysis, each cluster was subsampled to find the twenty closest molecules to the centroid of each cluster (Fig.~\ref{fig:clustering}b). While the subsampling omits molecules at the edges of the respective clusters, it allows for analysis of the nature of the molecules that exist in each cluster. The densely packed region of the latent space is now more visible, with over half of the clusters localized in a narrow slice of the bonding/structural principal component space on the right of the plot. 

Five subclusters (labeled C1--C5 and indicated in Fig.~\ref{fig:clustering}b) were chosen for detailed analysis, to establish trends in the types of molecules that were being predicted and the features that increase or reduce the predicted P value. Statistics for the molecules in these subclusters are shown in Table~\ref{tab:clustering-stats}. Subclusters C1 and C2 show high average P values. They are both composed of highly conjugated molecules with numerous aromatic rings. These contained a variety of heteroatomic functional groups, including alcohols and aromatic amines, as previously noted in Fig.~\ref{fig:biasing}c, and both subclusters contained very few molecules with halogen substituents. The main difference between molecules in these subclusters was their overall size -- molecules in C1 were generally larger and contained more aromatic rings.

Subcluster C5 also exhibits a large average P value, although it differed from subclusters C1 and C2 due to all of its molecules being much smaller and centred around a single highly substituted benzene ring. Molecules in this subcluster contain a high proportion of aromatic amine groups, in addition to other oxygen- and nitrogen-containing groups. Again, there were very few halogenated molecules present. This is in direct contrast to the molecules of subcluster C4, which were also based around a single benzene ring but were predicted to have a very low P value. These rings were characterized by being less heavily substituted than those in C5 and contained a comparatively high proportion of halogens and nitro groups, the latter of which were not found in any high-P value clusters. It is notable that these subclusters, and indeed all of those in the previously noted high-density region of the latent space heat map, were based around substituted benzene molecules.

Finally, molecules within subclusters C3 and C2 are structurally very similar when judged from their vicinity in the principal component latent space. However, molecules in subcluster C3 exhibit much lower P values than molecules in C2. While C3 molecules contain aromatic rings, all molecules lacked conjugation between these rings due to aliphatic joining chains. Compared to the other high-P value subclusters, their rings were also significantly less substituted, and molecules were less heteroatomic overall.

We can conclude that molecules with high predicted P values fall into one of two categories: either they are large, conjugated aromatic systems, or they are smaller, highly substituted benzene rings. In both cases, the presence of oxygen and nitrogen-based substituents (particularly amines) was desired, while halogenation and nitro groups lead to lower P values.

To establish how the presence of each of these types of molecules varied over the biasing iterations, each analyzed subcluster's respective full cluster was separated out into a percentage contribution to each iteration, as shown in Fig.~\ref{fig:clustering}c. While C1, C2, C3 and C4 all contributed less to each iteration as biasing proceeded, C5 contributed significantly more, indicating that G-SchNet was consistently biased towards molecules similar to those in subcluster C5. This is sensible when the multi-property biasing task that was undertaken is considered, as the molecules in subcluster C5 were smaller and chemically simpler than those in subclusters C1 and C2, thereby receiving a lower SCScore since they would be simpler to synthesize. Since molecules in subcluster C5 have a relatively high P value and a relatively low SCScore, they were prioritized; molecules in subclusters C1 and C2 were too complex, yielding a higher SCScore, while molecules in subclusters C3 and C4 were simpler but had a low predicted P value, so molecules from these clusters did not fulfil the multi-property biasing criteria.

\subsection{Identification of candidate molecules}
\label{sec:best_candidates}

We selected generated molecules with $\mathrm{P}\geq4.25$ (based on predictions by the retrained PaiNN predictor) and employed AiZynthFinder to perform retrosynthetic planning. From the 1011 molecules satisfying this selection criterion, only 34 were predicted to have retrosynthetic routes from purchasable precursors~\cite{GenhedenJC20_AiZynthFinder} based on a stock from compounds available within the ZINC~\cite{SterlingJCIM15_ZINC15} database; retrosynthetic paths for these molecules can be found in Fig.~S11--S17. Notably, all 34 molecules belong to clusters from which subclusters C2 and C5 were drawn (ESI, Table SVI).

\begin{figure}
    \centering
    \includegraphics[width=0.9\linewidth]{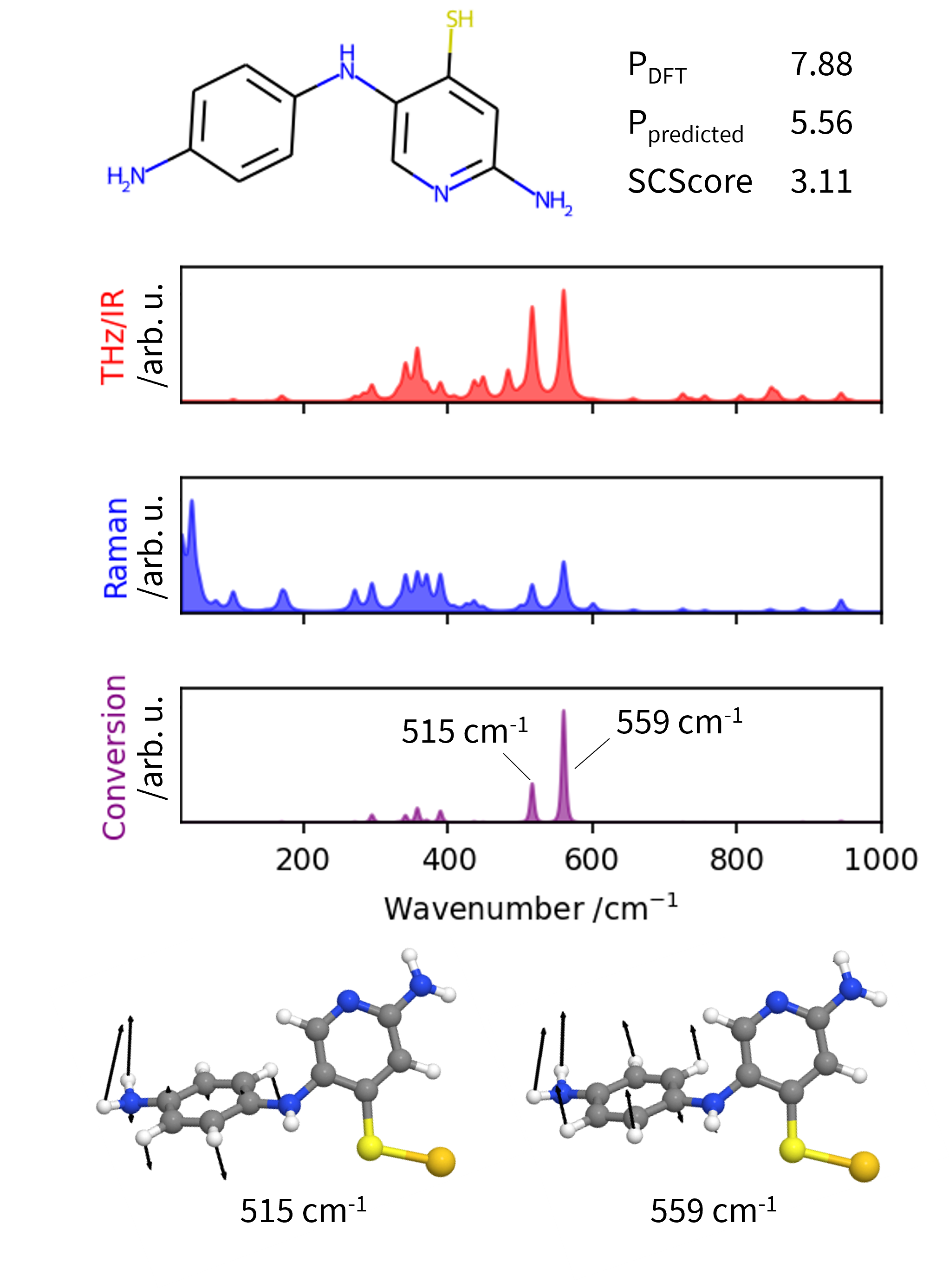}     
    \caption{Properties of the top candidate molecule, 2-amino-5-(4-aminophenylamino)pyridine-4-thiol, generated by G-SchNet. Density functional theory (DFT)-calculated (P\textsubscript{DFT}) and PaiNN-predicted (P\textsubscript{predicted}) P values, predicted SCScore, as well as DFT-calculated terahertz (THz)/infrared (IR) radiation absorption, Raman scattering and frequency upconversion spectra are shown. The two most intensive vibrational modes for frequency upconversion are also depicted.}
    \label{fig:best}
\end{figure}

To confirm the suitability of these molecules for THz radiation detection, their absorption, Raman scattering and frequency upconversion spectra were calculated, and their P values were determined using DFT. Figure~\ref{fig:best} shows the relevant properties and vibrational spectra of the top candidate, while vibrational spectra and properties of other candidate molecules with DFT-calculated P values above 5.20 are shown in the ESI (Fig.~S18--S21). The top candidate, 2-amino-5-(4-aminophenylamino)pyridine-4-thiol has a DFT-calculated P value of 7.88. Considering that the P value is a logarithmic quantity (Equation~(\ref{eq:P})), this is significantly higher than any of the molecules previously identified within commercial databases in \citet{Koczor-BendaPRX21}, where the top 5 candidates had P values between 5.30 and 6.18. For the fifth top molecule from \citet{Koczor-BendaPRX21}, 5-amino-2-mercaptobenzimidazole, \citet{redolat2024synthesis} developed a  functionalization technique to prepare self-assembled molecular monolayers in gold-based plasmonic nanocavities and successfully integrated these nanocavities on a silicon-based photonic chip. While frequency upconversion measurements are not yet available for this compound, our DFT simulations suggest about 14 times higher upconversion capability in the THz/mid-infrared range for the most active mode (559 cm$^{-1}$) of our top candidate compared to the most active mode (458 cm$^{-1}$) of 5-amino-2-mercaptobenzimidazole (see Fig.~S22 in ESI). 

The top molecule has two vibrational modes that are highly active in frequency upconversion, which are located at 515~\si{\per\cm} and 559~\si{\per\cm}. Both modes involve an out-of-plane (umbrella) motion of one of the amino groups that is coupled to out-of-plane motions of hydrogen atoms of the neighboring ring. This out-of-plane motion of the amino group is also responsible for the highest intensity peaks of other top candidates, as shown in the ESI (Fig.~S18--S21). This provides evidence that the aromatic amine functional group not only correlates with high P values, but is also directly involved in the upconversion process. The highly active mode appears in the 515--832~\si{\per\cm} spectral range for the top candidates, showing that the chemical environment and the coupling of the out-of-plane motion of the amino group with other vibrations of the molecule have a significant effect on the position of the peak. This can be advantageous for the tuning of narrowband THz radiation detectors operating at different frequencies.
We also note that within the top candidates, molecules with the same SMILES string were generated multiple times with different 3D structures in the different biasing iterations. As the SCScore and KRR-predicted P values depend only on 2D information, they remain the same for different conformers. However, the PaiNN-predicted P values for raw generated structures and DFT-calculated P values for structures that have undergone geometry optimization can differ, as shown in section S10 and Fig.~S24 of the ESI. This further highlights the benefits of working with property predictors that are based on 3D descriptors.

Of the 34 molecules listed in Table SVI, only one compound (generated three times as different conformers, all sharing the same SMILES string) was identified in the PubChem~\cite{PubChem, KimNAR25} database, Nc1cc(S)c(cc1N)N, which corresponds to 2,4,5-triaminobenzenethiol (Compound Identifier 67981805~\cite{CID_67981805}). The remaining 31 molecules were not found in PubChem, likely representing novel candidate structures for THz upconversion applications.

\section{Conclusions and Outlook}

Generative design of functional organic molecules can be biased towards certain properties by iteratively adapting the underlying training dataset. Here we do this to design candidate molecules for THz radiation detection by mixing molecules from an existing database with selected molecules created by the autoregressive generative deep learning model G-SchNet. This enables us to perform property-driven design of novel and synthesizable monothiolated molecules with high THz-to-visible upconversion efficiencies. By performing a comprehensive structural analysis on the dataset of generated molecules, we have revealed key chemical trends among generated molecules and identified functional groups that contribute to enhanced upconversion, such as aromatic amines. From the novel, generated molecules, we were able to select several candidates and provide potential retrosynthetic pathways from commercially available reactants. The top candidate molecule has a DFT-calculated THz upconversion efficiency of 7.88, which is significantly higher than any of the molecules previously identified from commercial databases. 

This work also revealed several practical challenges associated with property-driven generative design that require careful consideration when designing such workflows. 
First of all, we have seen that even unbiased molecular generation in G-SchNet creates a distribution of molecules that significantly differs from the training dataset in terms of elemental and functional group composition. If the model cannot capture the chemical space spanned by the data, this means that the ability of the property-driven design workflow to drive the generation in a directed way is limited. The performance of G-SchNet and other generative algorithms in this regard needs to be analysed in greater detail in the future.
Secondly, during sequential iterations of biasing with a changing training dataset, the ML-based property predictor that selects suitable molecules must continue to provide accurate predictions. We showed that GNN-based ML predictors, based on MACE and PaiNN models and 3D input structures, gave more accurate P values than predictors based on 2D molecular fingerprints. The figure of merit of THz upconversion efficiency, P, was shown to be a highly integrated quantity that is challenging to learn due to its dependence on low-lying vibrational modes. Careful validation revealed that contrary to previous work on the property-driven generative design of fundamental electronic gaps~\cite{WestermayrNCS23} none of the P predictors trained on the original data set were transferable to the newly generated molecules. Their prediction accuracy deteriorated during the iterative biasing workflow. Therefore, the PaiNN predictor had to be retrained based on new DFT training data. Uncertainty-based active learning during biasing iterations would not have been a robust strategy due to the lack of correlation between prediction accuracy and uncertainty in highly regularized GNNs. Therefore, active learning based on structural diversity sampling is likely a more robust choice to retain ML predictor performance throughout the iterative biasing procedure.

Significant future work will be needed to make property-driven generative design workflows more efficient and robust. To this end, constrained generation with (semi-)supervised generative models such as constrained G-SchNet~\cite{GebauerNC22_cG-SchNet} that can constrain specific functional groups or diffusion models able to perform inpainting tasks will likely be beneficial. This would reduce the portion of generated molecules that are discarded during the workflow due to the absence of a thiol group. The question of whether generative models faithfully represent the structural and functional group distribution of the underlying training dataset requires further attention. Commonly, generative models are only assessed on their ability to generate valid and unique molecules, which is insufficient when aiming to employ models for directed exploration of chemical space.

Both the property-driven design workflow and the novel candidate molecules we have identified in this study will contribute to advancing the discovery of functional organic materials for nanosensor applications such as THz radiation detection. Our results highlight the potential of generative models to not only expand the chemical space of viable molecules but also to guide future experimental and computational efforts in the molecular design of plasmonic nanocavities. 


\section{Conflicts of interest}
There are no conflicts to declare.

\section{Data availability}
Data for this article, including molecular databases in ASE database format, DFT-optimized best candidate molecules, and ASE databases for xTB calculations are available online:  \url{https://doi.org/10.6084/m9.figshare.28539995.v3}~\cite{figshare}. The repository also contains the trained ML models, and Jupyter Notebooks and scripts associated with the generation, prediction, and analysis workflows described in this work. Code for the extraction of bonding features from molecular databases and obtaining the principal components of the structural/bonding descriptors has been released in our GSchNetTools package, available at \url{https://github.com/maurergroup/GSchNetTools}. The SCScore model used in this work is publicly available at \url{https://github.com/connorcoley/scscore}, and files pertaining to retrosynthetic planning with AiZynthFinder are publicly available at \url{https://figshare.com/articles/dataset/AiZynthFinder_a_fast_robust_and_flexible_open-source_software_for_retrosynthetic_planning/12334577}. 

\section{Acknowledgements}
The authors thank the Research Development Fund of the University of Warwick, Wellcome Leap as part of the Quantum for Bio Program, the EPSRC Centre for Doctoral Training in Modelling of Heterogeneous Systems [EP/S022848/1], the UKRI Future Leaders Fellowship programme [MR/X023109/1], and a UKRI Frontier research grant [EP/X014088/1] for funding this work. Computing resources were provided by the Scientific Computing Research Technology Platform of the University of Warwick for access to Avon; the EPSRC-funded HPC Midlands+ consortium [EP/T022108/1] for access to Sulis; and the EPSRC-funded Northern Ireland High Performance Computing service [EP/T022175/1] for access to Kelvin2. We also thank Niklas Gebauer (Machine Learning Group, Technische Universit{\"a}t Berlin) for help with the schnetpack-gschnet software. 




\begin{mcitethebibliography}{67}
\providecommand*{\natexlab}[1]{#1}
\providecommand*{\mciteSetBstSublistMode}[1]{}
\providecommand*{\mciteSetBstMaxWidthForm}[2]{}
\providecommand*{\mciteBstWouldAddEndPuncttrue}
  {\def\EndOfBibitem{\unskip.}}
\providecommand*{\mciteBstWouldAddEndPunctfalse}
  {\let\EndOfBibitem\relax}
\providecommand*{\mciteSetBstMidEndSepPunct}[3]{}
\providecommand*{\mciteSetBstSublistLabelBeginEnd}[3]{}
\providecommand*{\EndOfBibitem}{}
\mciteSetBstSublistMode{f}
\mciteSetBstMaxWidthForm{subitem}
{(\emph{\alph{mcitesubitemcount}})}
\mciteSetBstSublistLabelBeginEnd{\mcitemaxwidthsubitemform\space}
{\relax}{\relax}

\bibitem[Tonouchi(2007)]{TonouchiNP07}
M.~Tonouchi, \emph{Nature Photon.}, 2007, \textbf{1}, 97--105\relax
\mciteBstWouldAddEndPuncttrue
\mciteSetBstMidEndSepPunct{\mcitedefaultmidpunct}
{\mcitedefaultendpunct}{\mcitedefaultseppunct}\relax
\EndOfBibitem
\bibitem[Dhillon \emph{et~al.}(2017)Dhillon, Vitiello, Linfield, Davies,
  Hoffmann, Booske, Paoloni, Gensch, Weightman, Williams, Castro-Camus,
  Cumming, Simoens, Escorcia-Carranza, Grant, Lucyszyn, Kuwata-Gonokami,
  Konishi, Koch, Schmuttenmaer, Cocker, Huber, Markelz, Taylor, Wallace,
  Axel~Zeitler, Sibik, Korter, Ellison, Rea, Goldsmith, Cooper, Appleby, Pardo,
  Huggard, Krozer, Shams, Fice, Renaud, Seeds, St{\"o}hr, Naftaly, Ridler,
  Clarke, Cunningham, and Johnston]{DhillonJPD17}
S.~S. Dhillon, M.~S. Vitiello, E.~H. Linfield, A.~G. Davies, M.~C. Hoffmann,
  J.~Booske, C.~Paoloni, M.~Gensch, P.~Weightman, G.~P. Williams,
  E.~Castro-Camus, D.~R.~S. Cumming, F.~Simoens, I.~Escorcia-Carranza,
  J.~Grant, S.~Lucyszyn, M.~Kuwata-Gonokami, K.~Konishi, M.~Koch, C.~A.
  Schmuttenmaer, T.~L. Cocker, R.~Huber, A.~G. Markelz, Z.~D. Taylor, V.~P.
  Wallace, J.~Axel~Zeitler, J.~Sibik, T.~M. Korter, B.~Ellison, S.~Rea,
  P.~Goldsmith, K.~B. Cooper, R.~Appleby, D.~Pardo, P.~G. Huggard, V.~Krozer,
  H.~Shams, M.~Fice, C.~Renaud, A.~Seeds, A.~St{\"o}hr, M.~Naftaly, N.~Ridler,
  R.~Clarke, J.~E. Cunningham and M.~B. Johnston, \emph{J. Phys. D: Appl.
  Phys.}, 2017, \textbf{50}, 043001\relax
\mciteBstWouldAddEndPuncttrue
\mciteSetBstMidEndSepPunct{\mcitedefaultmidpunct}
{\mcitedefaultendpunct}{\mcitedefaultseppunct}\relax
\EndOfBibitem
\bibitem[Roelli \emph{et~al.}(2016)Roelli, Galland, Piro, and
  Kippenberg]{RoelliNN16}
P.~Roelli, C.~Galland, N.~Piro and T.~J. Kippenberg, \emph{Nature Nanotech.},
  2016, \textbf{11}, 164--169\relax
\mciteBstWouldAddEndPuncttrue
\mciteSetBstMidEndSepPunct{\mcitedefaultmidpunct}
{\mcitedefaultendpunct}{\mcitedefaultseppunct}\relax
\EndOfBibitem
\bibitem[Roelli \emph{et~al.}(2020)Roelli, Martin-Cano, Kippenberg, and
  Galland]{RoelliPRX20}
P.~Roelli, D.~Martin-Cano, T.~J. Kippenberg and C.~Galland, \emph{Phys. Rev.
  X}, 2020, \textbf{10}, 031057\relax
\mciteBstWouldAddEndPuncttrue
\mciteSetBstMidEndSepPunct{\mcitedefaultmidpunct}
{\mcitedefaultendpunct}{\mcitedefaultseppunct}\relax
\EndOfBibitem
\bibitem[Redolat \emph{et~al.}(2024)Redolat, Camarena-P{\'e}rez, Griol, Lozano,
  G{\'o}mez-G{\'o}mez, V{\'a}zquez-Lozano, Miele, Baumberg, Mart{\'\i}nez, and
  Pinilla-Cienfuegos]{redolat2024synthesis}
J.~Redolat, M.~Camarena-P{\'e}rez, A.~Griol, M.~S. Lozano, M.~I.
  G{\'o}mez-G{\'o}mez, J.~E. V{\'a}zquez-Lozano, E.~Miele, J.~J. Baumberg,
  A.~Mart{\'\i}nez and E.~Pinilla-Cienfuegos, \emph{Nano Letters}, 2024,
  \textbf{24}, 3670--3677\relax
\mciteBstWouldAddEndPuncttrue
\mciteSetBstMidEndSepPunct{\mcitedefaultmidpunct}
{\mcitedefaultendpunct}{\mcitedefaultseppunct}\relax
\EndOfBibitem
\bibitem[Xomalis \emph{et~al.}(2021)Xomalis, Zheng, Chikkaraddy, Koczor-Benda,
  Miele, Rosta, Vandenbosch, Martínez, and Baumberg]{xomalis_detecting_2021}
A.~Xomalis, X.~Zheng, R.~Chikkaraddy, Z.~Koczor-Benda, E.~Miele, E.~Rosta,
  G.~A.~E. Vandenbosch, A.~Martínez and J.~J. Baumberg, \emph{Science}, 2021,
  \textbf{374}, 1268--1271\relax
\mciteBstWouldAddEndPuncttrue
\mciteSetBstMidEndSepPunct{\mcitedefaultmidpunct}
{\mcitedefaultendpunct}{\mcitedefaultseppunct}\relax
\EndOfBibitem
\bibitem[Chen \emph{et~al.}(2021)Chen, Roelli, Hu, Verlekar, Amirtharaj,
  Barreda, Kippenberg, Kovylina, Verhagen, Martínez, and
  Galland]{chen_continuous-wave_2021}
W.~Chen, P.~Roelli, H.~Hu, S.~Verlekar, S.~P. Amirtharaj, A.~I. Barreda, T.~J.
  Kippenberg, M.~Kovylina, E.~Verhagen, A.~Martínez and C.~Galland,
  \emph{Science}, 2021, \textbf{374}, 1264--1267\relax
\mciteBstWouldAddEndPuncttrue
\mciteSetBstMidEndSepPunct{\mcitedefaultmidpunct}
{\mcitedefaultendpunct}{\mcitedefaultseppunct}\relax
\EndOfBibitem
\bibitem[Neubrech \emph{et~al.}(2017)Neubrech, Huck, Weber, Pucci, and
  Giessen]{NeubrechCR17}
F.~Neubrech, C.~Huck, K.~Weber, A.~Pucci and H.~Giessen, \emph{Chem. Rev.},
  2017, \textbf{117}, 5110--5145\relax
\mciteBstWouldAddEndPuncttrue
\mciteSetBstMidEndSepPunct{\mcitedefaultmidpunct}
{\mcitedefaultendpunct}{\mcitedefaultseppunct}\relax
\EndOfBibitem
\bibitem[Stiles \emph{et~al.}(2008)Stiles, Dieringer, Shah, and
  Van~Duyne]{StilesARAC08}
P.~L. Stiles, J.~A. Dieringer, N.~C. Shah and R.~P. Van~Duyne, \emph{Annu. Rev.
  Anal. Chem.}, 2008, \textbf{1}, 601--626\relax
\mciteBstWouldAddEndPuncttrue
\mciteSetBstMidEndSepPunct{\mcitedefaultmidpunct}
{\mcitedefaultendpunct}{\mcitedefaultseppunct}\relax
\EndOfBibitem
\bibitem[Humbert \emph{et~al.}(2019)Humbert, Noblet, Dalstein, Busson, and
  Barbillon]{HumbertMaterials19}
C.~Humbert, T.~Noblet, L.~Dalstein, B.~Busson and G.~Barbillon,
  \emph{Materials}, 2019, \textbf{12}, 836\relax
\mciteBstWouldAddEndPuncttrue
\mciteSetBstMidEndSepPunct{\mcitedefaultmidpunct}
{\mcitedefaultendpunct}{\mcitedefaultseppunct}\relax
\EndOfBibitem
\bibitem[Koczor-Benda \emph{et~al.}(2021)Koczor-Benda, Boehmke, Xomalis, Arul,
  Readman, Baumberg, and Rosta]{Koczor-BendaPRX21}
Z.~Koczor-Benda, A.~L. Boehmke, A.~Xomalis, R.~Arul, C.~Readman, J.~J. Baumberg
  and E.~Rosta, \emph{Phys. Rev. X}, 2021, \textbf{11}, 041035\relax
\mciteBstWouldAddEndPuncttrue
\mciteSetBstMidEndSepPunct{\mcitedefaultmidpunct}
{\mcitedefaultendpunct}{\mcitedefaultseppunct}\relax
\EndOfBibitem
\bibitem[Koczor-Benda \emph{et~al.}(2022)Koczor-Benda, Roelli, Galland, and
  Rosta]{Koczor-BendaJPCA22}
Z.~Koczor-Benda, P.~Roelli, C.~Galland and E.~Rosta, \emph{J. Phys. Chem. A},
  2022, \textbf{126}, 4657--4663\relax
\mciteBstWouldAddEndPuncttrue
\mciteSetBstMidEndSepPunct{\mcitedefaultmidpunct}
{\mcitedefaultendpunct}{\mcitedefaultseppunct}\relax
\EndOfBibitem
\bibitem[G{\'o}mez-Bombarelli \emph{et~al.}(2016)G{\'o}mez-Bombarelli,
  Aguilera-Iparraguirre, Hirzel, Duvenaud, Maclaurin, Blood-Forsythe, Sik~Chae,
  Einzinger, Ha, Wu, Markopoulos, Jeon, Kang, Miyazaki, Numata, Kim, Huang,
  Ik~Hong, Baldo, Adams, and Aspuru-Guzik]{Gomez-BombarelliNM16}
R.~G{\'o}mez-Bombarelli, J.~Aguilera-Iparraguirre, T.~D. Hirzel, D.~Duvenaud,
  D.~Maclaurin, M.~A. Blood-Forsythe, H.~Sik~Chae, M.~Einzinger, D.-G. Ha,
  T.~Wu, G.~Markopoulos, S.~Jeon, H.~Kang, H.~Miyazaki, M.~Numata, S.~Kim,
  W.~Huang, S.~Ik~Hong, M.~Baldo, R.~P. Adams and A.~Aspuru-Guzik, \emph{Nature
  Mater.}, 2016, \textbf{15}, 1120--1127\relax
\mciteBstWouldAddEndPuncttrue
\mciteSetBstMidEndSepPunct{\mcitedefaultmidpunct}
{\mcitedefaultendpunct}{\mcitedefaultseppunct}\relax
\EndOfBibitem
\bibitem[Sahu \emph{et~al.}(2019)Sahu, Yang, Ye, Ma, Fang, and Ma]{SahuJMCA19}
H.~Sahu, F.~Yang, X.~Ye, J.~Ma, W.~Fang and H.~Ma, \emph{J. Mater. Chem. A},
  2019, \textbf{7}, 17480--17488\relax
\mciteBstWouldAddEndPuncttrue
\mciteSetBstMidEndSepPunct{\mcitedefaultmidpunct}
{\mcitedefaultendpunct}{\mcitedefaultseppunct}\relax
\EndOfBibitem
\bibitem[Saeki and Kranthiraja(2020)]{SaekiJJAP20}
A.~Saeki and K.~Kranthiraja, \emph{Jpn. J. Appl. Phys.}, 2020, \textbf{59},
  SD0801\relax
\mciteBstWouldAddEndPuncttrue
\mciteSetBstMidEndSepPunct{\mcitedefaultmidpunct}
{\mcitedefaultendpunct}{\mcitedefaultseppunct}\relax
\EndOfBibitem
\bibitem[Chechik and Stirling(1999)]{Checkik99}
V.~Chechik and C.~J.~M. Stirling, in \emph{{Patai's Chemistry of Functional
  Groups}}, ed. Z.~Rappoport, Wiley, 1999, ch. {Gold–Thiol Self-Assembled
  Monolayers}\relax
\mciteBstWouldAddEndPuncttrue
\mciteSetBstMidEndSepPunct{\mcitedefaultmidpunct}
{\mcitedefaultendpunct}{\mcitedefaultseppunct}\relax
\EndOfBibitem
\bibitem[Westermayr \emph{et~al.}(2023)Westermayr, Gilkes, Barrett, and
  Maurer]{WestermayrNCS23}
J.~Westermayr, J.~Gilkes, R.~Barrett and R.~J. Maurer, \emph{Nat. Comput.
  Sci.}, 2023, \textbf{3}, 139--148\relax
\mciteBstWouldAddEndPuncttrue
\mciteSetBstMidEndSepPunct{\mcitedefaultmidpunct}
{\mcitedefaultendpunct}{\mcitedefaultseppunct}\relax
\EndOfBibitem
\bibitem[Gebauer \emph{et~al.}(2022)Gebauer, Gastegger, Hessmann, M{\"u}ller,
  and Sch{\"u}tt]{GebauerNC22_cG-SchNet}
N.~W.~A. Gebauer, M.~Gastegger, S.~S.~P. Hessmann, K.-R. M{\"u}ller and K.~T.
  Sch{\"u}tt, \emph{Nat. Commun.}, 2022, \textbf{13}, 973\relax
\mciteBstWouldAddEndPuncttrue
\mciteSetBstMidEndSepPunct{\mcitedefaultmidpunct}
{\mcitedefaultendpunct}{\mcitedefaultseppunct}\relax
\EndOfBibitem
\bibitem[Joshi \emph{et~al.}(2021)Joshi, Gebauer, Bontha, Khazaieli, James,
  Brown, and Kumar]{JoshiJPCB21_3D-Scaffold}
R.~P. Joshi, N.~W.~A. Gebauer, M.~Bontha, M.~Khazaieli, R.~M. James, J.~B.
  Brown and N.~Kumar, \emph{J. Phys. Chem. B}, 2021, \textbf{125},
  12166--12176\relax
\mciteBstWouldAddEndPuncttrue
\mciteSetBstMidEndSepPunct{\mcitedefaultmidpunct}
{\mcitedefaultendpunct}{\mcitedefaultseppunct}\relax
\EndOfBibitem
\bibitem[Sanchez-Lengeling and Aspuru-Guzik(2018)]{Sanchez-LengelingScience18}
B.~Sanchez-Lengeling and A.~Aspuru-Guzik, \emph{Science}, 2018, \textbf{361},
  360--365\relax
\mciteBstWouldAddEndPuncttrue
\mciteSetBstMidEndSepPunct{\mcitedefaultmidpunct}
{\mcitedefaultendpunct}{\mcitedefaultseppunct}\relax
\EndOfBibitem
\bibitem[G{\'o}mez-Bombarelli \emph{et~al.}(2018)G{\'o}mez-Bombarelli, Wei,
  Duvenaud, Miguel Hern{\'a}ndez-Lobato, S{\'a}nchez-Lengeling, Sheberla,
  Aguilera-Iparraguirre, Hirzel, Adams, and
  Aspuru-Guzik]{Gomez-BombarelliACSCS18}
R.~G{\'o}mez-Bombarelli, J.~N. Wei, D.~Duvenaud, J.~Miguel
  Hern{\'a}ndez-Lobato, B.~S{\'a}nchez-Lengeling, D.~Sheberla,
  J.~Aguilera-Iparraguirre, T.~D. Hirzel, R.~P. Adams and A.~Aspuru-Guzik,
  \emph{ACS Cent. Sci.}, 2018, \textbf{4}, 268--276\relax
\mciteBstWouldAddEndPuncttrue
\mciteSetBstMidEndSepPunct{\mcitedefaultmidpunct}
{\mcitedefaultendpunct}{\mcitedefaultseppunct}\relax
\EndOfBibitem
\bibitem[Meyers \emph{et~al.}(2021)Meyers, Fabian, and Brown]{MeyersDDT21}
J.~Meyers, B.~Fabian and N.~Brown, \emph{Drug Discov. Today}, 2021,
  \textbf{26}, 2707--2715\relax
\mciteBstWouldAddEndPuncttrue
\mciteSetBstMidEndSepPunct{\mcitedefaultmidpunct}
{\mcitedefaultendpunct}{\mcitedefaultseppunct}\relax
\EndOfBibitem
\bibitem[Ar{\'u}s-Pous \emph{et~al.}(2020)Ar{\'u}s-Pous, Patronov, Bjerrum,
  Tyrchan, Reymond, Chen, and Engkvist]{Arús-PousJC20}
J.~Ar{\'u}s-Pous, A.~Patronov, E.~J. Bjerrum, C.~Tyrchan, J.-L. Reymond,
  H.~Chen and O.~Engkvist, \emph{J. Cheminform.}, 2020, \textbf{12}, 38\relax
\mciteBstWouldAddEndPuncttrue
\mciteSetBstMidEndSepPunct{\mcitedefaultmidpunct}
{\mcitedefaultendpunct}{\mcitedefaultseppunct}\relax
\EndOfBibitem
\bibitem[Kong \emph{et~al.}(2022)Kong, Hu, Zhang, and Tin]{KongFP22}
W.~Kong, Y.~Hu, J.~Zhang and Q.~Tin, \emph{Front. Pharmacol.}, 2022,
  \textbf{13}, 1046524\relax
\mciteBstWouldAddEndPuncttrue
\mciteSetBstMidEndSepPunct{\mcitedefaultmidpunct}
{\mcitedefaultendpunct}{\mcitedefaultseppunct}\relax
\EndOfBibitem
\bibitem[Gebauer \emph{et~al.}(2019)Gebauer, Gastegger, and
  Sch{\"u}tt]{GebauerNeurIPS19_G-SchNet}
N.~W.~A. Gebauer, M.~Gastegger and K.~T. Sch{\"u}tt, in \emph{{Advances in
  Neural Information Processing Systems 32}}, ed. H.~Wallach, H.~Larochelle,
  A.~Beygelzimer, F.~d'Alch{\'e} Buc, E.~Fox and R.~Garnett, NeurIPS
  Proceedings, 2019, ch. {Symmetry-adapted generation of 3d point sets for the
  targeted discovery of molecules}\relax
\mciteBstWouldAddEndPuncttrue
\mciteSetBstMidEndSepPunct{\mcitedefaultmidpunct}
{\mcitedefaultendpunct}{\mcitedefaultseppunct}\relax
\EndOfBibitem
\bibitem[Westermayr and Marquetand(2021)]{westermayr_machine_2021}
J.~Westermayr and P.~Marquetand, \emph{Chem. Rev.}, 2021, \textbf{121},
  9873--9926\relax
\mciteBstWouldAddEndPuncttrue
\mciteSetBstMidEndSepPunct{\mcitedefaultmidpunct}
{\mcitedefaultendpunct}{\mcitedefaultseppunct}\relax
\EndOfBibitem
\bibitem[eMo()]{eMolecules}
\emph{{eMolecules database}}, \url{https://www.emolecules.com}, (accessed 01
  March 2020)\relax
\mciteBstWouldAddEndPuncttrue
\mciteSetBstMidEndSepPunct{\mcitedefaultmidpunct}
{\mcitedefaultendpunct}{\mcitedefaultseppunct}\relax
\EndOfBibitem
\bibitem[Weininger(1988)]{SMILES}
D.~Weininger, \emph{J. Chem. Inf. Comput. Sci.}, 1988, \textbf{28},
  31--36\relax
\mciteBstWouldAddEndPuncttrue
\mciteSetBstMidEndSepPunct{\mcitedefaultmidpunct}
{\mcitedefaultendpunct}{\mcitedefaultseppunct}\relax
\EndOfBibitem
\bibitem[Halgren(1996)]{halgren1996merck}
T.~A. Halgren, \emph{J. Comput. Chem.}, 1996, \textbf{17}, 490--519\relax
\mciteBstWouldAddEndPuncttrue
\mciteSetBstMidEndSepPunct{\mcitedefaultmidpunct}
{\mcitedefaultendpunct}{\mcitedefaultseppunct}\relax
\EndOfBibitem
\bibitem[Landrum()]{RDKit}
G.~Landrum, \emph{{RDKit: Open-source cheminformatics}},
  \url{http://www.rdkit.org/}, (accessed November 13, 2024)\relax
\mciteBstWouldAddEndPuncttrue
\mciteSetBstMidEndSepPunct{\mcitedefaultmidpunct}
{\mcitedefaultendpunct}{\mcitedefaultseppunct}\relax
\EndOfBibitem
\bibitem[Bart{\'o}k \emph{et~al.}(2013)Bart{\'o}k, Kondor, and
  Cs{\'a}nyi]{SOAP}
A.~P. Bart{\'o}k, R.~Kondor and G.~Cs{\'a}nyi, \emph{Phys. Rev. B}, 2013,
  \textbf{87}, 184115\relax
\mciteBstWouldAddEndPuncttrue
\mciteSetBstMidEndSepPunct{\mcitedefaultmidpunct}
{\mcitedefaultendpunct}{\mcitedefaultseppunct}\relax
\EndOfBibitem
\bibitem[Himanen \emph{et~al.}(2020)Himanen, J{\"a}ger, Morooka,
  Federici~Canova, Ranawat, Gao, Rinke, and Foster]{dscribe}
L.~Himanen, M.~O.~J. J{\"a}ger, E.~V. Morooka, F.~Federici~Canova, Y.~S.
  Ranawat, D.~Z. Gao, P.~Rinke and A.~S. Foster, \emph{Comput. Phys. Commun.},
  2020, \textbf{247}, 106949\relax
\mciteBstWouldAddEndPuncttrue
\mciteSetBstMidEndSepPunct{\mcitedefaultmidpunct}
{\mcitedefaultendpunct}{\mcitedefaultseppunct}\relax
\EndOfBibitem
\bibitem[Pedregosa \emph{et~al.}(2011)Pedregosa, Varoquaux, Gramfort, Michel,
  Thirion, Grisel, Blondel, Prettenhofer, Weiss, Dubourg, Vanderplas, Passos,
  Cournapeau, Brucher, Perrot, and Duchesnay]{scikit-learn}
F.~Pedregosa, G.~Varoquaux, A.~Gramfort, V.~Michel, B.~Thirion, O.~Grisel,
  M.~Blondel, P.~Prettenhofer, R.~Weiss, V.~Dubourg, J.~Vanderplas, A.~Passos,
  D.~Cournapeau, M.~Brucher, M.~Perrot and {\'E}.~Duchesnay, \emph{J. Mach.
  Learn. Res.}, 2011, \textbf{23}, 2825--2830\relax
\mciteBstWouldAddEndPuncttrue
\mciteSetBstMidEndSepPunct{\mcitedefaultmidpunct}
{\mcitedefaultendpunct}{\mcitedefaultseppunct}\relax
\EndOfBibitem
\bibitem[Bannwarth \emph{et~al.}(2019)Bannwarth, Ehlert, and Grimme]{GFN2-xTB}
C.~Bannwarth, S.~Ehlert and S.~Grimme, \emph{J. Chem. Theory Comput.}, 2019,
  \textbf{15}, 1652--1671\relax
\mciteBstWouldAddEndPuncttrue
\mciteSetBstMidEndSepPunct{\mcitedefaultmidpunct}
{\mcitedefaultendpunct}{\mcitedefaultseppunct}\relax
\EndOfBibitem
\bibitem[Gebauer and Sch{\"u}tt()]{schnetpack-gschnet}
N.~Gebauer and K.~T. Sch{\"u}tt, \emph{{Conditional G-SchNet extension for
  SchNetPack 2.0 - A generative neural network for 3d molecules}},
  \url{https://github.com/atomistic-machine-learning/schnetpack-gschnet},
  (accessed November 13, 2024)\relax
\mciteBstWouldAddEndPuncttrue
\mciteSetBstMidEndSepPunct{\mcitedefaultmidpunct}
{\mcitedefaultendpunct}{\mcitedefaultseppunct}\relax
\EndOfBibitem
\bibitem[Sch{\"u}tt \emph{et~al.}(2023)Sch{\"u}tt, Hessmann, Gebauer, Lederer,
  and Gastegger]{SchüttJCP23_SchNetPack2.0}
K.~T. Sch{\"u}tt, S.~S.~P. Hessmann, N.~W.~A. Gebauer, J.~Lederer and
  M.~Gastegger, \emph{J. Chem. Phys.}, 2023, \textbf{158}, 144801\relax
\mciteBstWouldAddEndPuncttrue
\mciteSetBstMidEndSepPunct{\mcitedefaultmidpunct}
{\mcitedefaultendpunct}{\mcitedefaultseppunct}\relax
\EndOfBibitem
\bibitem[Sch{\"u}tt \emph{et~al.}(2017)Sch{\"u}tt, Kindermans, Sauceda,
  Chmiela, Tkatchenko, and M{\"u}ller]{SchüttNeurIPS17_SchNet}
K.~T. Sch{\"u}tt, P.-J. Kindermans, H.~E. Sauceda, S.~Chmiela, A.~Tkatchenko
  and K.-R. M{\"u}ller, in \emph{{Advances in Neural Information Processing
  Systems 30}}, ed. I.~Guyon, U.~Von~Luxburg, S.~Bengio, H.~Wallach, R.~Fergus,
  S.~Vishwanathan and R.~Garnett, NeurIPS Proceedings, 2017, ch. {SchNet: A
  continuous-filter convolutional neural network for modeling quantum
  interactions}\relax
\mciteBstWouldAddEndPuncttrue
\mciteSetBstMidEndSepPunct{\mcitedefaultmidpunct}
{\mcitedefaultendpunct}{\mcitedefaultseppunct}\relax
\EndOfBibitem
\bibitem[Coley \emph{et~al.}(2018)Coley, Rogers, Green, and Jensen]{SCScore}
C.~W. Coley, L.~Rogers, W.~H. Green and K.~F. Jensen, \emph{J. Chem. Inf.
  Model}, 2018, \textbf{58}, 252--261\relax
\mciteBstWouldAddEndPuncttrue
\mciteSetBstMidEndSepPunct{\mcitedefaultmidpunct}
{\mcitedefaultendpunct}{\mcitedefaultseppunct}\relax
\EndOfBibitem
\bibitem[Lawson \emph{et~al.}(2014)Lawson, Swienty-Busch, G{\'e}oui, and
  Evans]{Reaxys}
A.~J. Lawson, J.~Swienty-Busch, T.~G{\'e}oui and D.~Evans, \emph{{The Making of
  Reaxys—Towards Unobstructed Access to Relevant Chemistry Information}},
  American Chemical Society, 2014, pp. 127--148\relax
\mciteBstWouldAddEndPuncttrue
\mciteSetBstMidEndSepPunct{\mcitedefaultmidpunct}
{\mcitedefaultendpunct}{\mcitedefaultseppunct}\relax
\EndOfBibitem
\bibitem[O'Boyle \emph{et~al.}(2011)O'Boyle, Banck, James, Morley,
  Vandermeersch, and Hutchison]{Open_Babel}
N.~M. O'Boyle, M.~Banck, C.~A. James, C.~Morley, T.~Vandermeersch and G.~R.
  Hutchison, \emph{J. Cheminform.}, 2011, \textbf{3}, 33\relax
\mciteBstWouldAddEndPuncttrue
\mciteSetBstMidEndSepPunct{\mcitedefaultmidpunct}
{\mcitedefaultendpunct}{\mcitedefaultseppunct}\relax
\EndOfBibitem
\bibitem[Hohenberg and Kohn(1964)]{Hohenberg-Kohn}
P.~Hohenberg and W.~Kohn, \emph{Phys. Rev.}, 1964, \textbf{136},
  B864--B871\relax
\mciteBstWouldAddEndPuncttrue
\mciteSetBstMidEndSepPunct{\mcitedefaultmidpunct}
{\mcitedefaultendpunct}{\mcitedefaultseppunct}\relax
\EndOfBibitem
\bibitem[Kohn and Sham(1965)]{Kohn-Sham}
W.~Kohn and L.~Sham, \emph{Phys. Rev.}, 1965, \textbf{140}, A1133--A1138\relax
\mciteBstWouldAddEndPuncttrue
\mciteSetBstMidEndSepPunct{\mcitedefaultmidpunct}
{\mcitedefaultendpunct}{\mcitedefaultseppunct}\relax
\EndOfBibitem
\bibitem[Becke(1988)]{becke1988density}
A.~D. Becke, \emph{Phys. Rev. A}, 1988, \textbf{38}, 3098\relax
\mciteBstWouldAddEndPuncttrue
\mciteSetBstMidEndSepPunct{\mcitedefaultmidpunct}
{\mcitedefaultendpunct}{\mcitedefaultseppunct}\relax
\EndOfBibitem
\bibitem[Lee \emph{et~al.}(1988)Lee, Yang, and Parr]{lee1988development}
C.~Lee, W.~Yang and R.~G. Parr, \emph{Phys. Rev. B}, 1988, \textbf{37},
  785\relax
\mciteBstWouldAddEndPuncttrue
\mciteSetBstMidEndSepPunct{\mcitedefaultmidpunct}
{\mcitedefaultendpunct}{\mcitedefaultseppunct}\relax
\EndOfBibitem
\bibitem[Grimme \emph{et~al.}(2010)Grimme, Antony, Ehrlich, and
  Krieg]{grimme2010consistent}
S.~Grimme, J.~Antony, S.~Ehrlich and H.~Krieg, \emph{J. Chem. Phys.}, 2010,
  \textbf{132}, year\relax
\mciteBstWouldAddEndPuncttrue
\mciteSetBstMidEndSepPunct{\mcitedefaultmidpunct}
{\mcitedefaultendpunct}{\mcitedefaultseppunct}\relax
\EndOfBibitem
\bibitem[Weigend and Ahlrichs(2005)]{weigend2005balanced}
F.~Weigend and R.~Ahlrichs, \emph{Phys. Chem. Chem. Phys.}, 2005, \textbf{7},
  3297--3305\relax
\mciteBstWouldAddEndPuncttrue
\mciteSetBstMidEndSepPunct{\mcitedefaultmidpunct}
{\mcitedefaultendpunct}{\mcitedefaultseppunct}\relax
\EndOfBibitem
\bibitem[Griffiths \emph{et~al.}(2021)Griffiths, F{\"o}ldes, de~Nijs,
  Chikkaraddy, Wright, Deacon, Berta, Readman, Grys,
  Rosta,\emph{et~al.}]{griffiths2021resolving}
J.~Griffiths, T.~F{\"o}ldes, B.~de~Nijs, R.~Chikkaraddy, D.~Wright, W.~M.
  Deacon, D.~Berta, C.~Readman, D.-B. Grys, E.~Rosta \emph{et~al.},
  \emph{Nature Communications}, 2021, \textbf{12}, 6759\relax
\mciteBstWouldAddEndPuncttrue
\mciteSetBstMidEndSepPunct{\mcitedefaultmidpunct}
{\mcitedefaultendpunct}{\mcitedefaultseppunct}\relax
\EndOfBibitem
\bibitem[Boehmke~Amoruso \emph{et~al.}(2024)Boehmke~Amoruso, Boto, Elliot,
  de~Nijs, Esteban, F{\"o}ldes, Aguilar-Galindo, Rosta, Aizpurua, and
  Baumberg]{boehmke2024uncovering}
A.~Boehmke~Amoruso, R.~A. Boto, E.~Elliot, B.~de~Nijs, R.~Esteban,
  T.~F{\"o}ldes, F.~Aguilar-Galindo, E.~Rosta, J.~Aizpurua and J.~J. Baumberg,
  \emph{Nature Communications}, 2024, \textbf{15}, 6733\relax
\mciteBstWouldAddEndPuncttrue
\mciteSetBstMidEndSepPunct{\mcitedefaultmidpunct}
{\mcitedefaultendpunct}{\mcitedefaultseppunct}\relax
\EndOfBibitem
\bibitem[Wright \emph{et~al.}(2021)Wright, Lin, Berta, F{\"o}ldes, Wagner,
  Griffiths, Readman, Rosta, Reisner, and Baumberg]{wright2021mechanistic}
D.~Wright, Q.~Lin, D.~Berta, T.~F{\"o}ldes, A.~Wagner, J.~Griffiths,
  C.~Readman, E.~Rosta, E.~Reisner and J.~J. Baumberg, \emph{Nature Catalysis},
  2021, \textbf{4}, 157--163\relax
\mciteBstWouldAddEndPuncttrue
\mciteSetBstMidEndSepPunct{\mcitedefaultmidpunct}
{\mcitedefaultendpunct}{\mcitedefaultseppunct}\relax
\EndOfBibitem
\bibitem[Frisch \emph{et~al.}(2016)Frisch, Trucks, Schlegel, Scuseria, Robb,
  Cheeseman, Scalmani, Barone, Petersson, Nakatsuji, Li, Caricato, Marenich,
  Bloino, Janesko, Gomperts, Mennucci, Hratchian, Ortiz, Izmaylov, Sonnenberg,
  Williams-Young, Ding, Lipparini, Egidi, Goings, Peng, Petrone, Henderson,
  Ranasinghe, Zakrzewski, Gao, Rega, Zheng, Liang, Hada, Ehara, Toyota, Fukuda,
  Hasegawa, Ishida, Nakajima, Honda, Kitao, Nakai, Vreven, Throssell,
  Montgomery, Peralta, Ogliaro, Bearpark, Heyd, Brothers, Kudin, Staroverov,
  Keith, Kobayashi, Normand, Raghavachari, Rendell, Burant, Iyengar, Tomasi,
  Cossi, Millam, Klene, Adamo, Cammi, Ochterski, Martin, Morokuma, Farkas,
  Foresman, and Fox]{g16}
M.~J. Frisch, G.~W. Trucks, H.~B. Schlegel, G.~E. Scuseria, M.~A. Robb, J.~R.
  Cheeseman, G.~Scalmani, V.~Barone, G.~A. Petersson, H.~Nakatsuji, X.~Li,
  M.~Caricato, A.~V. Marenich, J.~Bloino, B.~G. Janesko, R.~Gomperts,
  B.~Mennucci, H.~P. Hratchian, J.~V. Ortiz, A.~F. Izmaylov, J.~L. Sonnenberg,
  D.~Williams-Young, F.~Ding, F.~Lipparini, F.~Egidi, J.~Goings, B.~Peng,
  A.~Petrone, T.~Henderson, D.~Ranasinghe, V.~G. Zakrzewski, J.~Gao, N.~Rega,
  G.~Zheng, W.~Liang, M.~Hada, M.~Ehara, K.~Toyota, R.~Fukuda, J.~Hasegawa,
  M.~Ishida, T.~Nakajima, Y.~Honda, O.~Kitao, H.~Nakai, T.~Vreven,
  K.~Throssell, J.~A. Montgomery, {Jr.}, J.~E. Peralta, F.~Ogliaro, M.~J.
  Bearpark, J.~J. Heyd, E.~N. Brothers, K.~N. Kudin, V.~N. Staroverov, T.~A.
  Keith, R.~Kobayashi, J.~Normand, K.~Raghavachari, A.~P. Rendell, J.~C.
  Burant, S.~S. Iyengar, J.~Tomasi, M.~Cossi, J.~M. Millam, M.~Klene, C.~Adamo,
  R.~Cammi, J.~W. Ochterski, R.~L. Martin, K.~Morokuma, O.~Farkas, J.~B.
  Foresman and D.~J. Fox, \emph{Gaussian˜16 {R}evision {C}.01}, 2016, Gaussian
  Inc. Wallingford CT\relax
\mciteBstWouldAddEndPuncttrue
\mciteSetBstMidEndSepPunct{\mcitedefaultmidpunct}
{\mcitedefaultendpunct}{\mcitedefaultseppunct}\relax
\EndOfBibitem
\bibitem[Sch{\"u}tt \emph{et~al.}(2021)Sch{\"u}tt, Unke, and
  Gastegger]{Schütt21ICML_PaiNN}
K.~Sch{\"u}tt, O.~Unke and M.~Gastegger, in \emph{{Proceedings of the 38th
  International Conference on Machine Learning}}, ed. M.~Meila and T.~Zhang,
  Proceedings of Machine Learning Research, 2021, ch. {Equivariant message
  passing for the prediction of tensorial properties and molecular
  spectra}\relax
\mciteBstWouldAddEndPuncttrue
\mciteSetBstMidEndSepPunct{\mcitedefaultmidpunct}
{\mcitedefaultendpunct}{\mcitedefaultseppunct}\relax
\EndOfBibitem
\bibitem[Batatia \emph{et~al.}(2022)Batatia, Kovacs, Simm, Ortner, and
  Csanyi]{BatatiaNeurIPS22_MACE}
I.~Batatia, D.~P. Kovacs, G.~Simm, C.~Ortner and G.~Csanyi, in \emph{{Advances
  in Neural Information Processing Systems 35}}, ed. S.~Koyejo, S.~Mohamed,
  A.~Agarwal, D.~Belgrave, K.~Cho and A.~Oh, NeurIPS Proceedings, 2022, ch.
  {MACE: Higher Order Equivariant Message Passing Neural Networks for Fast and
  Accurate Force Fields}\relax
\mciteBstWouldAddEndPuncttrue
\mciteSetBstMidEndSepPunct{\mcitedefaultmidpunct}
{\mcitedefaultendpunct}{\mcitedefaultseppunct}\relax
\EndOfBibitem
\bibitem[Himanen \emph{et~al.}(2020)Himanen, Jäger, Morooka, {Federici
  Canova}, Ranawat, Gao, Rinke, and Foster]{HIMANEN2020}
L.~Himanen, M.~O. Jäger, E.~V. Morooka, F.~{Federici Canova}, Y.~S. Ranawat,
  D.~Z. Gao, P.~Rinke and A.~S. Foster, \emph{Computer Physics Communications},
  2020, \textbf{247}, 106949\relax
\mciteBstWouldAddEndPuncttrue
\mciteSetBstMidEndSepPunct{\mcitedefaultmidpunct}
{\mcitedefaultendpunct}{\mcitedefaultseppunct}\relax
\EndOfBibitem
\bibitem[Zhang \emph{et~al.}(1997)Zhang, Ramakrishnan, and Livny]{BIRCH}
T.~Zhang, R.~Ramakrishnan and M.~Livny, \emph{Data Min. Knowl. Discov.}, 1997,
  \textbf{1}, 141--182\relax
\mciteBstWouldAddEndPuncttrue
\mciteSetBstMidEndSepPunct{\mcitedefaultmidpunct}
{\mcitedefaultendpunct}{\mcitedefaultseppunct}\relax
\EndOfBibitem
\bibitem[Schubert \emph{et~al.}(2017)Schubert, Sander, Ester, Kriegel, and
  Xu]{DBSCAN}
E.~Schubert, J.~Sander, M.~Ester, H.~P. Kriegel and X.~Xu, \emph{ACM Trans.
  Database Syst.}, 2017, \textbf{42}, 1--21\relax
\mciteBstWouldAddEndPuncttrue
\mciteSetBstMidEndSepPunct{\mcitedefaultmidpunct}
{\mcitedefaultendpunct}{\mcitedefaultseppunct}\relax
\EndOfBibitem
\bibitem[Genheden \emph{et~al.}(2020)Genheden, Thakkar, Chadimov{\'a}, Reymond,
  Engkvist, and Bjerrum]{GenhedenJC20_AiZynthFinder}
S.~Genheden, A.~Thakkar, V.~Chadimov{\'a}, J.-L. Reymond, O.~Engkvist and
  E.~Bjerrum, \emph{J. Cheminform.}, 2020, \textbf{12}, 70\relax
\mciteBstWouldAddEndPuncttrue
\mciteSetBstMidEndSepPunct{\mcitedefaultmidpunct}
{\mcitedefaultendpunct}{\mcitedefaultseppunct}\relax
\EndOfBibitem
\bibitem[Sterling and Irwin(2015)]{SterlingJCIM15_ZINC15}
T.~Sterling and J.~J. Irwin, \emph{J. Chem. Inf. Model.}, 2015, \textbf{55},
  2324--2337\relax
\mciteBstWouldAddEndPuncttrue
\mciteSetBstMidEndSepPunct{\mcitedefaultmidpunct}
{\mcitedefaultendpunct}{\mcitedefaultseppunct}\relax
\EndOfBibitem
\bibitem[Thakkar \emph{et~al.}(2020)Thakkar, Kogej, Reymond, Engkvist, and
  Bjerrum]{ThakkarCS20}
A.~Thakkar, T.~Kogej, J.-L. Reymond, O.~Engkvist and E.~J. Bjerrum, \emph{Chem.
  Sci.}, 2020, \textbf{11}, 154--168\relax
\mciteBstWouldAddEndPuncttrue
\mciteSetBstMidEndSepPunct{\mcitedefaultmidpunct}
{\mcitedefaultendpunct}{\mcitedefaultseppunct}\relax
\EndOfBibitem
\bibitem[Lowe(2017)]{Lowe17}
D.~Lowe, \emph{{Chemical reactions from US patents (1976-Sep2016)}}, 2017,
  \url{https://doi.org/10.6084/m9.figshare.5104873.v1}, (accessed November 13,
  2024)\relax
\mciteBstWouldAddEndPuncttrue
\mciteSetBstMidEndSepPunct{\mcitedefaultmidpunct}
{\mcitedefaultendpunct}{\mcitedefaultseppunct}\relax
\EndOfBibitem
\bibitem[Pub()]{PubChem}
\emph{{PubChem}}, \url{https://pubchem.ncbi.nlm.nih.gov/}, (accessed February
  12, 2025)\relax
\mciteBstWouldAddEndPuncttrue
\mciteSetBstMidEndSepPunct{\mcitedefaultmidpunct}
{\mcitedefaultendpunct}{\mcitedefaultseppunct}\relax
\EndOfBibitem
\bibitem[Kim \emph{et~al.}(2025)Kim, Chen, Cheng, Gindulyte, He, He, Li,
  Shoemaker, Thiessen, Yu, Zaslavsky, Zhang, and Bolton]{KimNAR25}
S.~Kim, J.~Chen, T.~Cheng, A.~Gindulyte, J.~He, S.~He, Q.~Li, B.~A. Shoemaker,
  P.~A. Thiessen, B.~Yu, L.~Zaslavsky, J.~Zhang and E.~E. Bolton, \emph{Nucleic
  Acids Res.}, 2025, \textbf{53}, D1516--D1525\relax
\mciteBstWouldAddEndPuncttrue
\mciteSetBstMidEndSepPunct{\mcitedefaultmidpunct}
{\mcitedefaultendpunct}{\mcitedefaultseppunct}\relax
\EndOfBibitem
\bibitem[Pub()]{PubChemPy}
\emph{{PubChemPy documentation}},
  \url{https://pubchempy.readthedocs.io/en/latest/}, (accessed February 12,
  2025)\relax
\mciteBstWouldAddEndPuncttrue
\mciteSetBstMidEndSepPunct{\mcitedefaultmidpunct}
{\mcitedefaultendpunct}{\mcitedefaultseppunct}\relax
\EndOfBibitem
\bibitem[Gebauer(2024)]{gebauer_autoregressive_2024}
N.~W.~A. Gebauer, \emph{PhD thesis}, Technische Universit{\"a}t Berlin,
  2024\relax
\mciteBstWouldAddEndPuncttrue
\mciteSetBstMidEndSepPunct{\mcitedefaultmidpunct}
{\mcitedefaultendpunct}{\mcitedefaultseppunct}\relax
\EndOfBibitem
\bibitem[Ruddigkeit \emph{et~al.}(2012)Ruddigkeit, van Deursen, Blum, and
  Reymond]{RuddigkeitJCIM12}
L.~Ruddigkeit, R.~van Deursen, L.~C. Blum and J.-L. Reymond, \emph{J. Chem.
  Inf. Model.}, 2012, \textbf{52}, 2864--2875\relax
\mciteBstWouldAddEndPuncttrue
\mciteSetBstMidEndSepPunct{\mcitedefaultmidpunct}
{\mcitedefaultendpunct}{\mcitedefaultseppunct}\relax
\EndOfBibitem
\bibitem[Ramakrishnan \emph{et~al.}(2014)Ramakrishnan, Dral, Rupp, and
  Anatole~von Lilienfeld]{RamakrishnanSD14}
R.~Ramakrishnan, P.~O. Dral, M.~Rupp and O.~Anatole~von Lilienfeld, \emph{Sci.
  Data}, 2014, \textbf{1}, 140022\relax
\mciteBstWouldAddEndPuncttrue
\mciteSetBstMidEndSepPunct{\mcitedefaultmidpunct}
{\mcitedefaultendpunct}{\mcitedefaultseppunct}\relax
\EndOfBibitem
\bibitem[CID()]{CID_67981805}
\emph{{PubChem Compound Summary for CID 67981805, 2,4,5-Triaminobenzenethiol}},
  \url{https://pubchem.ncbi.nlm.nih.gov/compound/2_4_5-Triaminobenzenethiol},
  (accessed February 12, 2025)\relax
\mciteBstWouldAddEndPuncttrue
\mciteSetBstMidEndSepPunct{\mcitedefaultmidpunct}
{\mcitedefaultendpunct}{\mcitedefaultseppunct}\relax
\EndOfBibitem
\bibitem[Koczor-Benda \emph{et~al.}(2025)Koczor-Benda, Chaudhuri, Gilkes,
  Bartucca, Li, and Maurer]{figshare}
Z.~Koczor-Benda, S.~Chaudhuri, J.~Gilkes, F.~Bartucca, L.~Li and R.~J. Maurer,
  \emph{{G-SchNet for THz Radiation Detection}}, 2025,
  \url{https://doi.org/10.6084/m9.figshare.28539995.v1}, (accessed March 10,
  2025)\relax
\mciteBstWouldAddEndPuncttrue
\mciteSetBstMidEndSepPunct{\mcitedefaultmidpunct}
{\mcitedefaultendpunct}{\mcitedefaultseppunct}\relax
\EndOfBibitem
\end{mcitethebibliography}
\providecommand{\noopsort}[1]{}\providecommand{\singleletter}[1]{#1}%
\providecommand*{\mcitethebibliography}{\thebibliography}
\csname @ifundefined\endcsname{endmcitethebibliography}
{\let\endmcitethebibliography\endthebibliography}{}

\end{document}


\thispagestyle{plain}
\maketitle
\tableofcontents
\newpage 

\section{Training Database Information}
Table~\ref{tab:Filtering_Statistics} details the sizes of the databases for each biasing iteration, before and after filtering subject to constraints relating to connectivity (a path should exist between any two atoms over bonds), uniqueness (no two structures should possess the same SMILES~\cite{SMILES} representation), and a sanity check based on RDKit~\cite{RDKit} to check atomic valencies. Molecules were also filtered to ensure the presence of only one thiol group, which would act as the linker to the gold nanoantenna in a terahertz radiation detector.

The iterative training procedure employed in this work slightly differs from the methodology outlined by \citet{WestermayrNCS23} They performed iterative biasing by using a previously trained G-SchNet model and retraining it only on the small subset of molecules. Herein, we retrain G-SchNet from scratch in each iteration with the modified training dataset.

\begin{table}[H]
    \centering
    \begin{tabular}{c|ccc}
        Iteration & Number of Generated Molecules & Number of Filtered Molecules & Number of Filtered Monothiols \\ \hline
        0 & \num{62592} & \num{38688} & \num{25890} \\
        1 & \num{84592} & \num{56687} & \num{41285} \\
        2 & \num{90238} & \num{59110} & \num{46028} \\
        3 & \num{92712} & \num{58343} & \num{45986} \\
        4 & \num{92918} & \num{57339} & \num{45043} \\
        5 & \num{93426} & \num{55942} & \num{46426} \\
        6 & \num{95702} & \num{56358} & \num{46799} \\
    \end{tabular}
    \caption{Sizes of databases of generated molecules for each biasing iteration, before and after filtering subject to constraints relating to connectivity, uniqueness, atomic valencies, and an additional filter for monothiolated molecules.}
    \label{tab:Filtering_Statistics}
\end{table}

Table~\ref{tab:Training_Database_Sizes} details the sizes of the training databases used to train G-SchNet models within each biasing iteration. 

\begin{table}[H]
    \centering
    \begin{tabular}{c|c}
        Iteration & Database Size \\ \hline
        0 & \num{30000} \\
        1 & \num{31146} \\
        2 & \num{35990} \\
        3 & \num{41236} \\
        4 & \num{45939} \\
        5 & \num{50405} \\
        6 & \num{54,739} \\
    \end{tabular}
    \caption{Sizes (number of molecules) of the training databases for each biasing iteration.}
    \label{tab:Training_Database_Sizes}
\end{table}

\section{P Value Definition}
\citet{Koczor-BendaPRX21} defined the target property P to the logarithm of the orientation-averaged upconversion intensity ($I_m^c$) summed for the 1--\SI{30}{\tera\hertz} frequency window:
\begin{equation}
    \mathrm{P} = \log{\bigg(\sum_{m\in M} \langle I_m^c \rangle\bigg)},
    \label{eq:P}
\end{equation}
where $M$ is the set of vibrational normal modes of the molecule in the 1--\SI{30}{\tera\hertz} (30--\SI{1000}{\per\cm}) frequency range. $I_m^c$ is defined as \cite{Koczor-BendaPRX21}:
\begin{equation}
    I_m^c = C\dfrac{(\overline{\nu}^\mathrm{aS}+\overline{\nu}_m)^4}{\overline{\nu}_m} \Big\langle \left|\underline{e}\underline{\mu}'_m\right|^2 \left|\underline{e}\doubleunderline{\alpha}'_m\underline{e}\right|^2 \Big\rangle
    \label{eq:I_m^c}
\end{equation}
where $C$ is a constant scaling factor ($6.026324\times10^{-42}$), $\overline{\nu}^\mathrm{aS}$ and $\overline{\nu}_m$ are the wavenumbers of the visible laser and the normal mode, respectively, $\underline{\mu}'_m$ is the dipole derivative vector, and $\doubleunderline{\alpha}'_m$ is the polarizability derivative tensor. The aligned terahertz and Raman in-out field polarization vectors are all denoted by $\underline{e}$. The visible laser wavelength was set to 785 nm and the temperature to 298.15 K, similarly to \citet{Koczor-BendaPRX21,Koczor-BendaJPCA22}. The scaling factor and the analytical formula for calculating the matrix elements in Equation \ref{eq:I_m^c} are detailed in \citet{Koczor-BendaPRX21}.

\section{Hyperparameter Optimization for PaiNN and MACE}
\begin{figure}[H]
    \centering
    \includegraphics[width=6.5in]{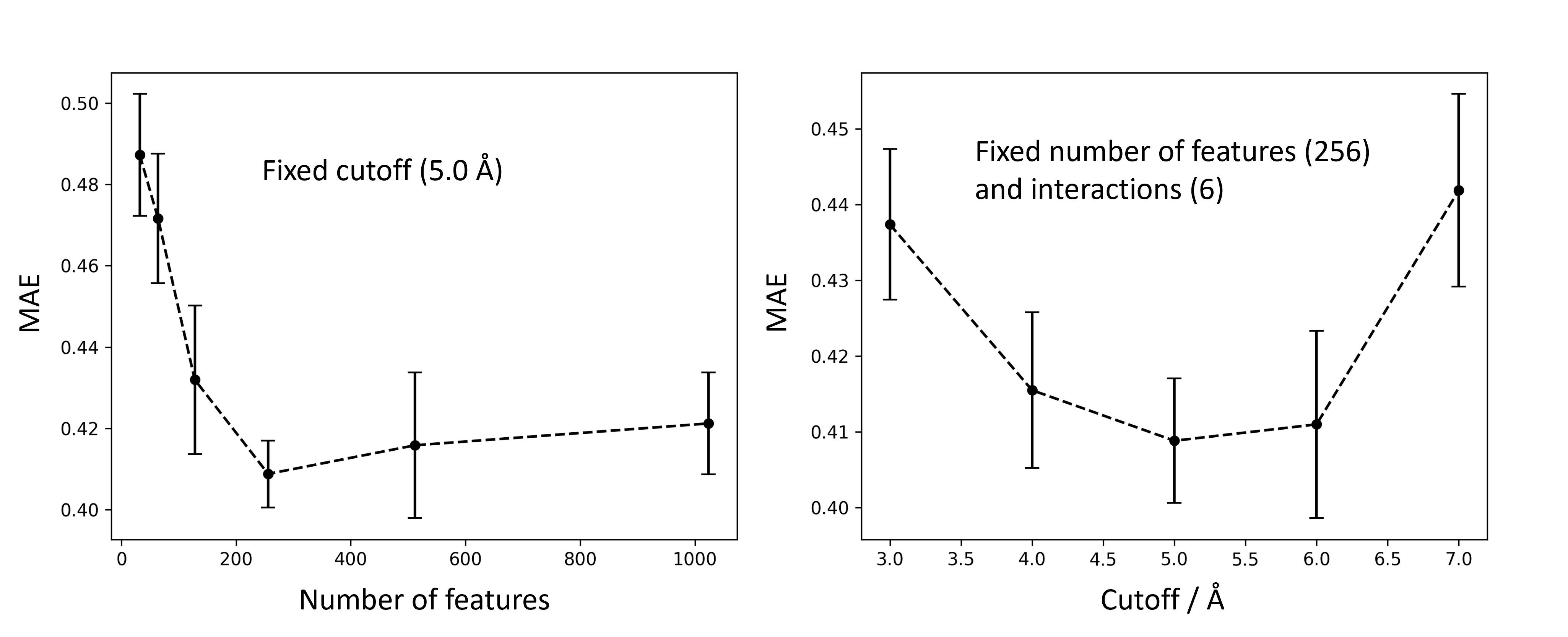}
    \caption{Hyperparameter optimization for PaiNN. For number of features, only combinations with number of interaction values specified in Table \ref{tab:painn_params} were tested. The mean and standard deviation of 5-fold cross-validation results are shown for each point.}
    \label{fig:painn_hyperparam}
\end{figure}

\begin{table}[]
    \centering
    \begin{tabular}{c|cccccc}
        Features	&32	&64	&128	&256	&612	&1024\\	
        \hline
        Interactions &3 	&4 	&5	&6	&7	 &8
    \end{tabular}
    \caption{Combination of hyperparameter values tested for number of features and interactions for the PaiNN model.}
    \label{tab:painn_params}
\end{table}

\begin{figure}[H]
    \centering
    \includegraphics[width=6.5in]{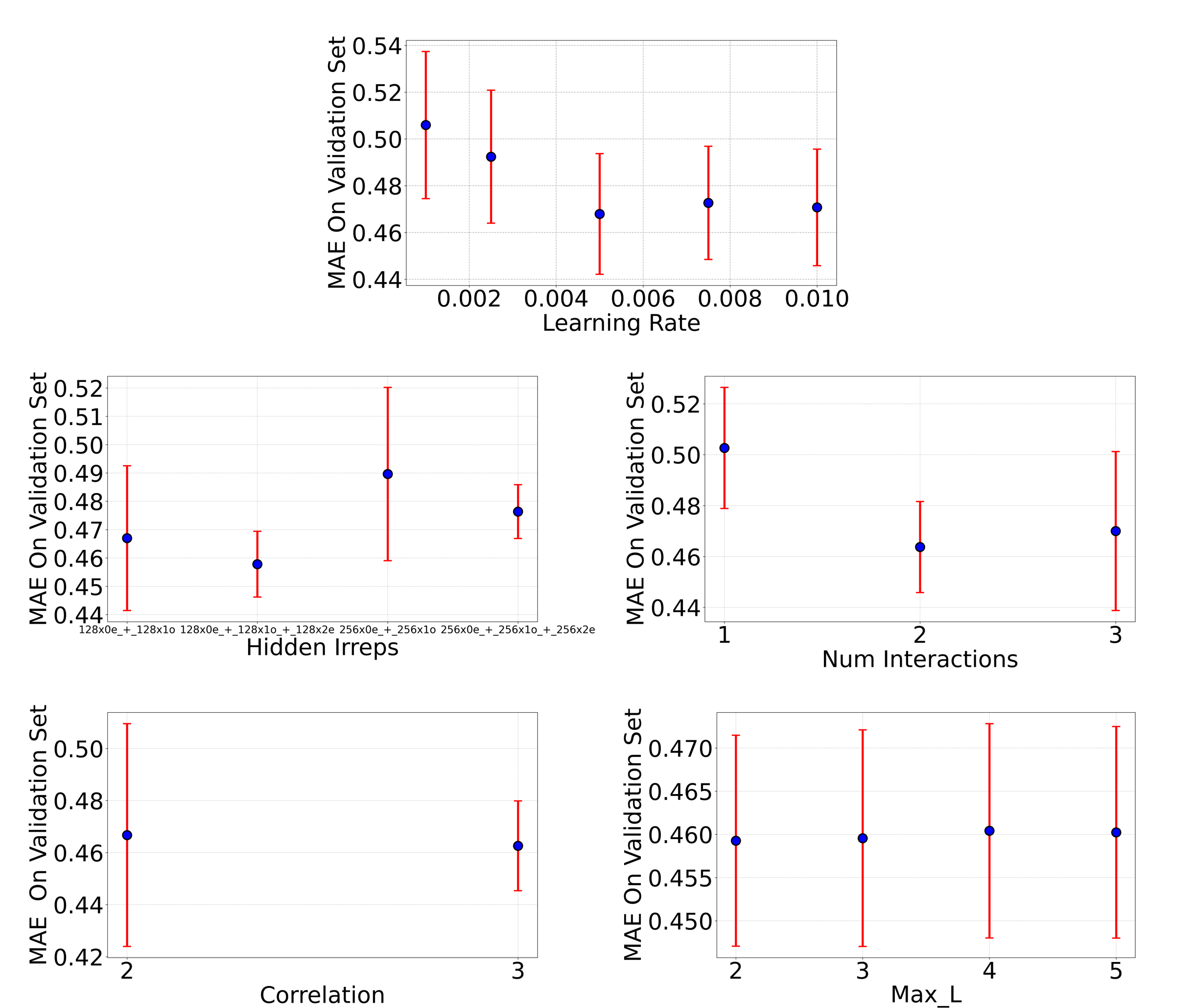}
    \caption{Hyperparameter optimization for MACE with 5-fold cross-validation.}
    \label{fig:mace_hyperparam}
\end{figure}

\begin{table}
    \centering
    \begin{tabular}{c|c}
        Hyperparameters & Best value \\
        \hline
        \makecell{Learning Rate(lr)} & 0.005 \\
        \makecell{Hidden Irreps(hidden\_irreps)} & 128x0e+128x1o+128x2e \\
        \makecell{Number of Interaction Layers(num\_interactions)} & 2 \\
        \makecell{Correlation Order(correlation)} & 3 \\
        \makecell{Angular Resolution(max\_L)} & 2 \\
    \end{tabular}
    \caption{The optimized values of 5 hyperparameters for MACE.}
    \label{tab:opt}
\end{table}

\begin{figure}[H]
    \centering
    \includegraphics[width=0.6\linewidth]{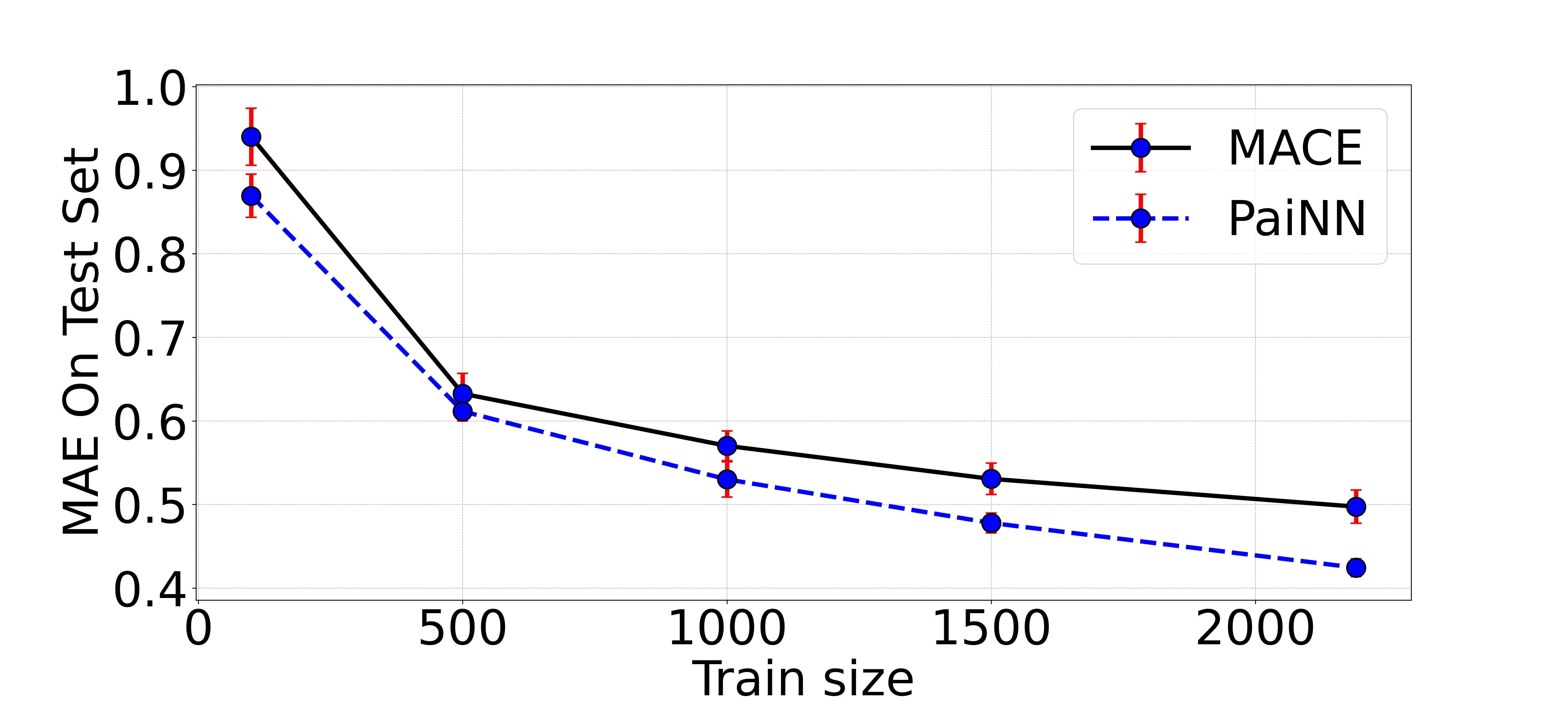}
    \caption{Mean absolute error (MAE) of MACE and PaiNN predictions for P values on the test set of the THz database as a function of training set size, using 5-fold cross-validation.}
    \label{fig:mace_painn_learning_curves}
\end{figure}

\section{Analysis of Generated Molecules}

\begin{figure}[H]
    \centering
    \includegraphics[width=7in]{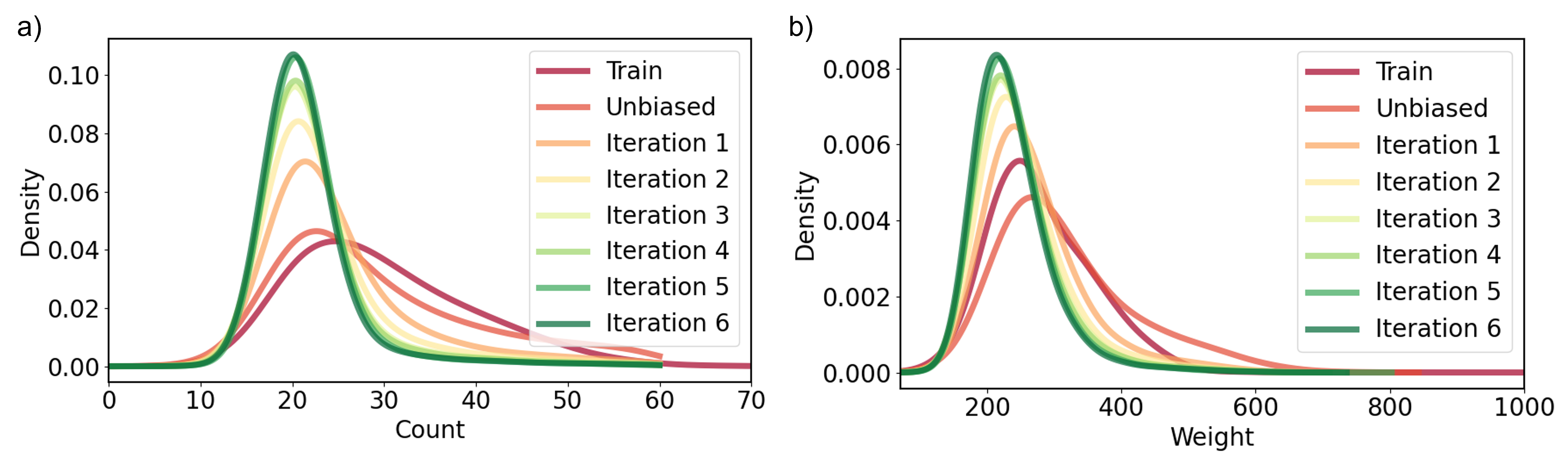}
    \caption{a) Number of atoms and b) molecular weight distribution of training and generated molecules.}
    \label{fig:nats}
\end{figure}

\begin{figure}[H]
    \centering
    \includegraphics[width=0.35\linewidth]{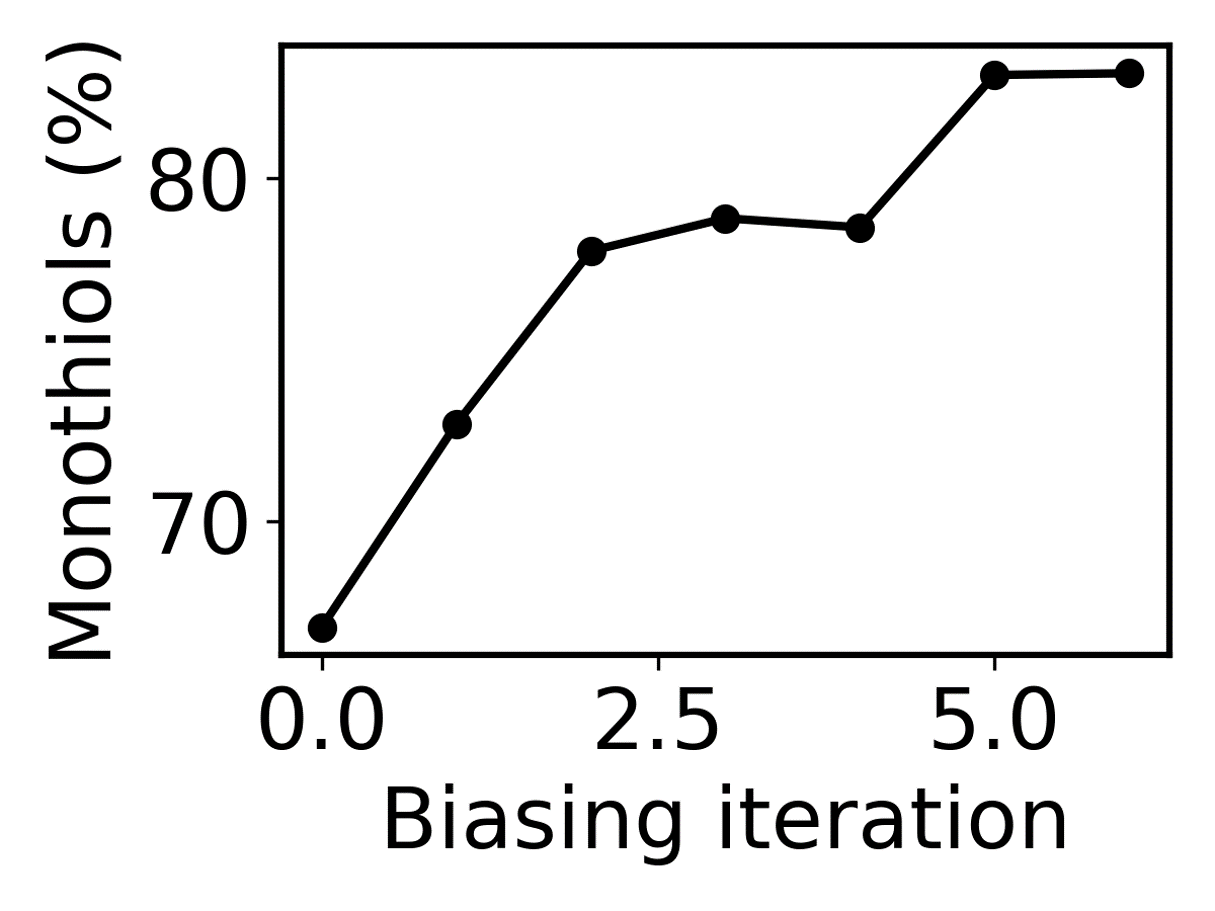}
    \caption{Proportion of molecules containing a single thiol group in the training set and the generated molecules.}
    \label{fig:thiols_biasing}
\end{figure}

\section{Accuracy of P Value Predictions}

Following \citet{Koczor-BendaPRX21}, 65\% of the molecules from the THz database were used for training (1,752), 20\% were used for testing (546) and the remaining molecules were used for validation (438). For the Thiol database and Iteration 6, converged DFT calculations were obtained for 131 and 197 molecules, respectively, out of the 200 randomly selected molecules. These are used as test sets for the KRR and PaiNN models trained on the THz training set.

\begin{figure}[h]
    \centering
    \includegraphics[width=7in]{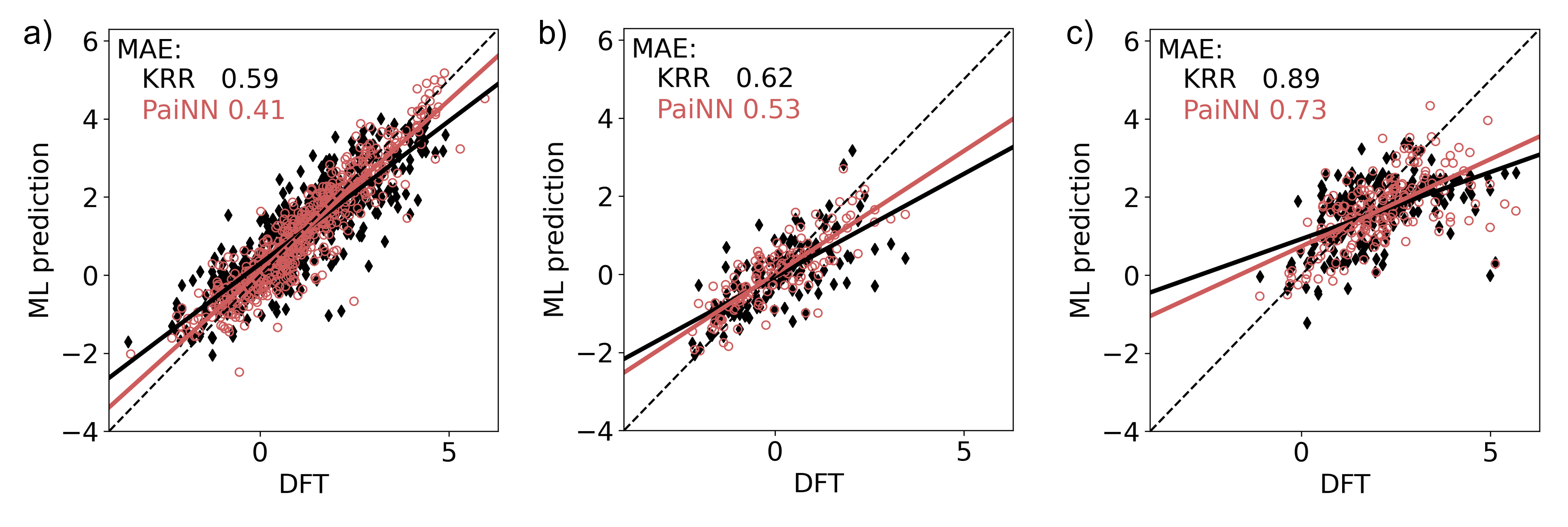}
    \caption{Accuracy of KRR (black diamonds) and PaiNN (pink circles) predictions for the P value for test molecules from the a) THz database b) Thiol database c) Molecules generated in Iteration 6. In the case of the PaiNN model, predictions are based on DFT-optimized molecular structures.}
    \label{fig:krr_painn}
\end{figure}

\section{Retraining the PaiNN Property Predictor}

\begin{figure}[H]
    \centering
    \includegraphics[width=4.5in]{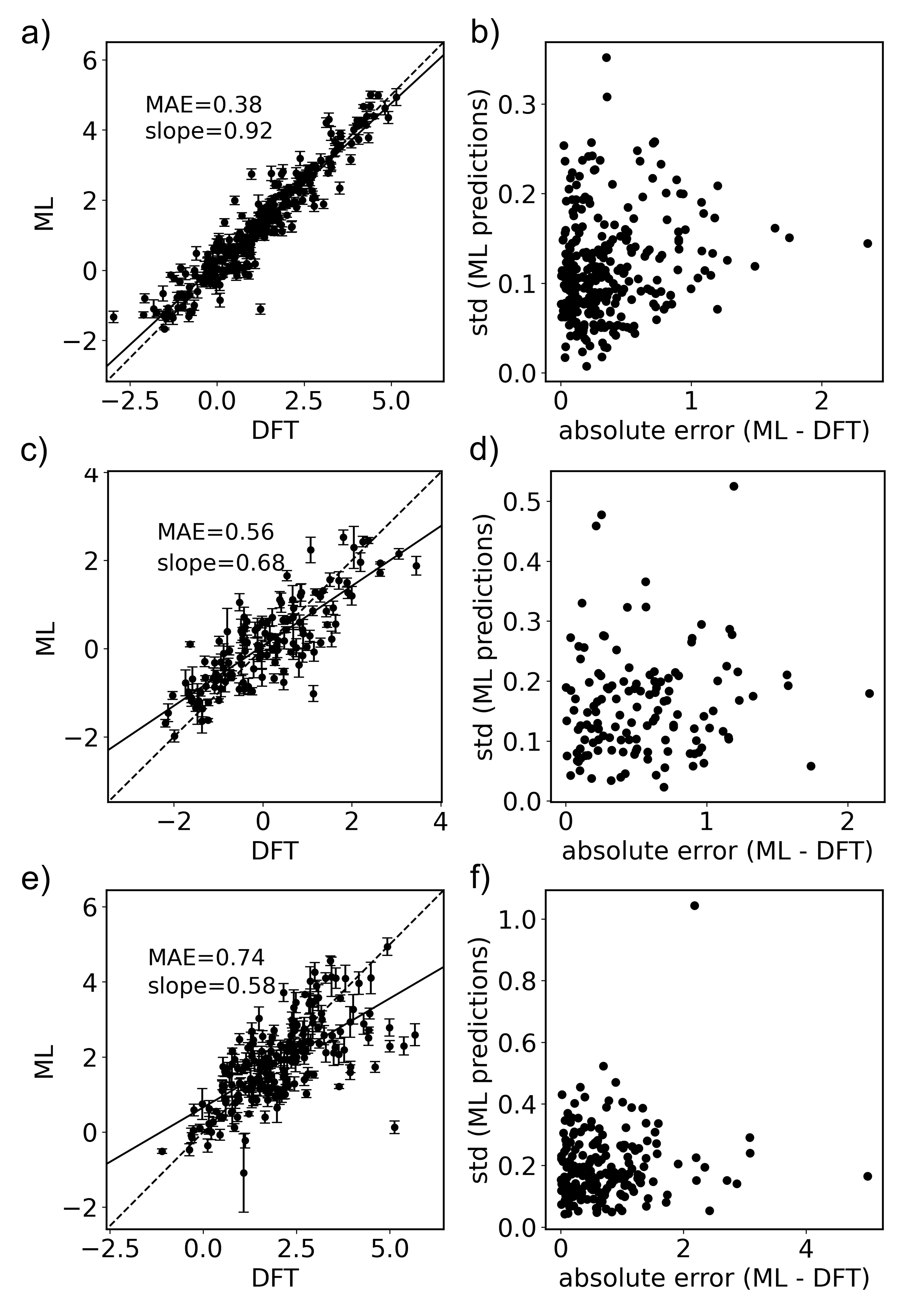}
    \caption{P value predictions of a committee of 5 PaiNN models trained on the THz set for test molecules from a) the THz database, c) the Thiol database, and e) Biasing Iteration 6. For all test molecules, the mean average and standard deviation of P values predicted by the 5 PaiNN models is shown. Standard deviation of the predictions versus absolute error of the mean of the predictions for test molecules from b) the THz database, d) the Thiol database, and f) Biasing Iteration 6. }
    \label{fig:painn_uncertainty}
\end{figure}

Next, a committee of 5 PaiNN models was trained using different train/validation sets (80\%/10\% split) and random seeds, and the same test set (10\%) across the 5 models (Figure \ref{fig:painn_uncertainty}). MAE values on the test sets in Figure  \ref{fig:painn_uncertainty} are slightly different from Figure \ref{fig:krr_painn} due to the different splitting of data. We find that the uncertainty of the predictions does not correlate with the accuracy of the predictions in any of these cases. The uncertainty of the committee tends to be very small compared to the range of P values and the prediction errors for most molecules. 

\begin{figure}[H]
    \centering
    \includegraphics[width=4.5in]{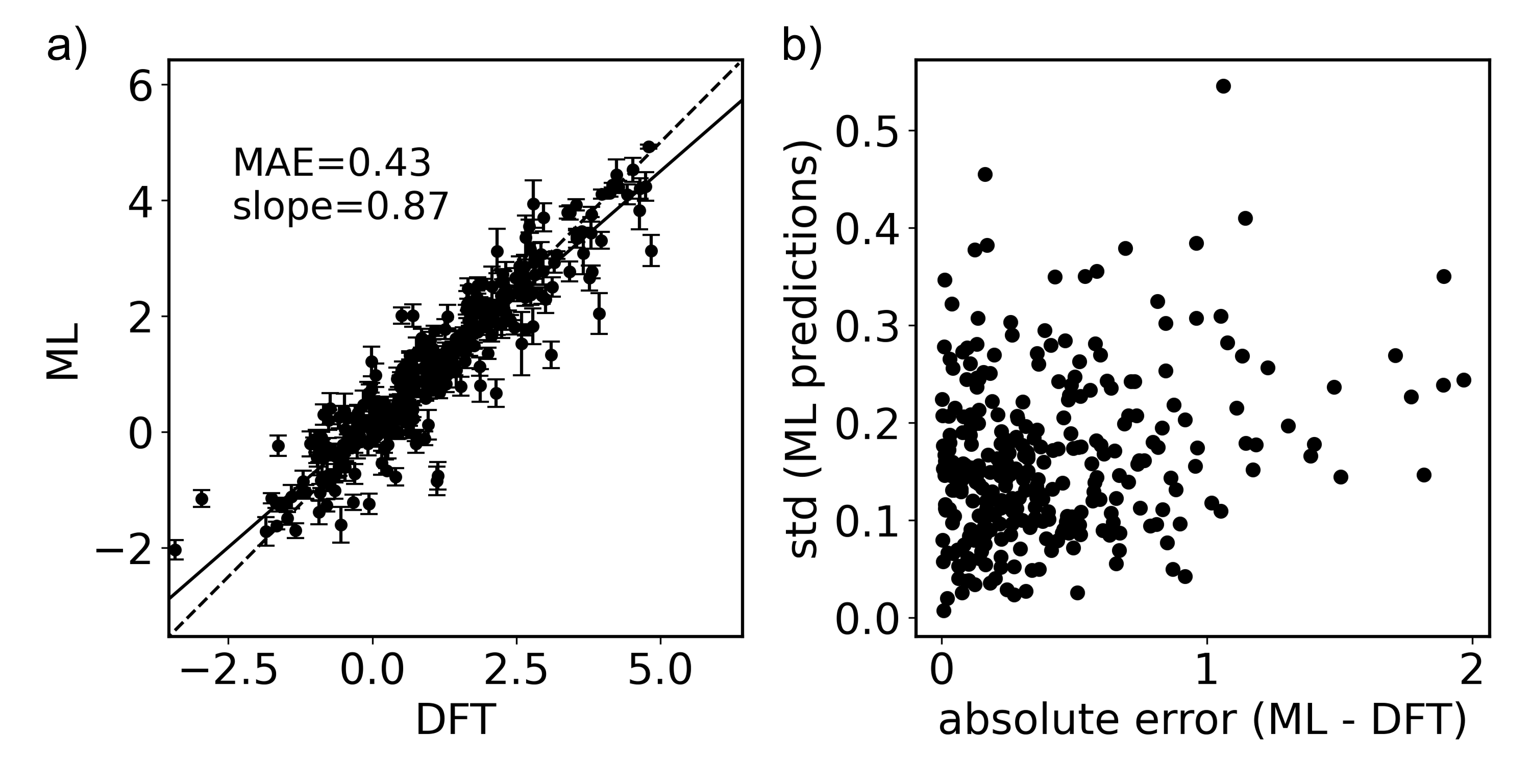}
    \caption{a) P value predictions on test molecules by a committee of 5 PaiNN models trained on the combined THz set plus randomly selected molecules from the Thiol database and Biasing Iteration 6. For all test molecules, the mean average and standard deviation of P values predicted by the 5 PaiNN models is shown. b) Standard deviation of the predictions versus absolute error of the mean of the predictions for test molecules.}
    \label{fig:painn_retrain}
\end{figure}

Finally, randomly selected molecules from the Thiol database and Iteration 6 were added to the training set, to improve the accuracy of predictions for generated molecules (Figure \ref{fig:painn_retrain}).

\section{Analysis of PaiNN Predictions for Generated Molecules}

\begin{figure}[H]
    \centering
    \includegraphics[width=\linewidth]{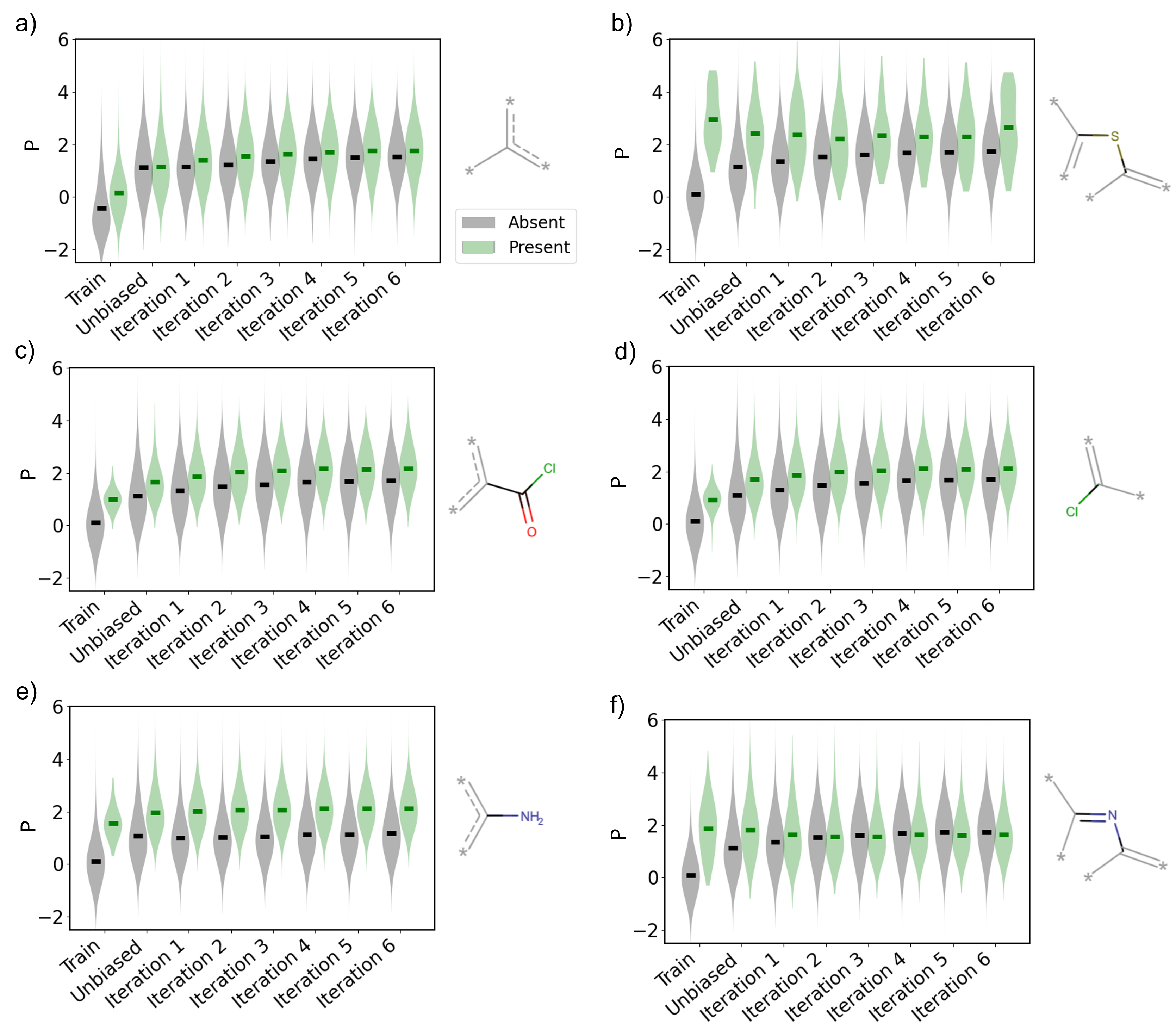}     
    \caption{Distribution and mean of P values predicted by PaiNN for molecules in which the functional group depicted is absent (gray) or present (green). }
    \label{fig:painn_P_funct}
\end{figure}

P values were predicted for all training and generated molecules using the retrained committee of PaiNN models, and the distribution of P values were plotted according to the absence or presence of specific functional groups identified by \citet{Koczor-BendaPRX21} as correlating with high P values (Fig. \ref{fig:painn_P_funct}). The presence of functional groups b), c), d) and e) show a correlation with higher P values, while the presence of a) and f) does not affect the P values apart from the training set. This suggests that the latter functional groups are themselves not correlated with higher P values, but they rather occur together with other molecular features that promote higher P values in the commercial thiol database (used for training).

\section{Chemical Space Mapping of Generated Molecules}

\begin{figure}[H]
    \centering
    \includegraphics[width=6in]{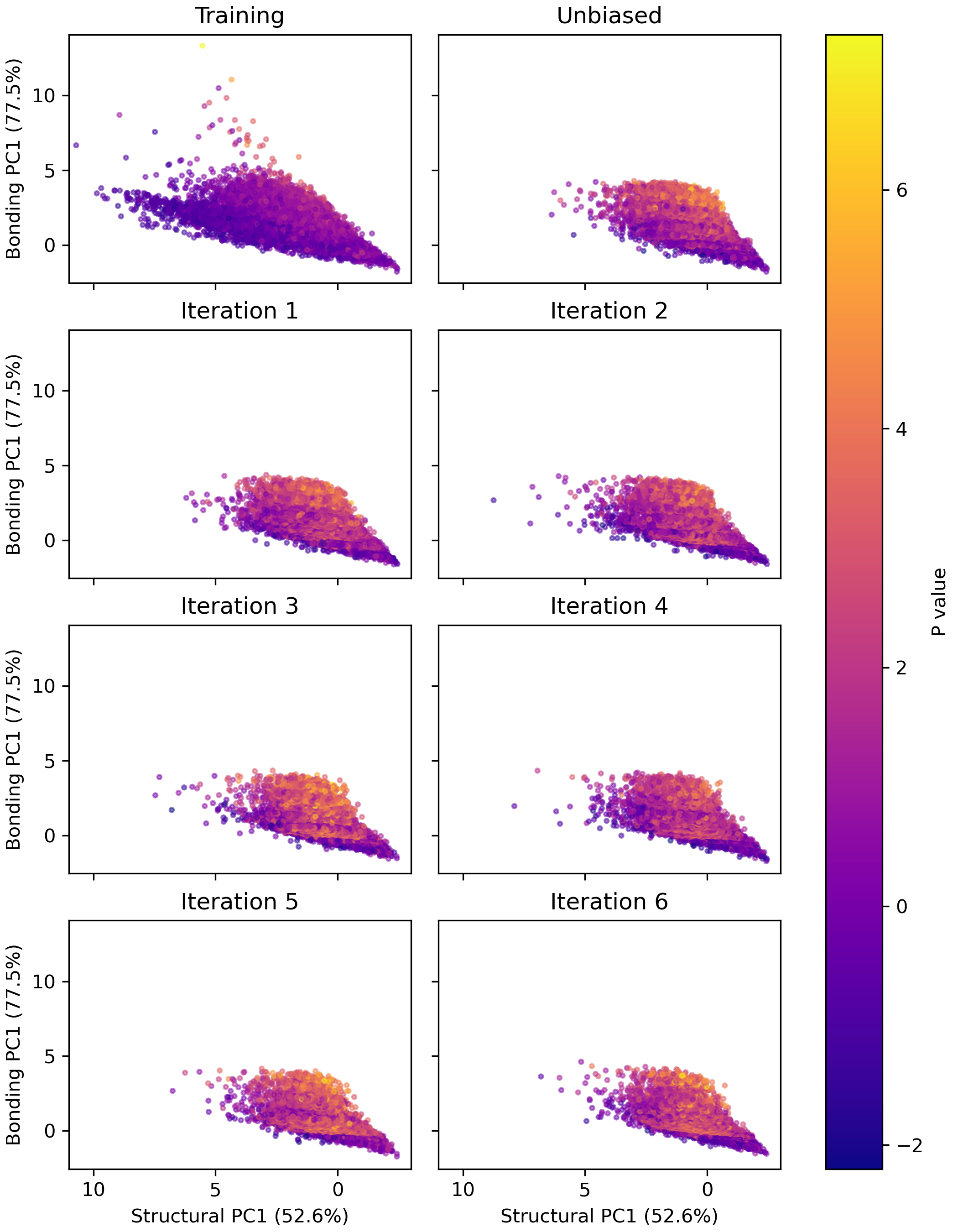}
    \caption{Latent chemical space plots of training molecules, and molecules generated in subsequent biasing iterations, colored by their PaiNN-predicted P values.}
    \label{fig:pca_per_iteration}
\end{figure}

\section{Retrosynthetic Paths and Vibrational Spectra of Candidate Molecules}
\begin{figure}[H]
    \centering
    \includegraphics[width=5.65in]{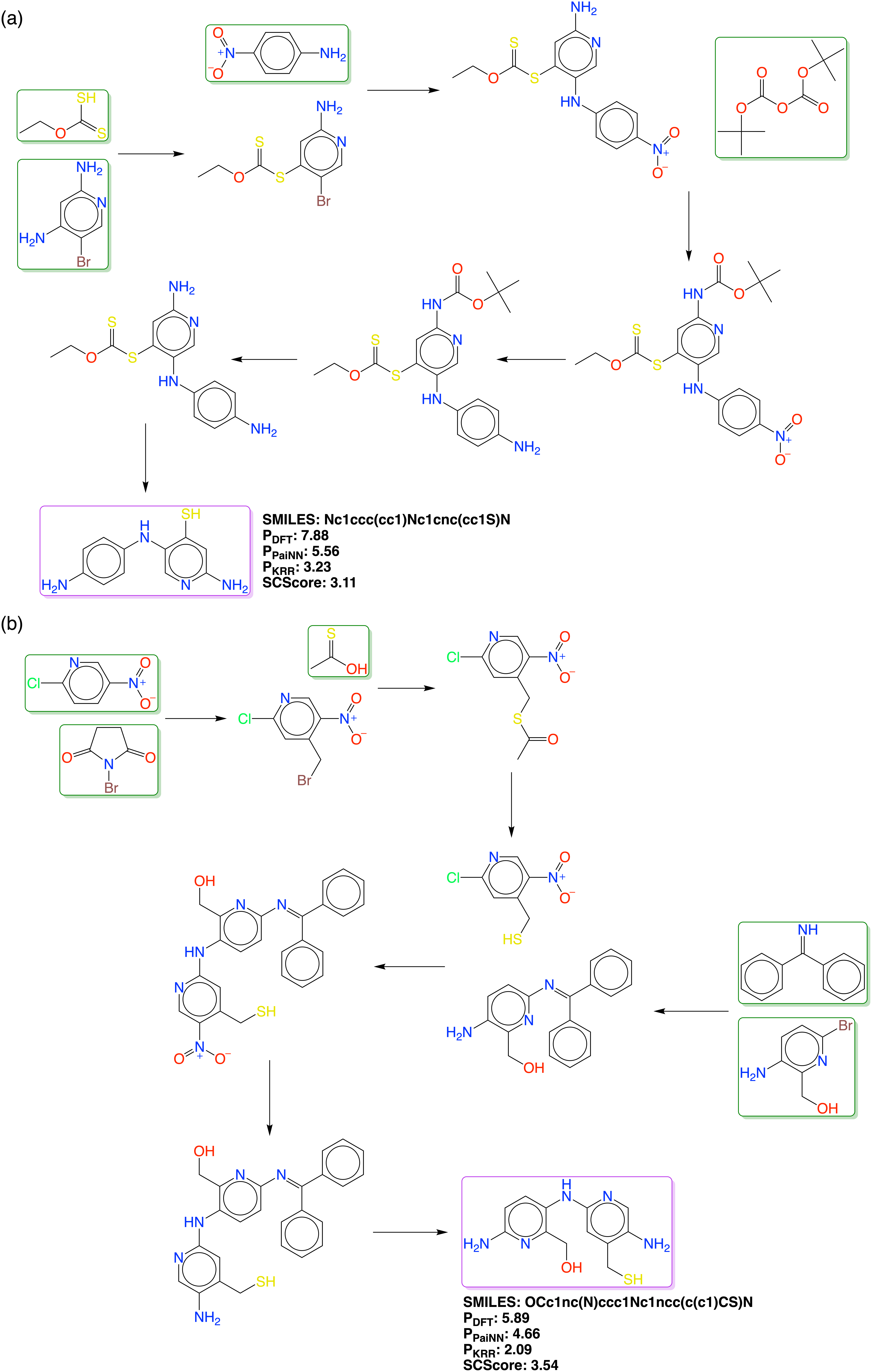}
    \caption{Top retrosynthetic paths for molecules with SMILES strings (a) Nc1ccc(cc1)Nc1cnc(cc1S)N and (b) OCc1nc(N)ccc1Nc1ncc(c(c1)CS)N. The final molecules are shown in purple boxes and precursors available within the ZINC database are shown in green boxes.}
    \label{fig:Retrosynthesis_1}
\end{figure}

\begin{figure}[H]
    \centering
    \includegraphics[width=6.2in]{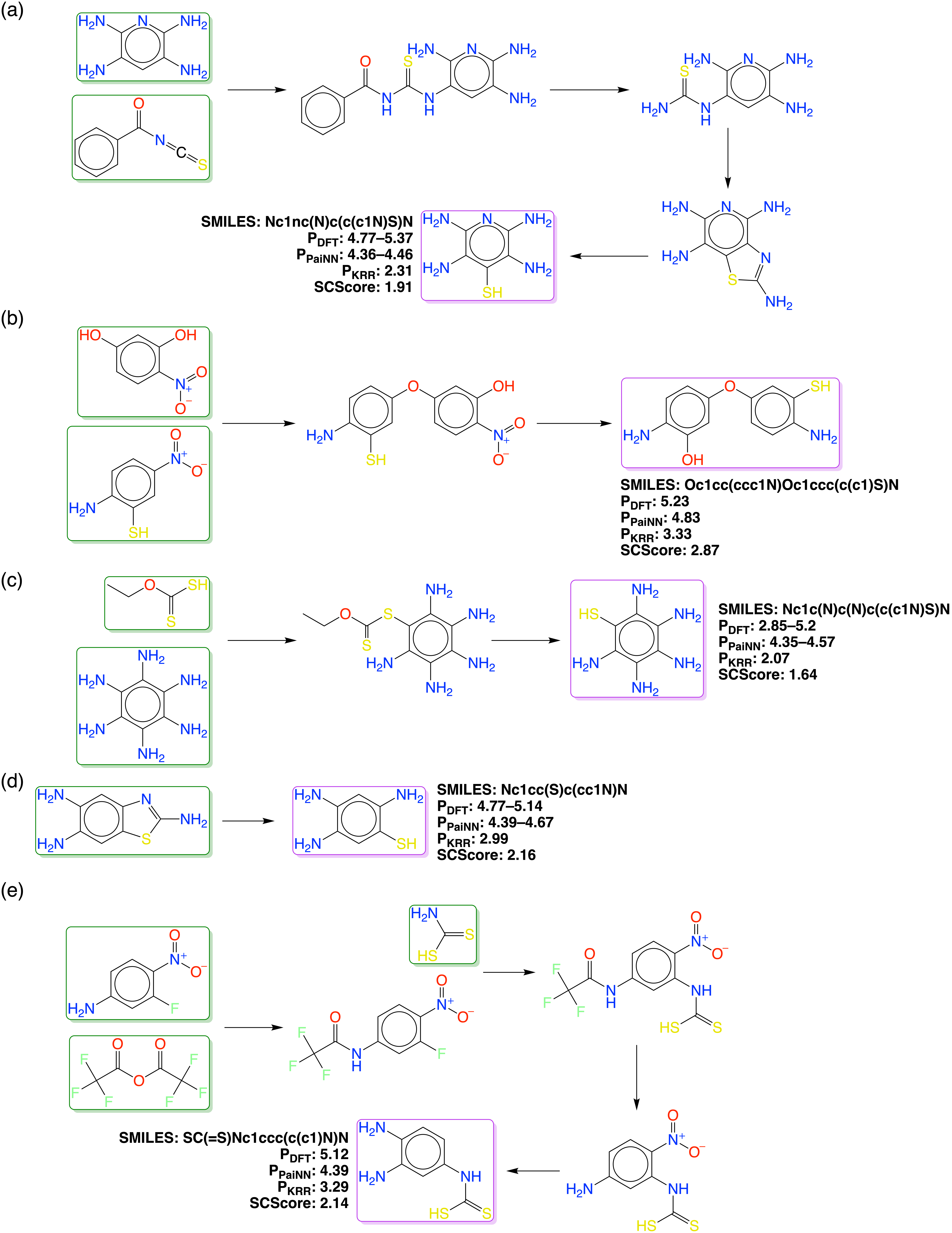}
    \caption{Top retrosynthetic paths for molecules with SMILES strings (a) Nc1nc(N)c(c(c1N)S)N, (b) Oc1cc(ccc1N)Oc1ccc(c(c1)S)N, (c) Nc1c(N)c(N)c(c(c1N)S)N, (d) Nc1cc(S)c(cc1N)N, and (e) SC(=S)Nc1ccc(c(c1)N)N. The final molecules are shown in purple boxes and precursors available within the ZINC database are shown in green boxes.}
    \label{fig:Retrosynthesis_2}
\end{figure}

\begin{figure}[H]
    \centering
    \includegraphics[width=6.2in]{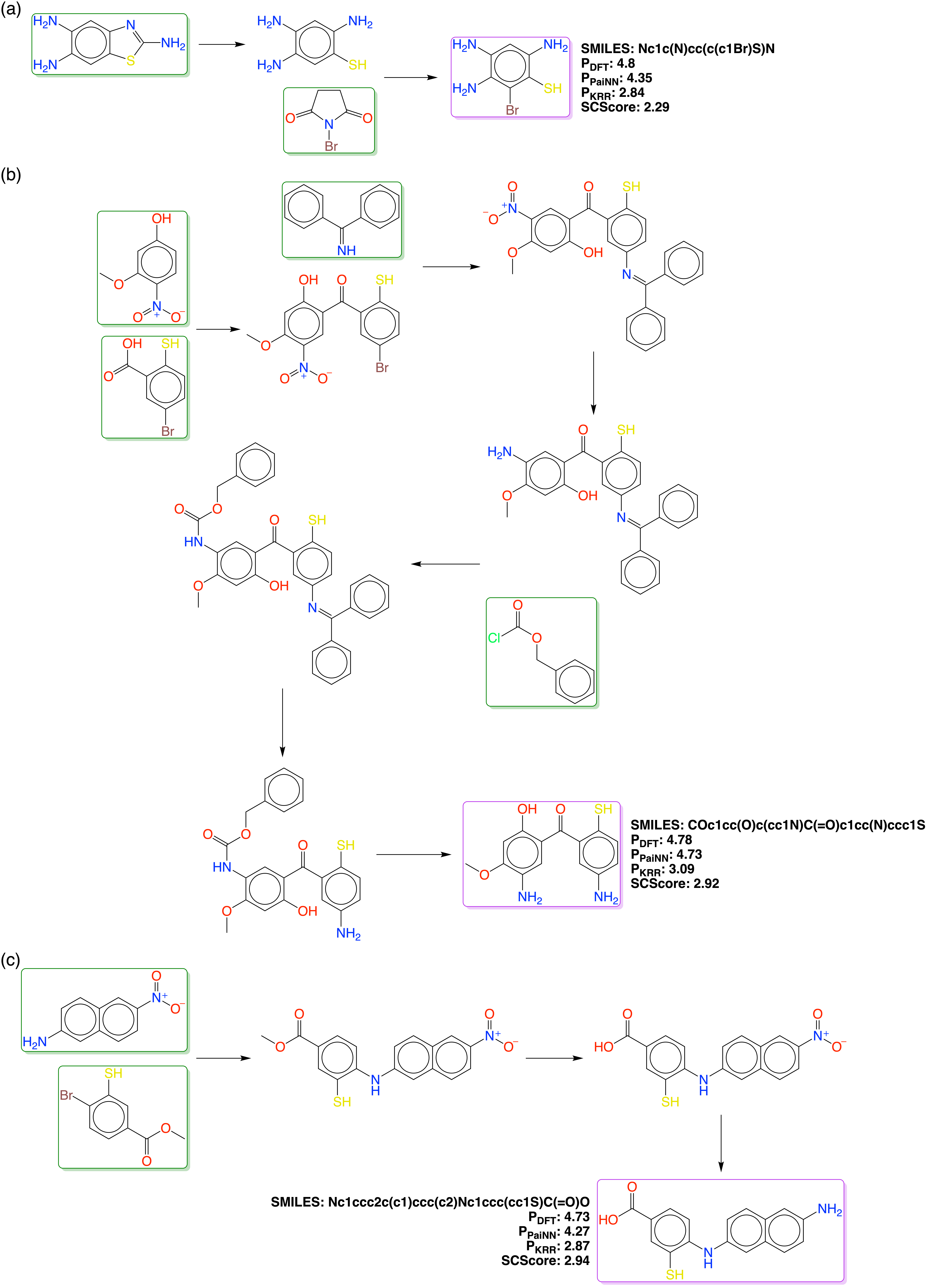}
    \caption{Top retrosynthetic paths for molecules with SMILES strings (a) Nc1c(N)cc(c(c1Br)S)N, (b) COc1cc(O)c(cc1N)C(=O)c1cc(N)ccc1S, and (c) Nc1ccc2c(c1)ccc(c2)Nc1ccc(cc1S)C(=O)O. The final molecules are shown in purple boxes and precursors available within the ZINC database are shown in green boxes.}
    \label{fig:Retrosynthesis_3}
\end{figure}

\begin{figure}[H]
    \centering
    \includegraphics[width=5.5in]{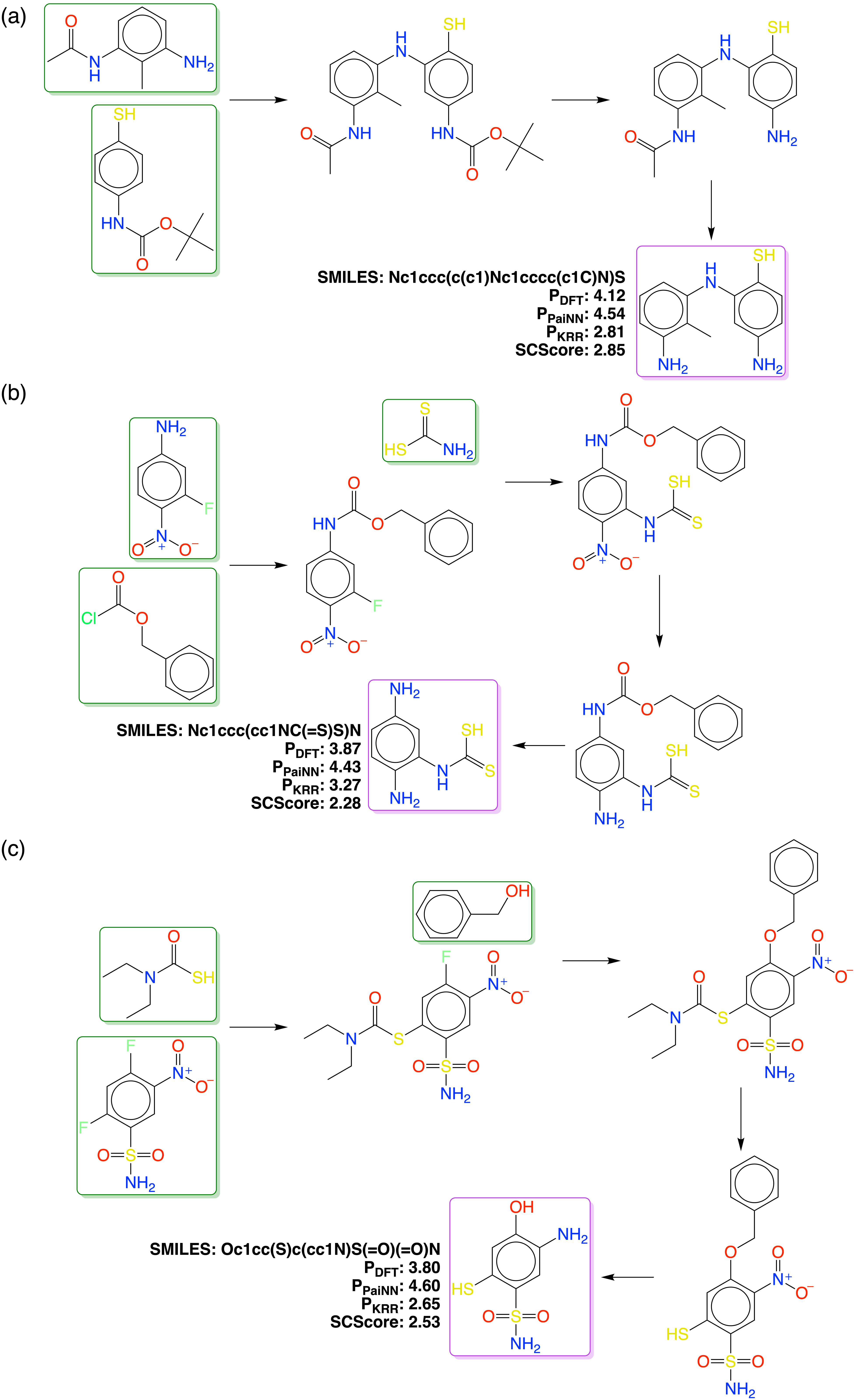}
    \caption{Top retrosynthetic paths for molecules with SMILES strings (a) Nc1ccc(c(c1)Nc1cccc(c1C)N)S, (b) Nc1ccc(cc1NC(=S)S)N, and (c) Oc1cc(S)c(cc1N)S(=O)(=O)N. The final molecules are shown in purple boxes and precursors available within the ZINC database are shown in green boxes.}
    \label{fig:Retrosynthesis_4}
\end{figure}

\begin{figure}[H]
    \centering
    \includegraphics[width=6in]{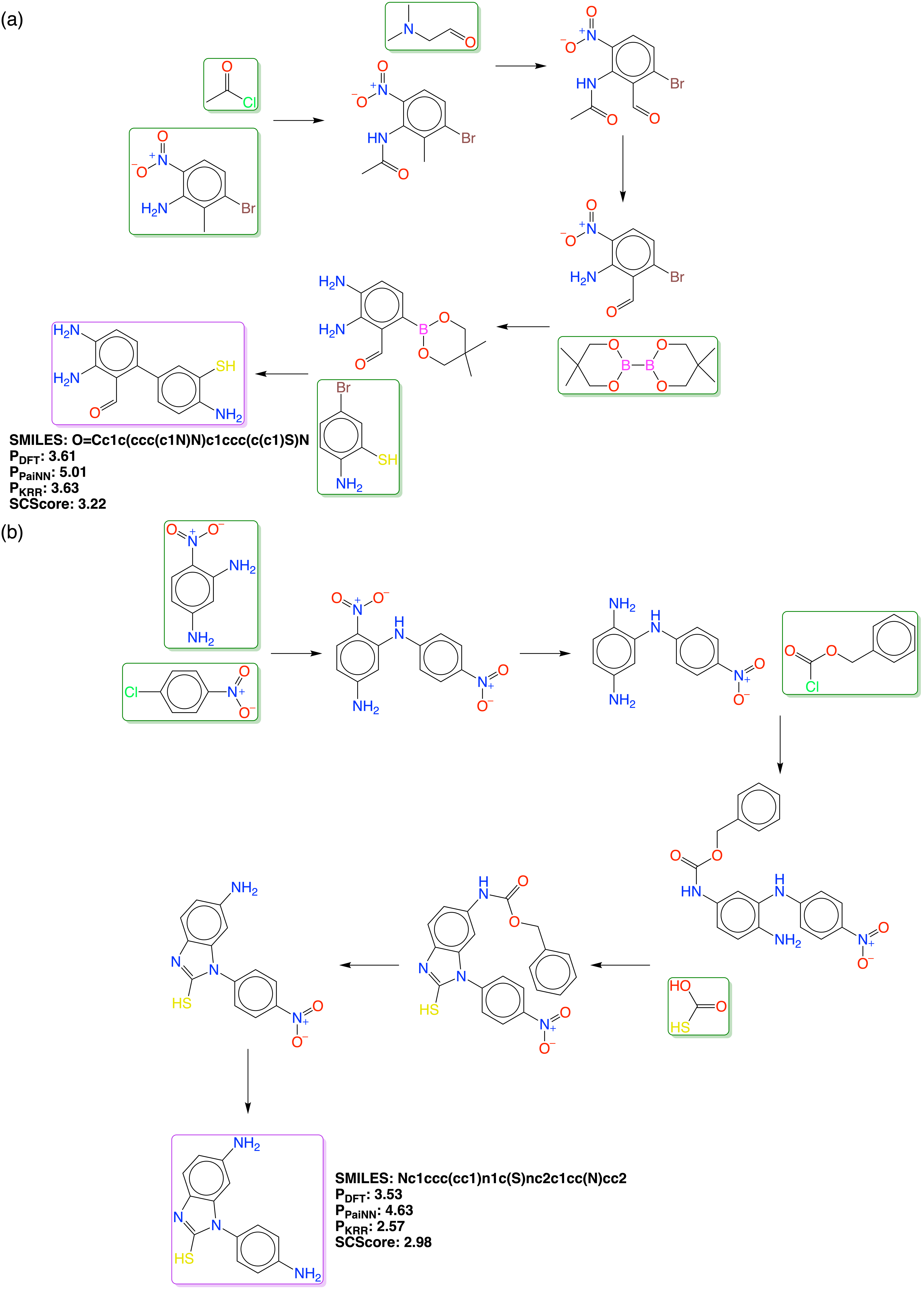}
    \caption{Top retrosynthetic paths for molecules with SMILES strings (a) O=Cc1c(ccc(c1N)N)c1ccc(c(c1)S)N and (b) Nc1ccc(cc1)n1c(S)nc2c1cc(N)cc2. The final molecules are shown in purple boxes and precursors available within the ZINC database are shown in green boxes.}
    \label{fig:Retrosynthesis_5}
\end{figure}

\begin{figure}[H]
    \centering
    \includegraphics[width=6.2in]{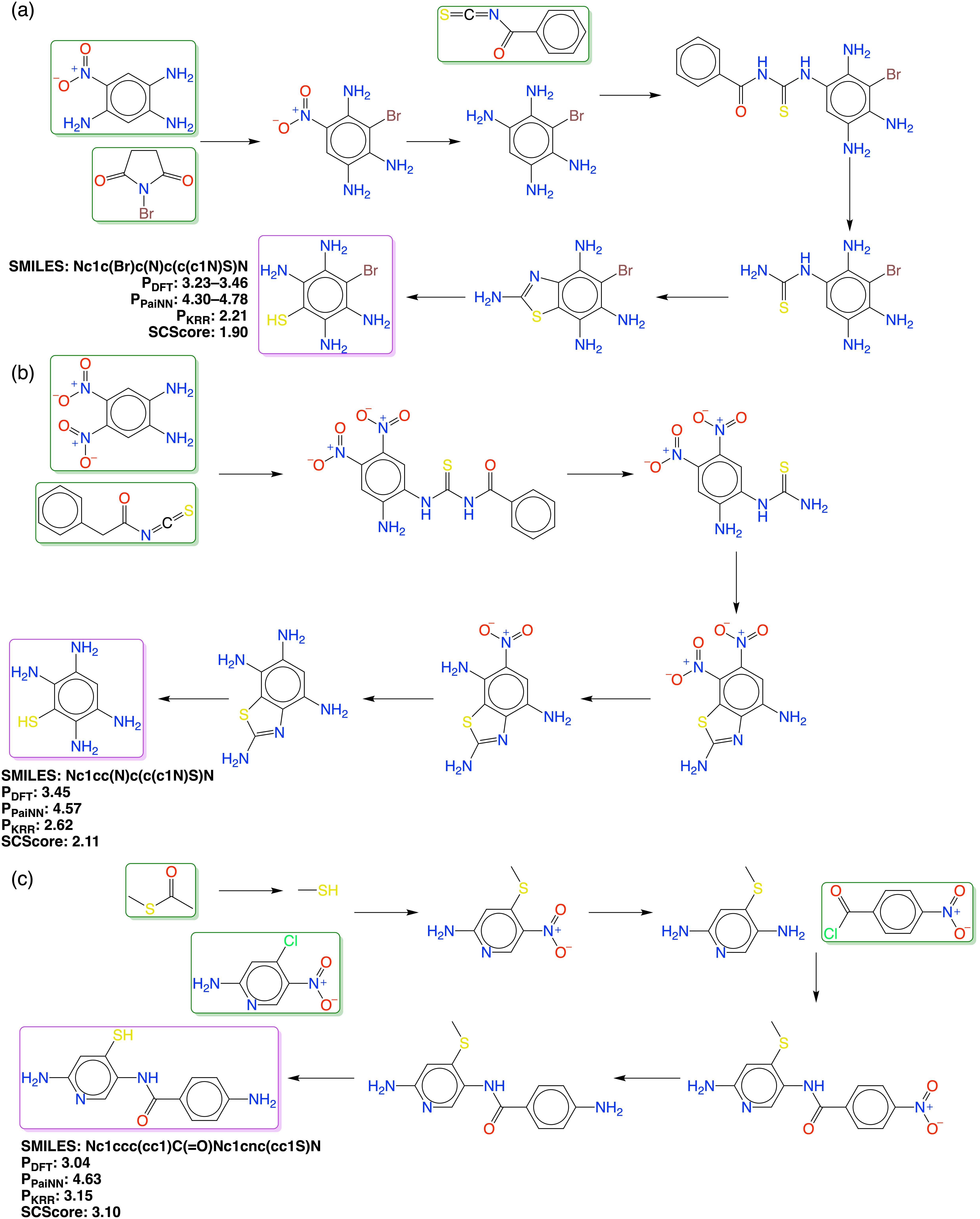}
    \caption{Top retrosynthetic paths for molecules with SMILES strings (a) Nc1c(Br)c(N)c(c(c1N)S)N, (b) Nc1cc(N)c(c(c1N)S)N, and (c) Nc1ccc(cc1)C(=O)Nc1cnc(cc1S)N. The final molecules are shown in purple boxes and precursors available within the ZINC database are shown in green boxes.}
    \label{fig:Retrosynthesis_6}
\end{figure}

\begin{figure}[H]
    \centering
    \includegraphics[width=6in]{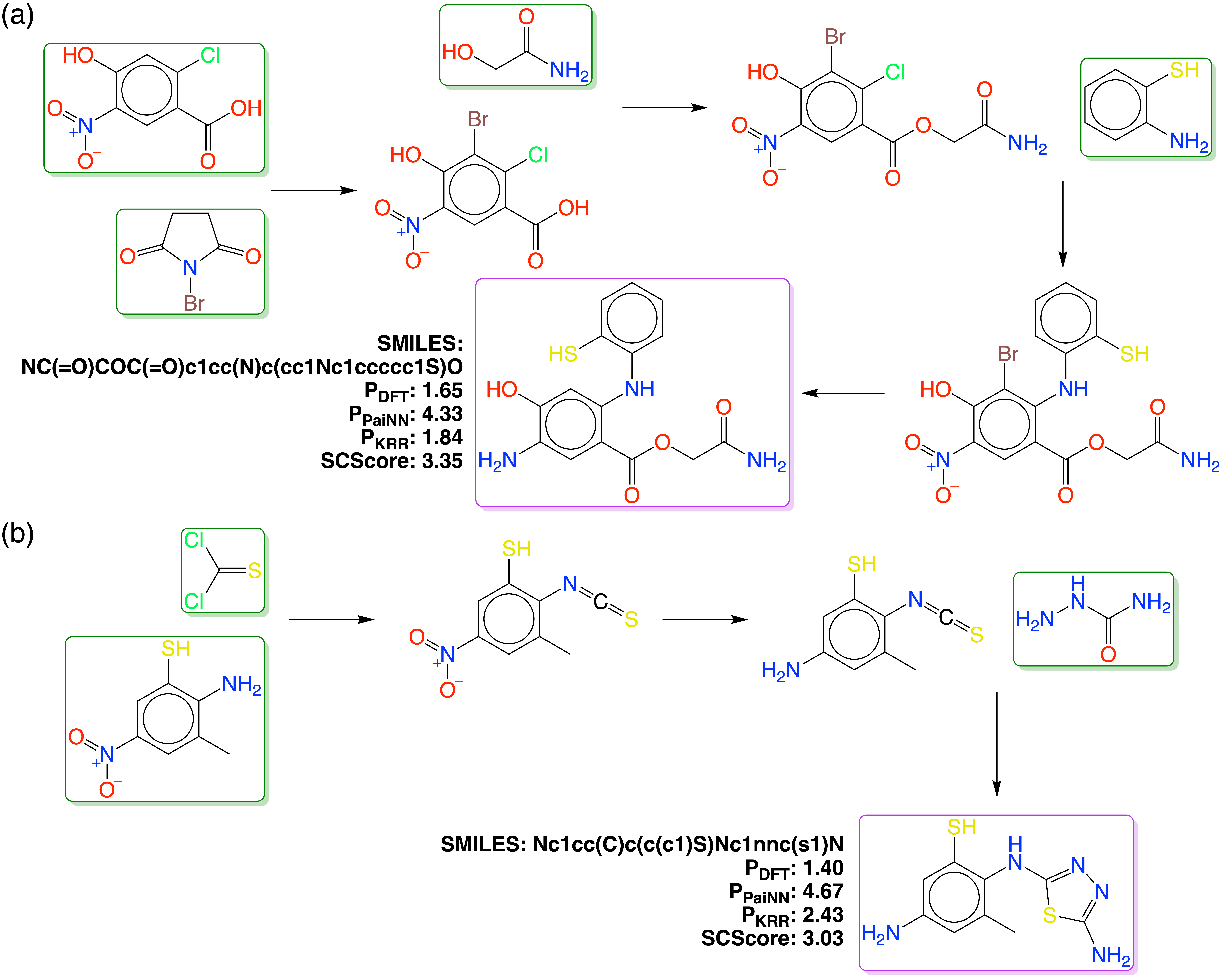}
    \caption{Top retrosynthetic paths for molecules with SMILES strings (a) NC(=O)COC(=O)c1cc(N)c(cc1Nc1ccccc1S)O and (b) Nc1cc(C)c(c(c1)S)Nc1nnc(s1)N. The final molecules are shown in purple boxes and precursors available within the ZINC database are shown in green boxes.}
    \label{fig:Retrosynthesis_7}
\end{figure}

Table~\ref{tab:Pmean≥4.25} details how many molecules satisfied the $\mathrm{P}_\mathrm{PaiNN}\geq4.25$ criterion for each database, as predicted from a committee of PaiNN models. Table~\ref{tab:best_mols} details the properties of the top generated molecules with solved retrosynthetic pathways using the AiZynthFinder~\cite{GenhedenJC20_AiZynthFinder} software and Figures~\ref{fig:Retrosynthesis_1}--\ref{fig:Retrosynthesis_7} show their top retrosynthesis paths based on a stock from compounds available within the ZINC~\cite{SterlingJCIM15_ZINC15} database and a policy~\cite{ThakkarCS20} trained on US patent office data~\cite{Lowe17}, as available within AiZynthFinder. \\

\begin{figure}[H]
    \centering
    \includegraphics[width=3in]{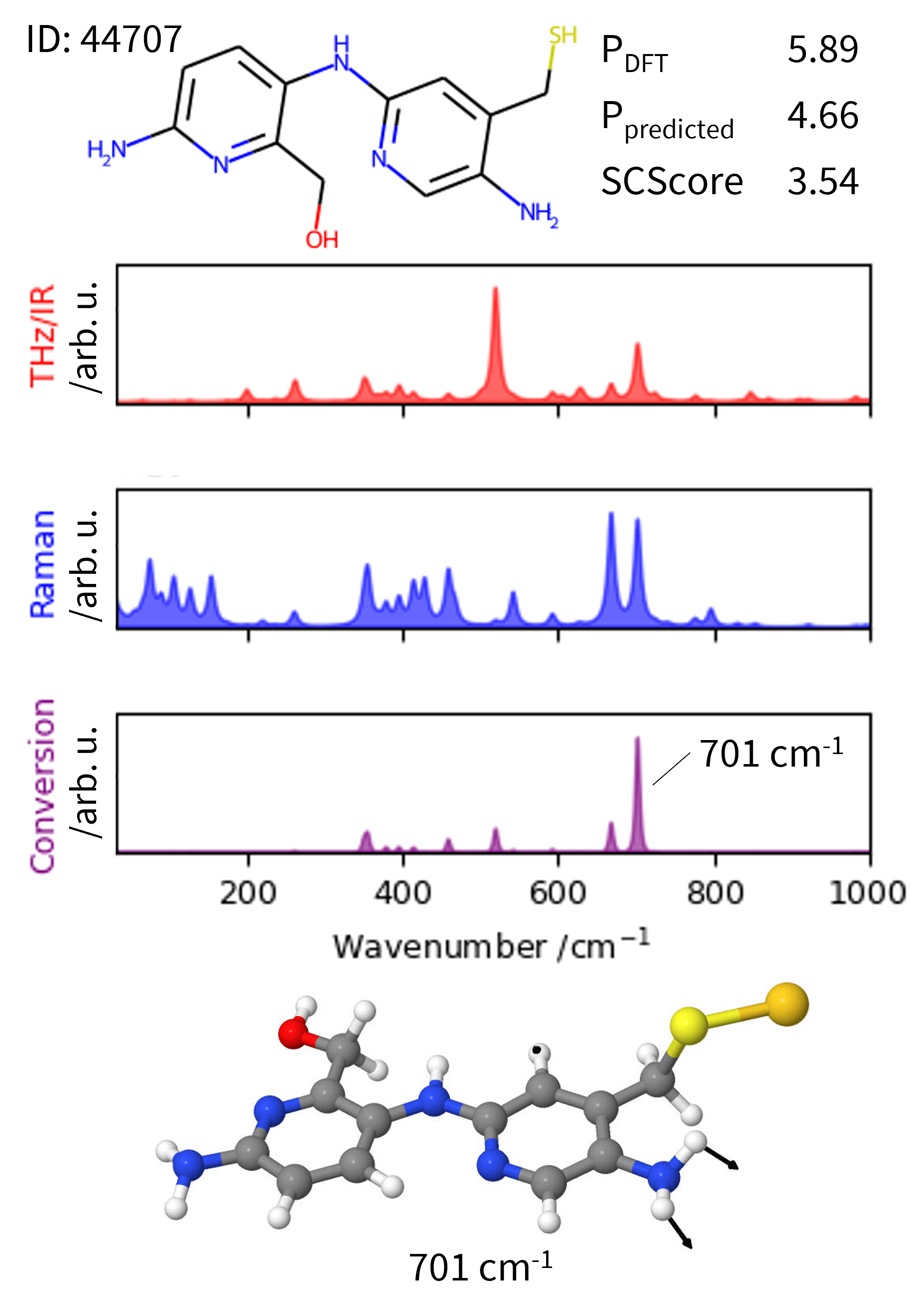}
    \caption{Vibrational spectra and properties of candidate molecule with ID 44707.}
    \label{fig:best_mols_1}
\end{figure}

\begin{figure}[H]
    \centering
    \includegraphics[width=6in]{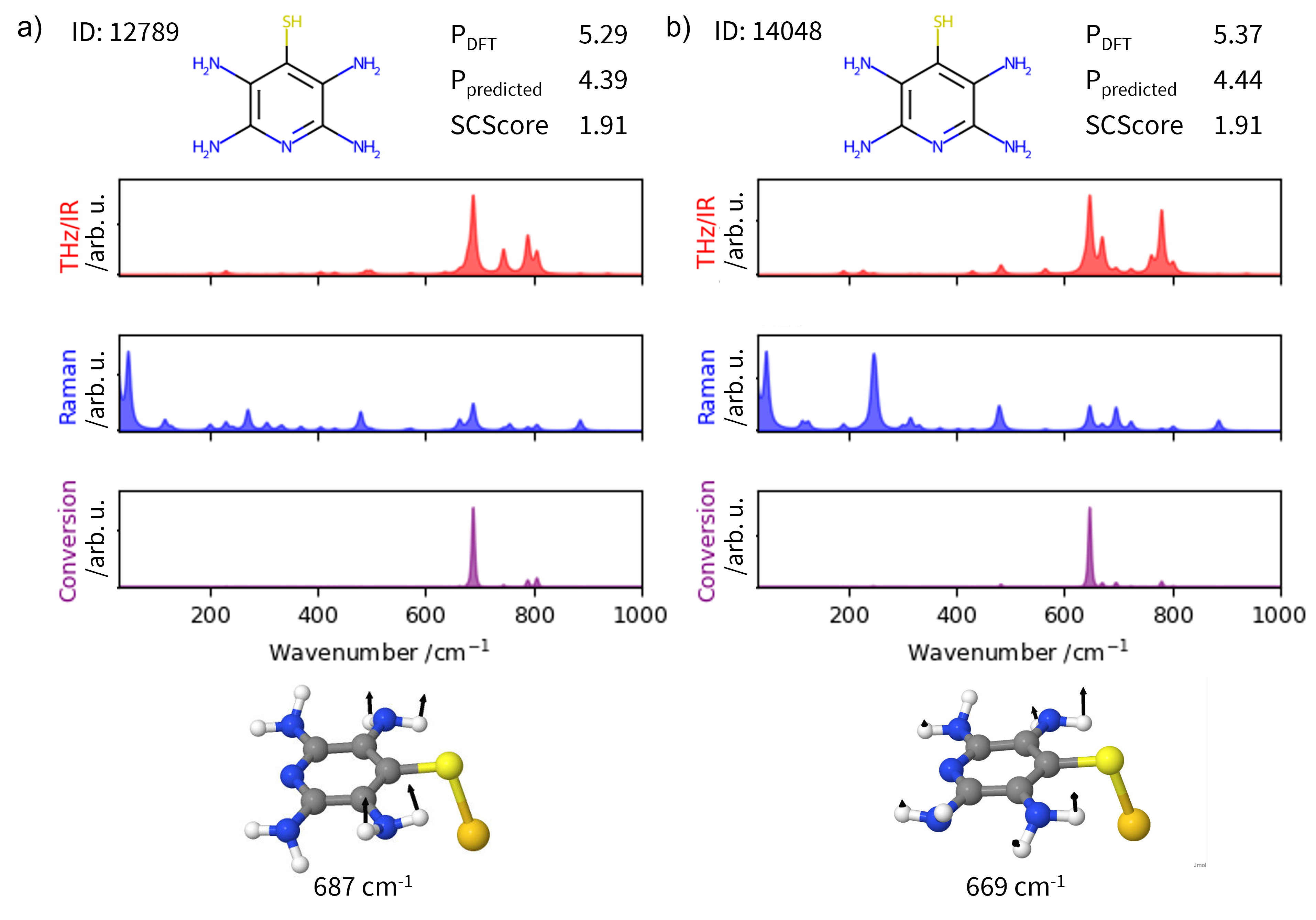}
    \caption{Vibrational spectra and properties of candidate molecules with IDs a) 12789 and b) 14048.}
    \label{fig:best_mols_2}
\end{figure}

\begin{figure}[H]
    \centering
    \includegraphics[width=3in]{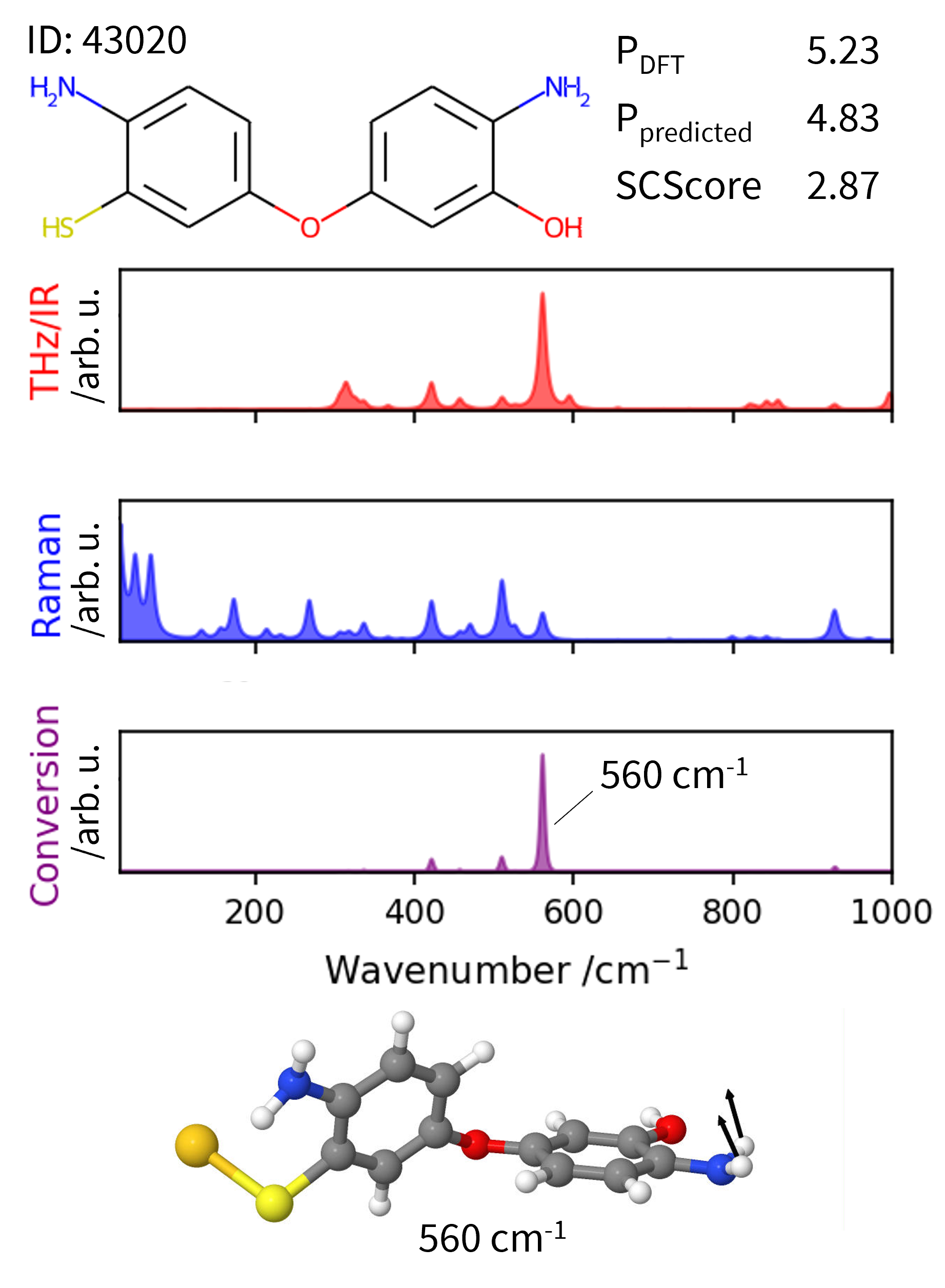}
    \caption{Vibrational spectra and properties of candidate molecule with ID 43020.}
    \label{fig:best_mols_3}
\end{figure}

\begin{figure}[H]
    \centering
    \includegraphics[width=3in]{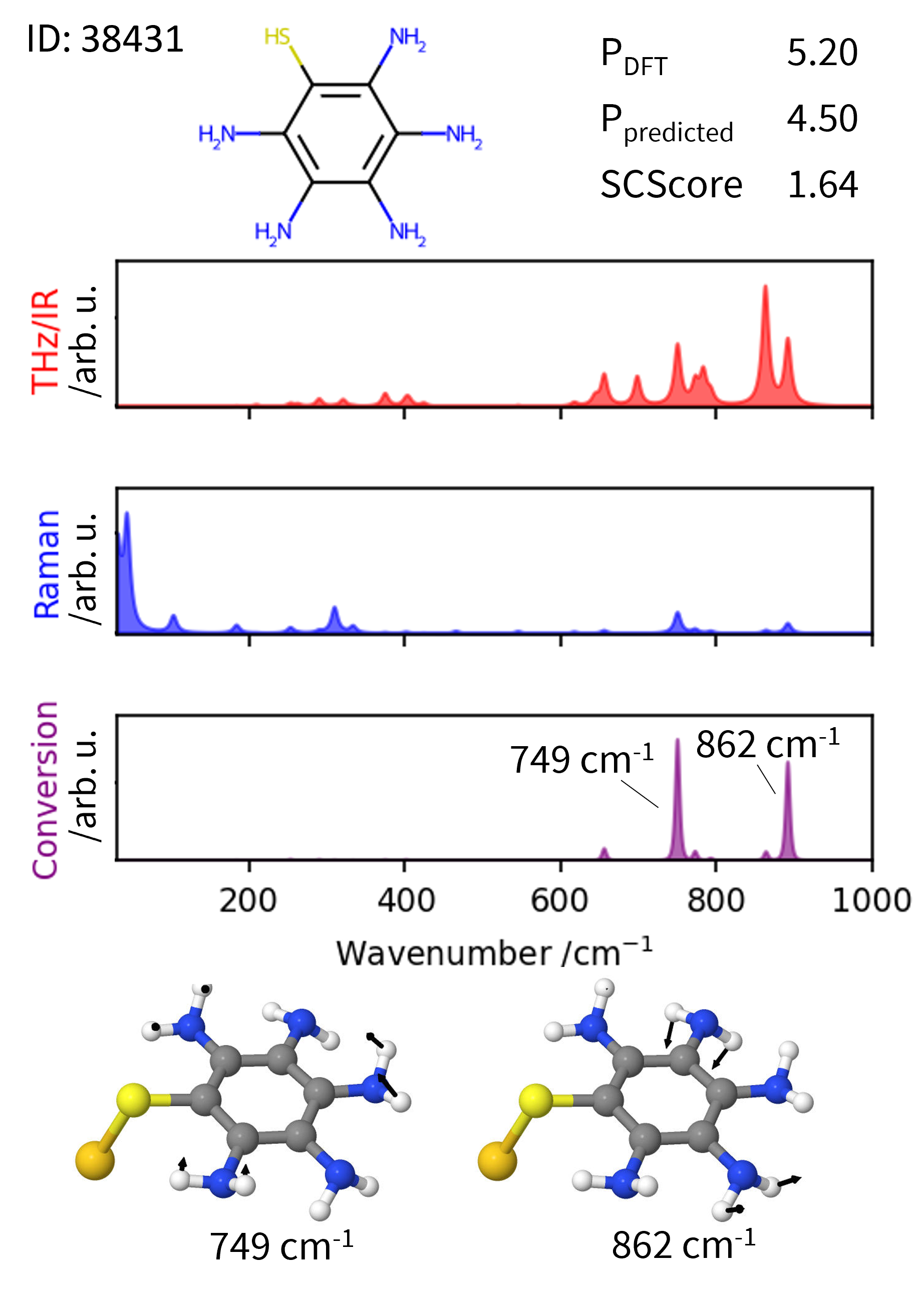}
    \caption{Vibrational spectra and properties of candidate molecule with ID 38431.}
    \label{fig:best_mols_4}
\end{figure}

Figures~\ref{fig:best_mols_1}--\ref{fig:best_mols_4} detail the vibrational spectra and properties of select candidate molecules, with their IDs taken from Table~\ref{tab:best_mols}.

\pagebreak
\begin{table}[H]
    \centering
    \begin{tabular}{c|c}
        Database & Number of Molecules \\ \hline
        Unbiased & 128 \\
        Bias 1 & 192 \\
        Bias 2 & 175 \\
        Bias 3 & 140 \\
        Bias 4 & 123 \\
        Bias 5 & 112 \\
        Bias 6 & 141 \\
    \end{tabular}
    \caption{Number of molecules within each database of generated molecules satisfying $\mathrm{P}_\mathrm{PaiNN}\geq4.25$.}
    \label{tab:Pmean≥4.25}
\end{table}

\begin{table}[H]
\centering
\begin{tabular}{l|ccccccl}
Molecule ID & Iteration & P$_\mathrm{DFT}$ & P$_\mathrm{PaiNN}$ & P$_\mathrm{KRR}$ & SCScore & Cluster & SMILES \\ \hline
27354         & 4 & 7.88 & 5.56   & 3.23  & 3.11 & C2  & Nc1ccc(cc1)Nc1cnc(cc1S)N              \\
44707         & 3 & 5.89 & 4.66   & 2.09  & 3.54 & C2  & OCc1nc(N)ccc1Nc1ncc(c(c1)CS)N         \\
14048         & 5 & 5.37 & 4.44   & 2.31  & 1.91 & C5  & Nc1nc(N)c(c(c1N)S)N                   \\
12789         & 6 & 5.29 & 4.39   & 2.31  & 1.91 & C5  & Nc1nc(N)c(c(c1N)S)N                   \\
13796         & 4 & 5.29 & 4.47   & 2.31  & 1.91 & C5  & Nc1nc(N)c(c(c1N)S)N                   \\
8518          & 2 & 5.29 & 4.36   & 2.31  & 1.91 & C5  & Nc1nc(N)c(c(c1N)S)N                   \\
43020         & 6 & 5.23 & 4.83   & 3.33  & 2.87 & C2  & Oc1cc(ccc1N)Oc1ccc(c(c1)S)N           \\
38431         & 4 & 5.2  & 4.5    & 2.07  & 1.64 & C5  & Nc1c(N)c(N)c(c(c1N)S)N                \\
34759         & 6 & 5.14 & 4.39   & 2.99  & 2.16 & C5  & Nc1cc(S)c(cc1N)N                      \\
9160          & 5 & 5.14 & 4.48   & 2.99  & 2.16 & C5  & Nc1cc(S)c(cc1N)N                      \\
18986         & 4 & 5.12 & 4.39   & 3.29  & 2.14 & C5  & SC(=S)Nc1ccc(c(c1)N)N                 \\
22423         & 6 & 4.99 & 4.35   & 2.07  & 1.64 & C5  & Nc1c(N)c(N)c(c(c1N)S)N                \\
26622         & 1 & 4.83 & 4.57   & 2.07  & 1.64 & C5  & Nc1c(N)c(N)c(c(c1N)S)N                \\
4030          & 2 & 4.8  & 4.35   & 2.84  & 2.29 & C5  & Nc1c(N)cc(c(c1Br)S)N                  \\
43713         & 5 & 4.78 & 4.73   & 3.09  & 2.92 & C2  & COc1cc(O)c(cc1N)C(=O)c1cc(N)ccc1S     \\
11249         & 3 & 4.77 & 4.46   & 2.31  & 1.91 & C5  & Nc1nc(N)c(c(c1N)S)N                   \\
4411          & 2 & 4.77 & 4.67   & 2.99  & 2.16 & C5  & Nc1cc(S)c(cc1N)N                      \\
23355         & 1 & 4.77 & 4.39   & 2.31  & 1.91 & C5  & Nc1nc(N)c(c(c1N)S)N                   \\
43114         & 5 & 4.73 & 4.27   & 2.87  & 2.94 & C2  & Nc1ccc2c(c1)ccc(c2)Nc1ccc(cc1S)C(=O)O \\
43881         & 4 & 4.12 & 4.54   & 2.81  & 2.85 & C2  & Nc1ccc(c(c1)Nc1cccc(c1C)N)S           \\
18261         & 6 & 3.87 & 4.43   & 3.27  & 2.28 & C5  & Nc1ccc(cc1NC(=S)S)N                   \\
23538         & 3 & 3.80 & 4.60   & 2.65  & 2.53 & C5  & Oc1cc(S)c(cc1N)S(=O)(=O)N             \\
31446         & 5 & 3.61 & 5.01   & 3.63  & 3.22 & C2  & O=Cc1c(ccc(c1N)N)c1ccc(c(c1)S)N       \\
27425         & 1 & 3.53 & 4.63   & 2.57  & 2.98 & C2  & Nc1ccc(cc1)n1c(S)nc2c1cc(N)cc2        \\
17848         & 6 & 3.46 & 4.35   & 2.21  & 1.90 & C5  & Nc1c(Br)c(N)c(c(c1N)S)N               \\
18617         & 4 & 3.46 & 4.78   & 2.21  & 1.90 & C5  & Nc1c(Br)c(N)c(c(c1N)S)N               \\
13331         & 2 & 3.45 & 4.57   & 2.62  & 2.11 & C5  & Nc1cc(N)c(c(c1N)S)N                   \\
25012         & 3 & 3.24 & 4.52   & 2.21  & 1.90 & C5  & Nc1c(Br)c(N)c(c(c1N)S)N               \\
12734         & 2 & 3.24 & 4.30   & 2.21  & 1.90 & C5  & Nc1c(Br)c(N)c(c(c1N)S)N               \\
19179         & 5 & 3.23 & 4.35   & 2.21  & 1.90 & C5  & Nc1c(Br)c(N)c(c(c1N)S)N               \\
31762         & 2 & 3.04 & 4.63   & 3.15  & 3.10 & C2  & Nc1ccc(cc1)C(=O)Nc1cnc(cc1S)N         \\
33480         & 3 & 2.85 & 4.36   & 2.07  & 1.64 & C5  & Nc1c(N)c(N)c(c(c1N)S)N                \\
45647         & 6 & 1.65 & 4.33   & 1.84  & 3.35 & C2  & NC(=O)COC(=O)c1cc(N)c(cc1Nc1ccccc1S)O \\
39452         & 3 & 1.40 & 4.67   & 2.43  & 3.03 & C2  & Nc1cc(C)c(c(c1)S)Nc1nnc(s1)N        
\end{tabular}
\caption{Properties of the top generated molecules with solved retrosynthetic pathways.}
\label{tab:best_mols}
\end{table}

\begin{figure}[H]
    \centering
    \includegraphics[width=7in]{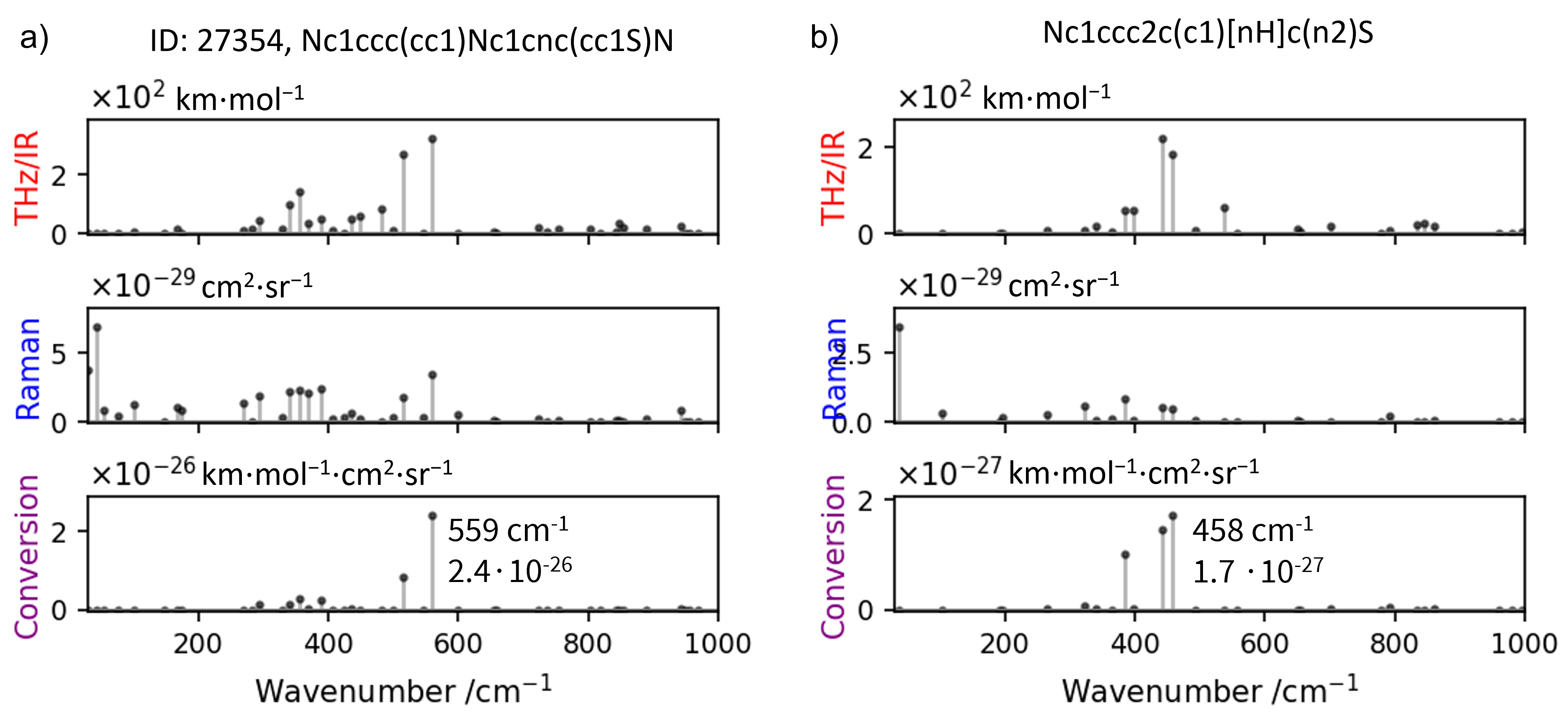}
    \caption{Comparison of vibrational properties between a) the top candidate molecule with ID 27354 and b) 5-Amino-2-mercaptobenzimidazole, available from Molecular Vibration Explorer \cite{Koczor-BendaJPCA22}. }
    \label{fig:5-A-2MBI}
\end{figure}

\section{Structural Validation of Generated Molecules}

Most 3D generative models of molecules have so far been trained on DFT-optimized structures~\cite{GebauerNeurIPS19_G-SchNet, GebauerNC22_cG-SchNet, WestermayrNCS23}. To assess the effect of using the semi-empirical xTB method instead of  DFT on the accuracy of the generated structures, subsequent xTB and DFT optimizations were performed on randomly selected molecules from the unbiased generation. As shown in Fig.~S22, the distribution of root-mean-square deviation (RMSD) values between generated and xTB-optimized structures is narrower and is centered at a smaller value than previously reported RMSD values between DFT-optimized structures and structures generated using G-SchNet (and trained on the OE62~\cite{StukeSD20_OE62} dataset).~\cite{WestermayrNCS23} This is likely due to the larger maximum size of generated molecules (100 atoms) set by \citet{WestermayrNCS23} compared to this work (60 atoms). The RMSD values of generated structures compared to DFT-optimized structures are somewhat larger than compared to xTB-optimized structures, as shown in Fig.~S23a, which is to be expected. Since the PaiNN predictor relies on 3D molecular structures, we are able to test how the structural relaxation of the generated molecules affects the property predictions. We compared predictions of the same PaiNN model on unrelaxed generated structures and the corresponding DFT-optimized structures for the randomly selected molecules from Iteration 6. As shown in Fig.~S23b, using the unrelaxed generated structures for P value prediction yields significant changes in the predicted values and would yield even further underestimation of high P values in the biasing workflow. 


\begin{figure}[H]
    \centering
    \includegraphics[width=4in]{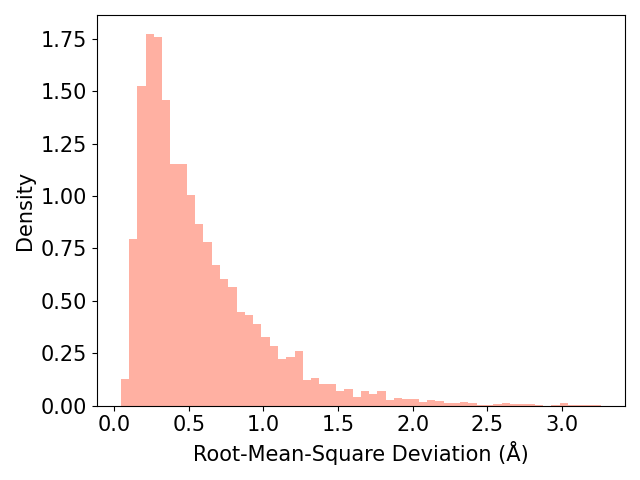}
    \caption{Distribution of the root-mean-square deviations of atomic positions between the molecules generated using an unbiased G-SchNet model and the same molecules subsequently optimized using the reference xTB method.}
    \label{fig:RMSD_Distribution}
\end{figure}

For unrelaxed generated structures, when exchanging the thiol-hydrogen to a gold atom, a fixed sulfur-gold bond length of \SI{2.88}{\angstrom} was applied, keeping all other internal coordinates unchanged from the generated structure.

\begin{figure}[H]
    \centering
    \includegraphics[width=5in]{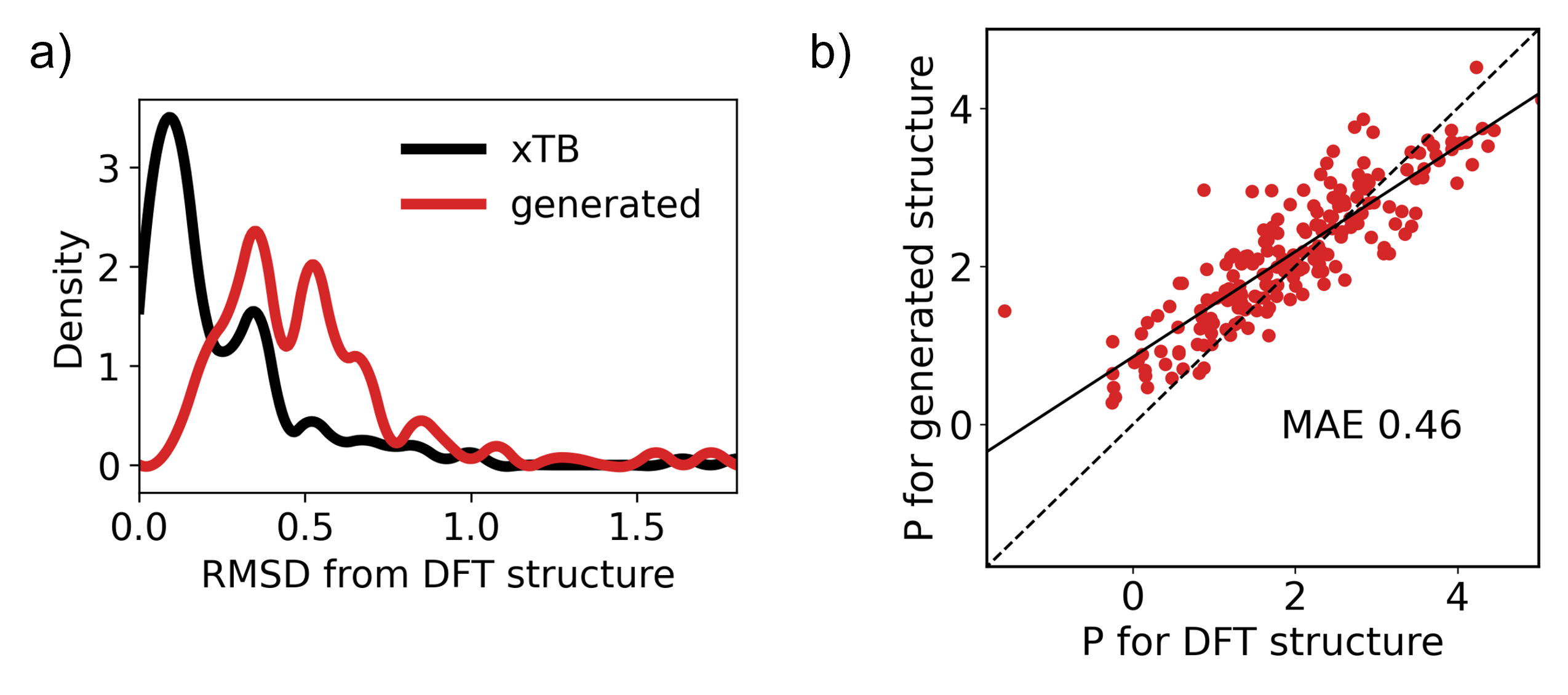}
    \caption{a) Distributions of the root-mean-square deviations (RMSDs) of atomic positions between DFT structures and the 3D structures generated using an unbiased G-SchNet model and the same molecules subsequently optimized using xTB. b) P values predicted by PaiNN based on the raw generated 3D structures versus DFT-optimized molecular structures.}
    \label{fig:RMSD_xTB_DFT}
\end{figure}

\pagebreak

\bibliographystyle{rsc}

\providecommand{\noopsort}[1]{}\providecommand{\singleletter}[1]{#1}%
\providecommand*{\mcitethebibliography}{\thebibliography}
\csname @ifundefined\endcsname{endmcitethebibliography}
{\let\endmcitethebibliography\endthebibliography}{}